\documentclass[11pt]{article}
\usepackage{amsmath}
\usepackage{mathrsfs}

\usepackage{epsfig,morefloats}
\usepackage{amsfonts}
\usepackage{natbib}
\usepackage{multirow}
\usepackage{enumitem}
\usepackage{amsthm}
\usepackage{amssymb}
\usepackage{color}
\usepackage{lscape}
\usepackage{rotating}
\usepackage{caption2}
\usepackage{subfigure}
\usepackage{graphicx}

\newcommand{\Date}[1]{\def\@Date{#1}}
\def\today{\number\day~\ifcase\month\or
 January\or February\or March\or April\or May\or June\or
 July\or August\or September\or October\or November\or December\fi~\number\year}

\def\be{\begin{equation}}
\def\ee{\end{equation}}
\def\bea{\begin{eqnarray}}
\def\eea{\end{eqnarray}}
\def\bd{\begin{displaymath}}
\def\ed{\end{displaymath}}
\def\bda{\begin{eqnarray*}}
\def\eda{\end{eqnarray*}}
\def\bsm{\begin{small}}
\def\esm{\end{small}}

\def\t0{\theta_0}

\def\nn{\nonumber}

\def\ha1{\hat \beta_1}

\def\bnt{\begin{enumerate}}
\def\ent{\end{enumerate}}
\def\T{{ \mathrm{\scriptscriptstyle T} }}

\def\AS{A\"{\i}t-Sahalia}

\def\bsc{\begin{scriptsize}}
\def\esc{\end{scriptsize}}

\newtheorem{theorem}{Theorem}

\newtheorem{lemma}{Lemma}
\newtheorem{cy}{Corollary}
\newtheorem{proposition}{Proposition}

\theoremstyle{definition}
\newtheorem{as}{Condition}
\newtheorem{ex}{Example}

\newtheorem{remark}{Remark}

\newcommand{\V}{\rm Var}

\makeatletter
\newcommand{\figcaption}{\def\@captype{figure}\caption}
\newcommand{\tabcaption}{\def\@captype{table}\caption}
\makeatother

\newcommand{\cov}{{\rm Cov}}
\newcommand{\diag}{{\rm diag}}

\newcommand{\supp}{\mbox{supp}}

\newcommand{\bA}{{\mathbf A}}

\newcommand{\bH}{{\mathbf H}}

\newcommand{\bL}{{\mathbf L}}

\newcommand{\bR}{{\mathbf R}}

\newcommand{\bV}{{\mathbf V}}
\newcommand{\bW}{{\mathbf W}}

\newcommand{\ba}{{\mathbf a}}

\newcommand{\bc}{{\mathbf c}}

\newcommand{\bg}{{\mathbf g}}

\newcommand{\bu}{{\mathbf u}}

\newcommand{\bx}{{\mathbf x}}
\newcommand{\by}{{\mathbf y}}
\newcommand{\bz}{{\mathbf z}}
\newcommand{\balpha} {\boldsymbol{\alpha}}

\newcommand{\bfeta}  {\boldsymbol{\eta}}

\newcommand{\bepsilonb}{\boldsymbol{\varepsilon}}
\newcommand{\bepsilonbb}{\boldsymbol{\epsilon}}
\newcommand{\bOmega}{\boldsymbol{\Omega}}

\newcommand{\bSigma}{\boldsymbol{\Sigma}}
\newcommand{\bDelta}{\boldsymbol{\Delta}}
\newcommand{\bgamma}{\boldsymbol{\gamma}}

\newcommand{\bvarsigma}{\mbox{\boldmath$\varsigma$}}

\newcommand{\btheta} {\boldsymbol{\theta}}
\newcommand{\bxi} {\boldsymbol{\xi}}
\newcommand{\bmu} {\boldsymbol{\mu}}

\newcommand{\bGamma} {\boldsymbol{\Gamma}}

\newcommand{\bD}{{\mathbf D}}
\newcommand{\bzero}{{\mathbf 0}}

\newcommand{\bUpsilon}{\boldsymbol{\Upsilon}}
\newcommand{\bPi}{\boldsymbol{\Pi}}
\newcommand{\bXi}{\boldsymbol{\Xi}}
\newcommand{\bchi}{\boldsymbol{\chi}}

\def\JRSSB{{\sl Journal of the Royal Statistical Society}, {\bf B}}

\def\BKA{{\sl Biometrika}}
\def\JASA{{\sl Journal of the American Statistical Association}}

\def\AS{{\sl The Annals of Statistics}}

\def\ET{{\sl Econometric Theory}}

\def\JE{{\sl Journal of Econometrics}}

\def\PTRF{{\sl Probability Theory and Related Fields}}

\def\ECA{{\sl Econometrica}}

\begin{document}

\normalsize

\title{\bf Confidence Regions for Entries of A Large Precision Matrix}

\author{Jinyuan Chang\\
School of Statistics\\
Southwestern University of Finance\\
 and Economics\\
Chengdu, Sichuan 611130, China\\
Email: changjinyuan@swufe.edu.cn \and Yumou Qiu\\ Department of Statistics\\
University of Nebraska-Lincoln\\ Lincoln, NE 68583-0963, USA\\ Email:
yumouqiu@unl.edu \and Qiwei Yao\footnote{Corresponding author. Tel.: +44
20 7955 6767.}\\ Department of Statistics\\
London School of Economics\\ London, WC2A 2AE, UK\\  E-mail: q.yao@lse.ac.uk \and Tao Zou\\ College of Business and Economics\\ The Australian National University\\ Acton, ACT 2601, Australia\\ E-mail: tao.zou@anu.edu.au}

\date{}
\maketitle

\begin{abstract}

We consider the statistical inference for high-dimensional precision matrices.
Specifically, we propose a data-driven procedure for constructing a class of
simultaneous confidence regions for a subset of the entries of a large
precision matrix.
The confidence regions can be applied to test for specific structures of a precision matrix, and to recover its nonzero components.
We first construct an estimator for the precision matrix via
penalized node-wise regression. We then develop the Gaussian approximation
to approximate the distribution of the maximum difference between
the estimated and the true precision coefficients. 
A computationally feasible parametric bootstrap algorithm is developed to
implement the proposed procedure.  The theoretical justification
is established under the setting which
allows temporal dependence among observations. Therefore the proposed procedure
is applicable to both independent and identically distributed data and time series data.
Numerical results with both simulated and real data
confirm the good performance of the proposed method.

\end{abstract}




\section{Introduction}

With an ever-increasing capacity of collecting and storing data, industry,
business and government offices all encounter the task of analyzing the
data of unprecedented size arisen from various practical fields such as
panel studies of economic, social and natural (such as weather)
phenomena, financial market analysis, genetic studies and communications
engineering. A significant feature of these data is that the number of
variables recorded on each individual is large or extremely large. Meanwhile,
in many empirical studies, observations taken over different
times are dependent with each other.
Therefore, many well-developed statistical inference methods for
independent and identically distributed (i.i.d.)\,data may no longer be
applicable. Those features of modern data bring both opportunities and challenges to statisticians and econometricians.

The entries of covariance matrix measure the marginal linear dependence of two
components of a random vector.
There is a large body of literature on
estimation and hypothesis testing of high-dimensional covariance matrices
with i.i.d.\,data, including \cite{Bj_2008a,Bj_2008b}, \cite{QC_2012},
\cite{CLX_2013}, \cite{ChangZhouZhouWang_2017} and references within.
In order to
capture the conditional dependence of two components of a random
vector conditionally
on all the others, the Gaussian graphical model (GGM) has been widely used.
Under GGM,  conditional independence of two components is equivalent to the fact that
the correspondent entry of the precision matrix (i.e. the inverse of the
covariance matrix) is zero.
Therefore, the conditional dependence among components of a random vector
can be well understood by investigating the structure of its
precision matrix. Beyond GGM, the bijection
relationship between the conditional dependence and the precision
matrix may not hold. Nevertheless, the precision matrix still plays an
important role in, among others,  linear regression
\citep{vandeGeeret_2014}, linear prediction and kriging, and partial correlation graphs \citep{Huang_2010}. See
also Examples 1--3 in Section 2 below.

Let $\bOmega$ denote a $p\times p$ precision matrix and $p$ be large.
With i.i.d.\,observations, \cite{YuanLin_2007} and \cite{Friedman_2008} adopted
graphical Lasso to estimate $\bOmega$ by maximizing the likelihood with an
$L_1$ penalty.
\cite{MB_2006} introduced a neighborhood selection procedure which
estimates $\bOmega$ by finding the nonzero regression coefficients of
each component on all the other components using Lasso \citep{Tibshirani_1996}
or Dantzig method \citep{CandesTao_2007}. Also see \cite{CLj_2011},
\cite{XueZou_2012} and \cite{SZ_2013} for other penalized estimation methods.
\cite{CXW_2013} investigated the
theoretical properties of the graphical Lasso estimator for $\bOmega$ with dependent
observations. Though these methods provide consistent estimators for
$\bOmega$ under some structural assumptions (for example, sparsity) imposed on $\bOmega$,  they cannot be used
for statistical inference directly 
due to the non-negligible estimation biases, caused
by the penalization,  which are of order slower than $n^{-1/2}$.

The bias issue has been successfully overcome with i.i.d.\,Gaussian observations by,
for example,
\cite{Liu_2013} based on $p$ node-wise regressions method.
Furthermore, \cite{Ren_2015} proposed a novel estimator for each entry of $\bOmega$ based on pairwise $L_1$ penalized regression,
and showed that their estimators achieved the minimax optimal  rate with
no bias terms.
In spite of $\frac{p(p-1)}{2}$ pairs among $p$ components, their method in practice
only requires at most $p(1+\bar{s})$ pairwise $L_1$ penalized
regressions, where $\bar{s}$ is the average size of the selected
node-wise regression models.

The major contribution of this paper is to construct the confidence regions
for subsets of the entries of $\bOmega$. To our best knowledge, this is the first
attempt of this kind. Furthermore we provide the asymptotic justification under
the setting which allows
dependent observations, and, hence, includes i.i.d.\,data as a special
case. See also Remark 2 in Section 3.2 below.
More precisely, let $\mathcal{S} \subset \{ 1,\ldots,p\}^2$ be a given
index set of interest, whose cardinality $|\mathcal{S}|$ can be finite or
grow with $p$.
Let $\bOmega_{\mathcal{S}}$ be the vector consisting of the entries of $\bOmega$ with their indices
in $\mathcal{S}$. We propose a class of data-driven confidence regions
$\{\mathcal{C}_{\mathcal{S},\alpha}\}_{0<\alpha<1}$ for $\bOmega_{\mathcal{S}}$
such that $\sup_{0<\alpha<1}|\mathbb{P}(\bOmega_{\mathcal{S}}\in\mathcal{C}_{\mathcal{S},\alpha})-\alpha|\rightarrow0$ when both $n, p \to \infty$, where $n$ denotes the
sample size. The potential application of $\mathcal{C}_{\mathcal{S},\alpha}$ is wide,
including, for example, testing for some specific structures of
$\bOmega$, and detecting and recovering
nonzero entries of $\bOmega$ consistently.

For any matrix $\bA=(a_{ij})$, let $|\bA|_\infty =
\max_{i,j}|a_{ij}|$ be its element-wise $L_\infty$-norm.
We proceed as follows.
 First we propose a bias corrected estimator
$\widehat{\bOmega}_{\mathcal{S}}$ for $\bOmega_{\mathcal{S}}$ via
penalized node-wise regressions, and develop an asymptotic expansion
for $n^{1/2}(\widehat{\bOmega}_{\mathcal{S}}-\bOmega_{\mathcal{S}})$
without assuming Gaussianity.
As the leading term in the asymptotic expansion is a partial sum,
we approximate the distribution of
$n^{1/2}|\widehat{\bOmega}_{\mathcal{S}}-\bOmega_{\mathcal{S}}|_\infty$
by that of the $L_\infty$-norm of a high-dimensional normal distributed random vector with mean zero and covariance
being an estimated long-run covariance matrix of an unobservable process.
This normal approximation, inspired by \cite{CCK_2013,CCK_2014},
paves the way for evaluating the probabilistic behavior of
$n^{1/2}|\widehat{\bOmega}_{\mathcal{S}}-\bOmega_{\mathcal{S}}|_\infty$
by parametric bootstrap.

It is worth pointing out that the kernel  estimator
for long-run covariances, initially proposed by \cite{Andrews_1991}
for the problem with fixed dimension (i.e.~$p$ fixed), also works under
our setting with $p\to \infty$ without requiring any structural assumptions on the
underlying long-run covariance matrix. Owning to the form of this kernel
estimator, the parametric bootstrap sampling can be implemented in an efficient
manner in terms of both computational complexity and the required storage
space; see Remark 4 in Section 3.2 below.


The rest of the paper is organized as follows. Section
\ref{se:background} introduces the problem to be solved and its
background. The proposed procedure and its theoretical properties are
presented in Section \ref{se:GA}. Section \ref{se:app} discusses the
applications of our results. Simulation studies and a real data analysis
are reported in Sections \ref{se:simulation} and \ref{se:case},
respectively. All the technical proofs are relegated to the Appendix.
We conclude this section by introducing some notation that is used throughout the paper.
We write $a_{n} \asymp b_{n}$
to mean  $0 < \liminf_{n\rightarrow\infty}|a_{n} / b_{n}| \leq\limsup_{n\rightarrow\infty}|a_n/b_n| < \infty$. We say $x_{n,j}=o_p(a_n)$ uniformly over $j\in\mathcal{J}$
if $\max_{j\in\mathcal{J}} |x_{n,j}/a_n| \xrightarrow{p} 0$ as $n\to\infty$. Let $|\cdot|_1$ and
$|\cdot|_0$ denote, respectively, the $L_1$- and $L_0$-norm of a vector.


\section{Preliminaries}\label{se:background}

Let 
$ \by_1,\ldots,\by_n$ be $n$ observations
from an $\mathbb{R}^{p}$-valued time series, where $\by_t = (y_{1, t}, \ldots, y_{p, t})^{\T}$ and each $\by_t$ has the constant first two moments, i.e. $\mathbb{E}(\by_t)=\bmu$ and $\cov(\by_t)=\bSigma$ for each $t$.
Let $\bOmega=\bSigma^{-1}$ be the precision matrix.
We assume that  $\{\by_t\}$ is
$\beta$-mixing in the sense that $\beta_k\rightarrow0$ as $k\rightarrow\infty$, where
\[
\beta_k=\sup_t\mathbb{E}\bigg\{\sup_{B\in\mathscr{F}_{t+k}^{\infty}}\big|\mathbb{P}(B|\mathscr{F}_{-\infty}^t)-\mathbb{P}(B)\big|\bigg\}.
\]
Here $\mathscr{F}_{-\infty}^t$ and $\mathscr{F}_{t+k}^{\infty}$ are the
$\sigma$-fields generated respectively by $\{\by_{u}\}_{u\leq t}$ and
$\{\by_u\}_{u\geq t+k}$. $\beta$-mixing is a mild condition
for time series. It is known that causal ARMA
processes with continuous innovation distributions, stationary Markov
chains under some mild conditions and stationary GARCH models with finite
second moments and continuous innovation distributions are all
$\beta$-mixing. We refer to Section 2.6 of \cite{FanYao_2003}
for the further details on $\beta$-mixing condition.

For a given index set $\mathcal{S}\subset \{1,\ldots,p\}^2$, recall
$\bOmega_{\mathcal{S}}$ denotes the vector consisting of the entries
of $\bOmega$ with their indices in $\mathcal{S}$.
 We are interested in constructing a class of confidence regions
$\{\mathcal{C}_{\mathcal{S},\alpha}\}_{0<\alpha<1}$ for  $\bOmega_{\mathcal{S}}$ such that
\begin{equation}\label{eq:cr1}
\sup_{0<\alpha<1}\big|\mathbb{P}(\bOmega_{\mathcal{S}}\in
\mathcal{C}_{\mathcal{S},\alpha})-\alpha\big|\rightarrow0~~~\textrm{as}~~n,
p \rightarrow\infty.
\end{equation}
We also allow $r\equiv |\mathcal{S}|$, the length of vector $\bOmega_{\mathcal{S}}$,
either to be fixed or to go to infinity together with $p$. The largest $r$ can be $p^2$.
We first give several motivating examples.

\begin{ex}
(High-dimensional linear regression)
Consider linear regression $z_{t} = \bx_{t}^\T\bgamma + \varepsilon_{t}$
with $\mathbb{E}(\bx_{t} \varepsilon_{t}) = \bzero$,
where $\bx_t $ 
consists of $m$ explanatory variables and $m$ is large, and
$\bgamma = (\gamma_{1}, \ldots, \gamma_{m})^{\T} =
\{\mathbb{E}(\bx_{t} \bx_{t}^{\T})\}^{-1}\mathbb{E}(\bx_{t} z_{t})$
are true regression coefficients. In order to
identify non-zero regression coefficients,
we test the hypotheses \be H_{0}: \gamma_{l} = 0 \mbox{ \ for all $l  \in
\mathcal{A}$~~~~~vs.~~~~~} H_{1}: \gamma_{l} \neq 0 \mbox{ \ for some $l
\in \mathcal{A}$}, \label{eq:testHDR}\ee
where $\mathcal{A} \subset \{1, \ldots, m\}$ is a given index set of interest.
Let $\by_{t} = (z_{t}, \bx_{t}^{\T})^{\T}$, and $\bOmega=(\omega_{j_1,j_2})_{p\times p}$
be the
precision matrix of $\by_t$. It can be shown that $(\omega_{1, 2},
\ldots, \omega_{1, p})^{\T} = -c \bgamma$, where $c = [ {\V}(z_t) -
\mathbb{E}(\bx^{\T}_tz_t)  \{\mathbb{E}(\bx_t \bx^{\T}_t)\}^{-1}
\mathbb{E}(\bx_t z_t) ]^{-1} > 0$. Thus,  (\ref{eq:testHDR}) can
be equivalently expressed as
\be H_{0}: \omega_{1, l} = 0 \mbox{ \ for all  $l \in
\mathcal{S}$~~~~~vs.~~~~~} H_{1}: \omega_{1, l} \neq 0 \mbox{ \ for some
$l \in \mathcal{S}$} \label{eq:testHDR1},\ee
where $\mathcal{S}=\{(1,l):l-1\in\mathcal{A}\}$. We reject $H_0$ at the significance
level $\alpha$ if  $
\mathcal{C}_{\mathcal{S},\alpha}$ does not contain the origin of $\mathbb{R}^r$
with $r= |\mathcal{A}|$.
\end{ex}

\begin{ex}
(Linear prediction and kriging)
In the context of predicting a random variable $z_t$ based on an observed $p$-dimensional vector
$\bx_t$, the best linear predictor in the sense of minimizing the mean squared
predictive error is $\cov(z_t, \bx_t) \bOmega \bx_t$, where $\bOmega$ is the precision
matrix of $\bx_t$. Here we assume the means of both $z_t$ and $\bx_t$ are zero, to
simplify the notation. We also assume that any redundant components of $\bx_t$
have been removed by applying the techniques described in Example 1 above.

To obtain a consistent estimate for $\bOmega$ when $p$ is large,
it is necessary to impose some structural assumptions on $\bOmega$. In the context
of kriging (i.e. linear prediction in the context of spatial or spatial-temporal
statistics), some lower-dimensional factor structures have been explored.
See \cite{HYZ_2017} and the references within.
Bandness/bandableness is another popular structural assumption often used in
estimating large covariance or precision matrices \citep{Bj_2008a}.
To investigate a banded structure for $\bOmega$, one may
test the hypotheses
\be H_{0}: \omega_{j_1,j_2} = 0 \mbox{ \ for any $| j_1-j_2 | > k$~~~~~vs.~~~~~} H_{1}: \omega_{j_1,j_2} \neq 0 \mbox{ \ for some $| j_1-j_2 | > k$}, \label{eq:banded}\ee
where $1 \le k< p$ is a prespecified integer. We reject $H_0$ if confidence
region $
\mathcal{C}_{\mathcal{S},\alpha}$ does not contain the origin $\mathbb{R}^r$, where
$\mathcal{S} = \{ (j_1,j_2): 1\le j_1, j_2\le p, \; j_2-j_1>k \}$ and $r=(p-k)(p-k-1)/2$.
\end{ex}

\begin{ex}
(Partial correlation network) Given a precision matrix
$\bOmega=(\omega_{j_1,j_2})_{p\times p}$, we can define an undirected
network $G=(V,E)$ where the vertex set $V=\{1,\ldots,p\}$ represents the
$p$ components of $\by$ and the edge set $E=\{(j_1,j_2)\in V\times
V:\omega_{j_1,j_2}\neq0, \; j_1 < j_2\}$ are the pairs of variables with non-zero
precision coefficients.  Let $\rho_{j_1,j_2} =
\mbox{Corr}(\varepsilon_{j_1},\varepsilon_{j_2})$ be the partial
correlation between the $j_1$-th and the $j_2$-th components of $\by$ for
any $j_1\neq j_2$, where $\varepsilon_{j_1}$ and $\varepsilon_{j_2}$ are
the errors of the best linear predictors of $y_{j_1}$ and $y_{j_2}$ given
$\by_{-(j_1,j_2)}=\{y_k: k\neq j_1,j_2 \}$, respectively.
From Lemma 1 of \cite{PengWangZhouZhu_2009},
it is known that
$\rho_{j_1,j_2}=-\frac{\omega_{j_1,j_2}}{\sqrt{\omega_{j_1,j_1}\omega_{j_2,j_2}}}$.
Therefore, the network $G = (V, E)$ also represents the partial
correlation graph of $\by$. The vertices $(j_1, j_2) \not\in E$ if and
only if $y_{j_1}$ and $y_{j_2}$ are partially uncorrelated.
The GGM assumes in addition that
$\by$ is multivariate normal. Then $\bOmega$ depicts the
conditional dependence among the $p$ vertices of the network,
i.e. $\omega_{j_1,j_2}$ is the conditional correlation between the $j_1$-th and
$j_2$-th vertices given all the others.

Neighborhood and community are two basic features in a network. The neighborhood of the $j$-th vertex, denoted by $\mathcal{N}_{j}$, is the set of all the vertices directly connected to it.
For most of the spatial data, it is believed that the partial correlation neighborhood is related to the spatial neighborhood.
Let ${\mathcal{N}}_{j}(k)$ be the set including the first $k$ closest vertices to the $j$-th vertex in the spatial domain.
It is of great interest to test $H_0:\mathcal{N}_j=\mathcal{N}_j(k)$ vs. $H_{1}: \mathcal{N}_j \neq \mathcal{N}_j(k)$ for some pre-specified positive constant $k$.
A community in a network is a group of vertices that have heavier connectivity within the group than outside the group. 
For graph estimation, we want to maximize the within-community connectivity and reduce the between-community connectivity.
Therefore, it is of practical importance to explore the connectivity between different communities. Assume the $p$ components of $\by$ are decomposed into $K$ disjoint communities $V_{1}, \ldots, V_{K}$. We are interested in recovering $\mathcal {D} = \{ (k_{1}, k_{2}) : \omega_{j_1,j_2} \neq 0 \mbox{ \ for some $j_1 \in V_{k_{1}}$ and $j_2 \in V_{k_{2}}$} \}$.
\end{ex}

\section{Main results}\label{se:GA}

\subsection{Estimation of $\bOmega$}

We first recall the relationship between a precision matrix and node-wise regressions. For a random vector $\by = (y_{1}, \ldots, y_{p})^\T$ with mean $\bmu=\bzero$ and covariance $\bSigma$, we consider $p$ node-wise regressions
\begin{equation}\label{eq:regression}
y_{j_1} = \sum_{j_2 \neq j_1}\alpha_{j_1,j_2}y_{j_2} + \epsilon_{j_1} ~~~ (j_1=1,\ldots,p).
\end{equation}
Let $\by_{-j_1} = \{y_{j_2} : j_2 \neq j_1\}$.
The regression error $\epsilon_{j_1}$ is uncorrelated with $\by_{-j_1}$
if and only if $\alpha_{j_1,j_2} =
-\frac{\omega_{j_1,j_2}}{\omega_{j_1,j_1}}$ for any $j_2 \neq j_1$. Under this
condition,
 $ {\cov}(\epsilon_{j_1}, \epsilon_{j_2}) =
\frac{\omega_{j_1,j_2}}{\omega_{j_1,j_1}\omega_{j_2,j_2}}$ for any $j_1$
and $j_2$. Let $\bepsilonbb = (\epsilon_{1}, \ldots, \epsilon_{p})^{\T}$
and $\bV = \textrm{Cov}(\bepsilonbb)= (v_{j_1,j_2})_{p\times p}$. Then
$\bOmega = \{\textrm{diag}(\bV)\}^{-1}\bV\{\textrm{diag}(\bV)\}^{-1}$; see Lemma 1
of \cite{PengWangZhouZhu_2009}.
This relationship between $\bOmega$ and $\bV$ provides a way to learn $\bOmega$ by the regression errors in (\ref{eq:regression}). 


Since the error vector $\bepsilonbb$ in (\ref{eq:regression}) is unobservable in practice, its ``proxy'' -- the residuals of the node-wise regressions -- can be used to estimate $\bV$.
Let
$\balpha_j=(\alpha_{j,1},\ldots,\alpha_{j,j-1},-1,\alpha_{j,j+1},
\ldots,\alpha_{j,p})^\T$. For each $j=1,\ldots,p$, we may fit  the high-dimensional
linear regression
 \begin{equation}\label{eq:regressionData}
 y_{j,t} = \sum_{k \neq j}\alpha_{j,k} y_{k,t} + \epsilon_{j,t}~~~ (t = 1, \ldots, n)
 \end{equation}
by Lasso \citep{Tibshirani_1996}, Dantzig estimation \citep{CandesTao_2007} or scaled Lasso \citep{SZ_2012}.
For the case $\bmu \neq \bzero$, the regression (\ref{eq:regressionData}) will be conducted on the centered data $\by_{t} - \bar{\by}$, where $\bar{\by} = n^{-1}\sum_{t=1}^{n}\by_{t}$ is the sample mean.
For simplicity, we adopt Lasso estimation.
Let $\widehat{\balpha}_j$ be the Lasso estimator of $\balpha_j$ defined as follows:
\begin{equation}\label{eq:bestimate}
\widehat{\balpha}_j= \arg\min_{\bgamma\in\Theta_j}\bigg[\frac{1}{n}\sum_{t=1}^n (\bgamma^\T\by_t )^2 + 2 \lambda_j|\bgamma|_1\bigg],
\end{equation}
where
$\Theta_j=\{\bgamma=(\gamma_1,\ldots,\gamma_p)^\T\in\mathbb{R}^p:\gamma_j=-1\}$ and $\lambda_{j}$ is the tuning parameter.
For each $t$, the residual
\begin{equation}\label{eq:epsilonest}
\widehat{\epsilon}_{j,t} =-\widehat{\balpha}_j^\T\by_t
\end{equation}
provides an estimate of $\epsilon_{j,t}$. Write $\widehat{\bepsilonbb}_{t} = (\widehat{\epsilon}_{1,t}, \ldots, \widehat{\epsilon}_{p,t})^\T$ and let $\widetilde{\bV}=(\widetilde{v}_{j_1,j_2})_{p\times p}$ be the sample covariance of $\{\widehat{\bepsilonbb}_{t}\}_{t=1}^{n}$, where $\widetilde{v}_{j_1,j_2} = n^{-1}\sum_{t=1}^{n}\widehat{\epsilon}_{j_1,t}\widehat{\epsilon}_{j_2,t}$. It is well known that $n^{-1}\sum_{t=1}^n\epsilon_{j_1,t}\epsilon_{j_2,t}$ is an unbiased estimator of $v_{j_1,j_2}$, however, replacing $\epsilon_{j_1,t}$ by $\widehat{\epsilon}_{j_1,t}$ will incur a bias term. Specifically, as shown in Lemma \ref{la:bias} in Appendix, under Conditions \ref{as:moment}--\ref{as:betamix} and some mild restrictions on the sparsity of $\bOmega$ and the growth rate of $p$ with respect to $n$, it holds that
\begin{equation}\label{eq:v}
\begin{split}
\widetilde{v}_{j_1,j_2}-\frac{1}{n}\sum_{t=1}^n\epsilon_{j_1,t}\epsilon_{j_2,t}=&-(\widehat{\alpha}_{j_1,j_2}-\alpha_{j_1,j_2})\bigg(\frac{1}{n}\sum_{t=1}^n\epsilon_{j_2,t}^2\bigg)\mathbb{I}(j_1\neq j_2)\\
&-(\widehat{\alpha}_{j_2,j_1}-\alpha_{j_2,j_1})\bigg(\frac{1}{n}\sum_{t=1}^n\epsilon_{j_1,t}^2\bigg)\mathbb{I}(j_1\neq j_2)\\
&+o_p\{(n\log p)^{-1/2}\}.
\end{split}
\end{equation}
Here the higher order term $o_p\{(n\log p)^{-1/2}\}$ is uniform over all $j_1$ and $j_2$. Since $n^{-1}\sum_{t=1}^n\epsilon_{j,t}^2$ is $n^{1/2}$-consistent for $v_{j,j}$, (\ref{eq:v}) implies that $\widetilde{v}_{j,j}$ is also $n^{1/2}$-consistent for $v_{j,j}$. However, for any $j_1\neq j_2$, due to the slow convergence rates of the Lasso estimators $\widehat{\alpha}_{j_1,j_2}$ and $\widehat{\alpha}_{j_2,j_1}$, $\widetilde{v}_{j_1,j_2}$ is no longer ${n}^{1/2}$-consistent for $v_{j_1,j_2}$. To eliminate the bias, we employ an estimator for $v_{j_1,j_2}$:
\begin{equation}\label{eq:hatv}
\widehat{v}_{j_1,j_2}= \left\{ \begin{aligned}
         -\frac{1}{n}\sum_{t=1}^n(\widehat{\epsilon}_{j_1,t}\widehat{\epsilon}_{j_2,t}+\widehat{\alpha}_{j_1,j_2}\widehat{\epsilon}_{j_2,t}^2+\widehat{\alpha}_{j_2,j_1}\widehat{\epsilon}_{j_1,t}^2),~~ &j_1\neq j_2; \\
                  \frac{1}{n}\sum_{t=1}^n\widehat{\epsilon}_{j_1,t}\widehat{\epsilon}_{j_2,t},~~~~~~~~~~~~~~&j_1=j_2.
                          \end{aligned} \right.
\end{equation}
By noticing that $\bOmega=\{\textrm{diag}(\bV)\}^{-1}\bV\{\textrm{diag}(\bV)\}^{-1}$, we estimate $\omega_{j_1,j_2}$ by
\begin{equation}\label{eq:estOmega}
\widehat{\omega}_{j_1,j_2}=\frac{\widehat{v}_{j_1,j_2}}{\widehat{v}_{j_1,j_1}\widehat{v}_{j_2,j_2}}
\end{equation}
for any $j_1$ and $j_2$. We need to point out that the asymptotic expansion (\ref{eq:v}) is still valid for other penalized methods such as Dantzig estimation \citep{CandesTao_2007} and scaled Lasso \citep{SZ_2012}. Hence, we can also estimate $v_{j_1,j_2}$ and $\omega_{j_1,j_2}$ as (\ref{eq:hatv}) and (\ref{eq:estOmega}), respectively, based on the residuals $\{\widehat{\bepsilonbb}_{t}\}_{t=1}^{n}$ obtained by other penalized methods. To study the theoretical properties of this estimator $\widehat{\omega}_{j_1,j_2}$, we need the following regularity conditions.

\begin{as}\label{as:moment}
There exist constants $K_1>0$, $K_2>1$, $0<\gamma_1\leq 2$ and $0<\gamma_2\leq 2$ independent of $p$ and $n$ such that for each $t=1,\ldots,n$,
\[
\max_{1\leq j\leq p}\mathbb{E}\{\exp(K_1|y_{j,t}|^{\gamma_1})\}\leq K_2~~~\textrm{and}~~~\max_{1\leq j\leq p}\mathbb{E}\{\exp(K_1|\epsilon_{j,t}|^{\gamma_2})\}\leq K_2.
\]
\end{as}

\begin{as}\label{as:cov}
The eigenvalues of $\bSigma$ are uniformly bounded away from zero and infinity.
\end{as}

\begin{as}\label{as:betamix}
There exist constants $K_3>0$ and $\gamma_3>0$ independent of $p$ and $n$ such that $\beta_k\leq \exp(-K_3k^{\gamma_3})$ for any positive $k$.
\end{as}

Condition \ref{as:moment} implies $\max_{1\leq j\leq p}\mathbb{P}(|y_{j,t}|\geq x)\leq K_2\exp(-K_1x^{\gamma_1})$ and $\max_{1\leq j\leq p}\mathbb{P}(|\epsilon_{j,t}|\geq x)\leq K_2\exp(-K_1x^{\gamma_2})$ for any $x>0$ and $t=1,\ldots,n$. It ensures the exponential
upper bounds for the tail probabilities of the statistics concerned (see for example Lemma 1 in Appendix), which makes our procedure work for $p$ diverging at some exponential rate of $n$.
Condition \ref{as:cov} implies the bounded eigenvalues of $\bSigma$ and $\bOmega$, which is commonly assumed in the literatures of high-dimensional data analysis.
Condition \ref{as:betamix} for the $\beta$-mixing coefficients of $\{\by_{t}\}$ is mild. Causal ARMA processes with continuous innovation
distributions are $\beta$-mixing with exponentially decaying $\beta_k$. So are stationary Markov chains
satisfying certain conditions. See Section 2.6.1 of \cite{FanYao_2003} and the references therein.
In fact, stationary GARCH models with finite second
moments and continuous innovation distributions are also $\beta$-mixing with exponentially decaying
$\beta_k$; see Proposition 12 of \cite{CarrascoChen_2002}. If we only require $\sup_t\max_{1\leq j\leq p} \mathbb{P}(|y_{j,t} | >
x) = O\{x^{-2(\nu+\iota)}\}$ and $\sup_t\max_{1\leq j\leq p} \mathbb{P}(|\epsilon_{j,t} | >
x) = O\{x^{-2(\nu+\iota)}\}$ for any $x > 0$ in Condition 1 and $\beta_k = O\{k^{-\nu(\nu+\iota)/(2\iota)}\}$ in Condition 3
for some $\nu > 2$ and $\iota> 0$, we can apply Fuk-Nagaev-type inequalities to construct the upper
bounds for the tail probabilities of the statistics if $p$
diverges at some polynomial rate of $n$. We refer to Section 3.2 of \cite{ChangGuoYao_2014} for
the implementation of Fuk-Nagaev-type inequalities in such a scenario. The $\beta$-mixing condition
can be replaced by the $\alpha$-mixing condition, under which we can justify the proposed method for
$p$ diverging at some polynomial rate of $n$ by using Fuk-Nagaev-type inequalities. However, it
remains an open problem to establish the relevant properties under $\alpha$-mixing for $p$ diverging at
some exponential rate of $n$.

\begin{proposition}\label{pro:1}
Let $s=\max_{1\leq j\leq p}|\balpha_j|_0$ and select the tuning parameter $\lambda_j$ in {\rm(\ref{eq:bestimate})} satisfying $\lambda_j \asymp (n^{-1}\log p)^{1/2}$ for each $j=1,\ldots,p$. Under Conditions {\rm \ref{as:moment}--\ref{as:betamix}}, if $s^2(\log p)^3n^{-1}=o(1)$ and $\log p=o(n^{\varrho_1})$ for a positive constant $\varrho_1$ specified in the proof of this proposition in Appendix, it holds that
\[
\widehat{\omega}_{j_1,j_2}-\omega_{j_1,j_2} = -\frac{\delta_{j_1,j_2}}{v_{j_1,j_1}v_{j_2,j_2}} + o_p\{(n\log p)^{-1/2}\},
\]
where $\delta_{j_1,j_2}=n^{-1}\sum_{t=1}^n(\epsilon_{j_1,t}\epsilon_{j_2,t}-v_{j_1,j_2})$ for any $j_1$ and $j_2$, and $o_p\{(n\log p)^{-1/2}\}$ is a uniform higher order term.
\end{proposition}

We see from Proposition \ref{pro:1} that $\widehat{\omega}_{j_1,j_2}$ is
centered at the true parameter $\omega_{j_1,j_2}$ with a standard deviation at the order $n^{-1/2}$.
Since $\alpha_{j_1,j_2}$ is proportional to $\omega_{j_1,j_2}$, it follows from
 $s^2(\log p)^3n^{-1}=o(1)$ that $\bOmega$ is sparse.
When the maximum number of nonzero elements in each row of $\bOmega$ is
of the order smaller than $n^{1/2}(\log p)^{-3/2}$,
Proposition \ref{pro:1} holds
even when $p$ is of an exponential rate of $n$. Similar to
the asymptotic expansion for $\widehat{\omega}_{j_1,j_2}$ in Proposition
1, \cite{Liu_2013} gave an asymptotic expansion for
$-\widehat{v}_{j_1,j_2}$ with $j_1\neq j_2$. More specifically, with
i.i.d.\,data, he showed that
$-\widehat{v}_{j_1,j_2}=-\frac{b_{j_1,j_2}\omega_{j_1,j_2}}
{\omega_{j_1,j_1}\omega_{j_2,j_2}}+\delta_{j_1,j_2}+R$ for
$\delta_{j_1,j_2}$ specified in Proposition 1 and
$b_{j_1,j_2}=\omega_{j_1,j_1}\widehat{v}_{j_1,j_1}+\omega_{j_2,j_2}
\widehat{v}_{j_2,j_2}-1$, where $R$ is a remainder term with the
convergence rate faster than $n^{-1/2}$. It follows from the central
limit theorem that
$-{n}^{1/2}c_{j_1,j_2}(\widehat{v}_{j_1,j_2}-\frac{b_{j_1,j_2}\omega_{j_1,j_2}}
{\omega_{j_1,j_1}\omega_{j_2,j_2}})$ converges to standard normal
distribution with some suitable scale $c_{j_1,j_2}$, which indicates that
$-{n}^{1/2}c_{j_1,j_2}\widehat{v}_{j_1,j_2}$ can be used as the testing
statistic to test $\omega_{j_1,j_2}=0$ or not. Notice that
$\widehat{v}_{j,j}=\omega_{j,j}^{-1}+O_p(n^{-1/2})$ which implies
$b_{j_1,j_2}=1+O_p(n^{-1/2})$. Hence, the magnitude of
$-{n}^{1/2}c_{j_1,j_2}\widehat{v}_{j_1,j_2}$ will be large if
$\omega_{j_1,j_2}\neq 0$. This indicates that the asymptotic expansion
given in \cite{Liu_2013} is enough for identifying non-zero entries of
$\bOmega$. However, it is not enough for constructing the confidence interval
for $\omega_{j_1,j_2}$ due to the fact that it does not contain the
asymptotic expansion of
$\widehat{\omega}_{j_1,j_2}$.

\subsection{Confidence regions}\label{se:cr}

Let $\bDelta = -n^{-1}\sum_{t=1}^n(\bepsilonbb_t\bepsilonbb_t^\T-\bV)$. It follows from Proposition 1 that
\begin{equation*}\label{eq:asympexp}
\widehat{\bOmega}-\bOmega = \bPi + \bUpsilon
\mbox{ \ for \ } \bPi = \{\textrm{diag}(\bV)\}^{-1}\bDelta\{\textrm{diag}(\bV)\}^{-1},
\end{equation*}
where $|\bUpsilon|_\infty=o_p\{(n\log p)^{-1/2}\}$. Restricted on a given index set $\mathcal{S}$ with $r=|\mathcal{S}|$, we have 
\begin{equation}\label{eq:asyp}
\widehat{\bOmega}_{\mathcal{S}}-\bOmega_{\mathcal{S}}=\bPi_{\mathcal{S}}+\bUpsilon_{\mathcal{S}}.
\end{equation}
Based on (\ref{eq:asyp}), we consider two kinds of confidence regions:
\begin{equation}\label{eq:cr2}
\begin{split}
\mathcal{C}_{\mathcal{S},\alpha,1}=&~\{\ba\in\mathbb{R}^r:n^{1/2}|\widehat{\bOmega}_{\mathcal{S}}-\ba|_\infty\leq q_{\mathcal{S},\alpha,1}\},\\
\mathcal{C}_{\mathcal{S},\alpha,2}=&~\{\ba\in\mathbb{R}^r:n^{1/2}|\widehat{\bD}^{-1}(\widehat{\bOmega}_{\mathcal{S}}-\ba)|_\infty\leq q_{\mathcal{S},\alpha,2}\},\\
\end{split}
\end{equation}
where $\widehat{\bD}$ is an $r\times r$ diagonal matrix,
specified in Remark \ref{re:stude} below,
of which the elements are the estimated
standard deviations of the $r$ components in
$n^{1/2}(\widehat{\bOmega}_{\mathcal{S}}-\bOmega_{\mathcal{S}})$. Here
$q_{\mathcal{S},\alpha,1}$ and $q_{\mathcal{S},\alpha,2}$ are two
critical values to be determined. $\mathcal{C}_{\mathcal{S},\alpha,1}$
and $\mathcal{C}_{\mathcal{S},\alpha,2}$ represent the so-called
``non-Studentized-type" and ``Studentized-type" confidence regions for
$\bOmega_{\mathcal{S}}$, respectively.
The Studentized-type confidence regions
perform better than the non-Studentized-type ones when the
heteroscedasticity exists, however, the performance of the
non-Studentized-type confidence regions is more stable when the sample size
$n$ is fairly small. See \cite{ChangZhouZhou_2014a}.

In the sequel, we mainly focus on estimating the critical value
$q_{\mathcal{S},\alpha,1}$ in (\ref{eq:cr2}), as $q_{\mathcal{S},\alpha,2}$ can be
estimated in the similar manner; see Remark \ref{re:stude} below.
To determine $q_{\mathcal{S},\alpha,1}$, we need to first characterize
the probabilistic behavior of
$n^{1/2}|\widehat{\bOmega}_{\mathcal{S}}-\bOmega_{\mathcal{S}}|_\infty$.
Since $\bUpsilon_{\mathcal{S}}$ is a higher order term,
$n^{1/2}|\widehat{\bOmega}_{\mathcal{S}}-\bOmega_{\mathcal{S}}|_\infty$
will behave similarly as $n^{1/2}|\bPi_{\mathcal{S}}|_\infty$ when $n$ is
large. 
Notice that each element of $n^{1/2}\bPi_{\mathcal{S}}$ is asymptotically normal distributed.
Following the idea of \cite{CCK_2013}, it can be proved that the limiting
behavior of ${n}^{1/2}|\bPi_{\mathcal{S}}|_\infty$ can be approximated by
that of the $L_\infty$-norm of a certain multivariate normal vector. See
Theorem \ref{tm:1} below. More specifically,
for each $t$, let $\bvarsigma_t$ be an $r$-dimensional vector whose $j$-th element is $\frac{\epsilon_{\chi_1(j),t}\epsilon_{\chi_2(j),t}-v_{\bchi(j)}}{v_{\chi_1(j),\chi_1(j)}v_{\chi_2(j),\chi_2(j)}}$ where $\bchi(\cdot)=\{\chi_1(\cdot),\chi_2(\cdot)\}$ is a bijective mapping from $\{1,\ldots,r\}$ to $\mathcal{S}$ such that $\bOmega_{\mathcal{S}}=\{\omega_{\bchi(1)},\ldots,\omega_{\bchi(r)}\}^\T$. Then, we have
\[
\bPi_{\mathcal{S}}=-\frac{1}{{n}}\sum_{t=1}^n\bvarsigma_t.
\]
Denote by $\bW$ the long-run covariance of $\{\bvarsigma_t\}_{t=1}^n$, namely,
\begin{equation}\label{eq:W}
\bW=\mathbb{E}\bigg\{\bigg(\frac{1}{{n}^{1/2}}\sum_{t=1}^n\bvarsigma_t\bigg)\bigg(\frac{1}{{n}^{1/2}}\sum_{t=1}^n\bvarsigma_t\bigg)^\T\bigg\}.
\end{equation}
Let $\bfeta_t=(\eta_{1,t},\ldots,\eta_{r,t})^\T$ where $\eta_{j,t}=\epsilon_{\chi_1(j),t}\epsilon_{\chi_2(j),t}-v_{\bchi(j)}$. Then $\bW$ specified in (\ref{eq:W}) can be written as
\begin{equation}\label{eq:w}
\bW=\bH \mathbb{E}\bigg\{\bigg(\frac{1}{{n}^{1/2}}\sum_{t=1}^n\bfeta_t\bigg)\bigg(\frac{1}{{n}^{1/2}}\sum_{t=1}^n\bfeta_t\bigg)^\T\bigg\}\bH
\end{equation}
where $\bH=\textrm{diag}\{v_{\chi_1(1),\chi_1(1)}^{-1}v_{\chi_2(1),\chi_2(1)}^{-1},\ldots,v_{\chi_1(r),\chi_1(r)}^{-1}v_{\chi_2(r),\chi_2(r)}^{-1}\}$.
To study the asymptotical distribution of the average of the temporally
dependent sequence $\{\bvarsigma_t\}_{t=1}^n$ and its long-run covariance
$\bW$, we introduce the following condition on $\{\bfeta_t\}_{t=1}^n$.

\begin{as}\label{as:block}
There exists constant $K_4>0$  such that
\[
\liminf_{b\rightarrow\infty}\inf_{1\leq \ell\leq n+1-b}\mathbb{E}\bigg(\bigg|\frac{1}{{b}^{1/2}}\sum_{t=\ell}^{\ell+b-1}\eta_{j,t}\bigg|^{2}\bigg)>K_4
\]
for each $j=1,\ldots,r$.
\end{as}

Condition 4 is for the validity of the Gaussian
approximation for dependent data. Under Conditions 1 and 3, Davydov
inequality \citep{Davydov_1968} entails
$\lim\sup_{b\rightarrow\infty}\sup_{1\leq \ell\leq
n+1-b}\mathbb{E}(|b^{-1/2}\sum_{t=\ell}^{\ell+b-1}\eta_{j,t}|^{2})<K_5$
for some universal constant $K_5>0$. Together with Condition 4, they
match the requirements of Gaussian approximation imposed on the long-run covariance of
$\{\eta_{j,t}\}_{t=\ell}^{\ell+b-1}$ for $j=1,\ldots,r$ and
$\ell=1,\ldots,n+1-b$. See Theorem B.1 of \cite{CCK_2014}. If
$\{\eta_{j,t}\}$ is stationary,
$\mathbb{E}(|b^{-1/2}\sum_{t=\ell}^{\ell+b-1}\eta_{j,t}|^{2})=\mathbb{E}
(\eta_{j,1}^2)+\sum_{k=1}^{b-1}(1-kb^{-1})\textrm{Cov}(\eta_{j,1},\eta_{j,1+k})$.
Under the stationarity assumption on each sequence $\{\eta_{j,t}\}$,
Condition 4 is equivalent to
$\sum_{k=0}^\infty\textrm{Cov}(\eta_{j,1},\eta_{j,1+k})>K_4$ for any
$j=1,\ldots,r$.
Now we are ready to state our main result.

\begin{theorem}\label{tm:1}
Let $\bxi\sim N(\bzero,\bW)$ for $\bW$ specified in {\rm(\ref{eq:W})}. Under the conditions of Proposition {\rm\ref{pro:1}} and Condition {\rm\ref{as:block}}, we have
\[
\sup_{x>0}\big|\mathbb{P}\big({n}^{1/2}|\widehat{\bOmega}_{\mathcal{S}}-\bOmega_{\mathcal{S}}|_\infty>x\big)-\mathbb{P}(|\bxi|_\infty>x)\big|\rightarrow0
\]
as $n\rightarrow\infty$, provided that $s^2(\log p)^3n^{-1}=o(1)$ and $\log p=o(n^{\varrho_2})$ where $s=\max_{1\leq j\leq p}|\balpha_j|_0$ and $\varrho_2$ is a positive constant specified in the proof of this theorem in Appendix.
\end{theorem}

\begin{remark}\label{re:1}
Theorem \ref{tm:1} shows that the Kolmogorov distance between the
distributions of
${n}^{1/2}|\widehat{\bOmega}_{\mathcal{S}}-\bOmega_{\mathcal{S}}|_\infty$
and $|\bxi|_\infty$ converges to zero. 
More specifically, as shown in the proof of Theorem \ref{tm:1} in Appendix, this convergence rate is $O(n^{-C})$ for some constant $C>0$ without requiring any structural assumption on the underlying covariance $\bW$. 
Note that ${n}^{1/2}|\widehat{\bOmega}_{\mathcal{S}}-\bOmega_{\mathcal{S}}|_\infty$
may converge weakly to  an extreme value distribution, which
however requires some more stringent assumptions on the structure of $\bW$. Furthermore the slow convergence to the
extreme value distribution, i.e. typically slower than $O(n^{-C})$, entails an less
accurate approximation than
that implied by Theorem 1. We need to point out that there is also a requirement imposed on the diverging rate of $r=|\mathcal{S}|$ such as $\log r=o(n^C)$ for some constant $C>0$ in the proof of Theorem \ref{tm:1}. Since $r\leq p^2$, such requirement is satisfied automatically when the requirements on $p$ in Theorem \ref{tm:1} are required.


\end{remark}

Theorem \ref{tm:1} provides a guideline to approximate the distribution of ${n}^{1/2}|\widehat{\bOmega}_{\mathcal{S}}-\bOmega_{\mathcal{S}}|_\infty$. To implement it in practice, we need to propose an estimator for $\bW$. Denote by $\bXi$ the matrix sandwiched by $\bH$'s on the right-hand side of (\ref{eq:w}), which is the long-run covariance of $\{\bfeta_t\}_{t=1}^n$. Notice that $\widehat{v}_{j,j}$ defined in (\ref{eq:hatv}) is $n^{1/2}$-consistent to $v_{j,j}$, we can estimate $\bH$ by
\begin{equation}\label{eq:hath}
\widehat{\bH}=\textrm{diag}\big\{\widehat{v}_{\chi_1(1),\chi_1(1)}^{-1}\widehat{v}_{\chi_2(1),\chi_2(1)}^{-1},\ldots,\widehat{v}_{\chi_1(r),\chi_1(r)}^{-1}\widehat{v}_{\chi_2(r),\chi_2(r)}^{-1}\big\}.
\end{equation}
Let $\widehat{\bfeta}_t=(\widehat{\eta}_{1,t},\ldots,\widehat{\eta}_{r,t})^\T$ for $\widehat{\eta}_{j,t}=\widehat{\epsilon}_{\chi_1(j),t}\widehat{\epsilon}_{\chi_2(j),t}-\widehat{v}_{\bchi(j)}$, and define
\[
\widehat{\bGamma}_k= \left\{ \begin{aligned}
         \frac{1}{n}\sum_{t=k+1}^n\widehat{\bfeta}_t\widehat{\bfeta}_{t-k}^\T,~~~ &k\geq0; \\
                  \frac{1}{n}\sum_{t=-k+1}^n\widehat{\bfeta}_{t+k}\widehat{\bfeta}_t^\T,~~&k<0.
                          \end{aligned} \right.
\]
Based on the $\widehat{\bGamma}_k$'s,
we propose a kernel estimator suggested by \cite{Andrews_1991} for $\bXi$ as
 \begin{equation}\label{eq:hatXi}
\widehat{\bXi}=\sum_{k=-n+1}^{n-1}\mathcal {K}\bigg(\frac{k}{S_n}\bigg)\widehat{\bGamma}_k
 \end{equation}
 where $S_n$ is the bandwidth, $\mathcal{K}(\cdot)$ is a symmetric kernel function that is continuous at 0 and satisfying $\mathcal{K}(0)=1$, $|\mathcal{K}(u)|\leq 1$ for any $u\in\mathbb{R}$, and $\int_{-\infty}^\infty\mathcal{K}^2(u)du<\infty$.  Given $\widehat{\bH}$ and $\widehat{\bXi}$ defined respectively in (\ref{eq:hath}) and (\ref{eq:hatXi}), an estimator for $\bW$ is given by
 \begin{equation}\label{eq:hatW}
\widehat{\bW}=\widehat{\bH}\widehat{\bXi}\widehat{\bH}.
 \end{equation}
Theorem \ref{tm:2} below shows that we can approximate the distribution of ${n}^{1/2}|\widehat{\bOmega}_{\mathcal{S}}-\bOmega_{\mathcal{S}}|_\infty$ by that of $|\widehat{\bxi}|_\infty$ for $\widehat{\bxi}\sim N(\bzero,\widehat{\bW})$.

\begin{remark}
\cite{Andrews_1991} systematically investigated the theoretical properties for
the kernel estimator for the long-run covariance matrix when $p$ is fixed.
It shows that the Quadratic Spectral kernel
 \[
 \mathcal{K}_{QS}(u)=\frac{25}{12\pi^2u^2}\bigg\{\frac{\sin(6\pi u/5)}{6\pi u/5}-\cos(6\pi u/5)\bigg\}
 \]
 is optimal 
in the sense of minimizing the asymptotic truncated mean square error.
In our numerical work, we adopt this quadratic spectral kernel with the
data-driven selected bandwidth
proposed in Section 6 of \cite{Andrews_1991},
though our theoretical analysis applies to general kernel functions. Both
our theoretical and simulation results show that this kernel estimator
$\widehat{\bXi}$ still works when $p$ is large in relation to $n$. There
also exist other estimation methods for long-run covariances,
including the estimation utilizing moving block bootstrap
\citep{Lahiri_2003,NordmanLahiri_2005}. Also see \cite{DenHanLevin_1997}
and \cite{Kieferetaj_2000}. Compared to those methods, an added advantage
of using the kernel estimator is the computational efficiency in terms of
both speed and storage space especially when $p$ is large; see See Remark \ref{re:gener}
below.
When the observations are i.i.d., a special case of our setting, $\bW$ as in
(\ref{eq:W}) is degenerated to $\mathbb{E}(\bvarsigma_t\bvarsigma_t^\T)$,
the marginal covariance of $\bvarsigma_t$. We can apply
$n^{-1}\sum_{t=1}^n\widehat{\bfeta}_t\widehat{\bfeta}_t^\T$ to estimate
$\bXi$, and then use
$\widehat{\bH}(n^{-1}\sum_{t=1}^n\widehat{\bfeta}_t
\widehat{\bfeta}_t^\T)\widehat{\bH}$ to estimate $\bW$ with
$\widehat{\bH}$ as in (\ref{eq:hath}).
\end{remark}

\begin{theorem}\label{tm:2}
Let $\widehat{\bxi}\sim N(\bzero,\widehat{\bW})$ for $\widehat{\bW}$ specified in {\rm(\ref{eq:hatW})}. Assume the kernel function $\mathcal{K}(\cdot)$ satisfy $|\mathcal{K}(x)|\asymp |x|^{-\tau}$ as $x\rightarrow\infty$ for some $\tau>1$, and the bandwidth $S_n\asymp n^{\rho}$ for some $0<\rho<\min\{\frac{\tau-1}{3\tau},\frac{\gamma_3}{2\gamma_3+1}\}$ and $\gamma_3$ in Condition {\rm\ref{as:betamix}}. Under the conditions of Theorem {\rm\ref{tm:1}},
it holds that
\[
\sup_{x>0}\big|\mathbb{P}\big({n}^{1/2}|\widehat{\bOmega}_{\mathcal{S}}-\bOmega_{\mathcal{S}}|_\infty>x\big)-\mathbb{P}\big(|\widehat{\bxi}|_\infty>x|\mathcal{Y}_n\big)\big|\xrightarrow{p}0
\]
as $n\rightarrow\infty$, provided that $s^2(\log p)n^{-1}\max\{S_n^2,(\log p)^2\}=o(1)$ and $\log p=o(n^{\varrho_3})$ where $s=\max_{1\leq j\leq p}|\balpha_j|_0$, $\varrho_3$ is a positive constant specified in the proof of this theorem in Appendix, and $\mathcal{Y}_n=\{\by_1,\ldots,\by_n\}$.
\end{theorem}

\begin{remark}
Theorem \ref{tm:2} is valid for any $\widehat{\bW}$ satisfying
$|\widehat{\bW}-\bW|_\infty=o_p(1)$; see \cite{CCK_2013}.
Different from the common practice in estimating large covariance
matrics, we construct $\widehat{\bW}$ in (\ref{eq:hatW}) without imposing
any structural assumptions on $\bW$.
\end{remark}

In practice, we approximate the distribution of $|\widehat{\bxi}|_\infty$
by Monto Carlo simulation.  Specifically, let $\widehat{\bxi}_1,\ldots,\widehat{\bxi}_M$ be i.i.d.\,$r$-dimensional random vectors drawn from $N(\bzero,\widehat{\bW})$. Then the conditional distribution of $|\widehat{\bxi}|_\infty$ given $\mathcal{Y}_n$ can be approximated by the empirical distribution of $\{|\widehat{\bxi}_1|_\infty,\ldots,|\widehat{\bxi}_M|_\infty\}$, namely,
\[
\widehat{F}_M(x)=\frac{1}{M}\sum_{m=1}^M\mathbb{I}\big\{|\widehat{\bxi}_m|_\infty\leq x\big\}.
\]
Then, $q_{\mathcal{S},\alpha,1}$ specified in (\ref{eq:cr2}) can be estimated by
\begin{equation}\label{eq:hatq}
\widehat{q}_{\mathcal{S},\alpha,1}=\inf\{x\in \mathbb{R}:\widehat{F}_M(x)\geq 1-\alpha\}.
\end{equation}
To improve computational efficiency, we propose the following  Kernel
based Multiplier Bootstrap (KMB) procedure to generate $\widehat{\bxi}
\sim N(\bzero,\widehat{\bW})$, which is much more efficient when $r$ is large.

\vspace{-10pt}
\begin{itemize}[leftmargin = 2.35cm, rightmargin=1cm]
\item[{\bf Step 1.}] Generate $\bg=(g_1,\ldots,g_n)^\T$ from $N(\bzero, \bA)$,
where $\bA$ is the $n\times n$ matrix with
$\mathcal{K}(|i-j|/S_n)$ as its $(i,j)$-th element.
\vspace{-5pt}
\item[{\bf Step 2.}] Let
$\widehat{\bxi}=n^{-1/2}\widehat{\bH}(\sum_{t=1}^ng_t\widehat{\bfeta}_t)$,
where $\widehat{\bH}$ is defined in (\ref{eq:hath}).
\end{itemize}

\begin{remark}\label{re:gener}
The standard approach to draw a random vector $\widehat{\bxi}\sim
N(\bzero,\widehat{\bW})$ consists of three steps: (i) perform the
Cholesky decomposition on the $r\times r$ matrix
$\widehat{\bW}=\bL^\T\bL$, (ii) generate $r$ independent standard normal
random variables $\bz=(z_1,\ldots,z_r)^\T$, (iii) perform transformation
$\widehat{\bxi}=\bL^\T\bz$. Thus, it requires to store matrix
$\widehat{\bW}$ and $\{\widehat{\bfeta}_t\}_{t=1}^n$, which amounts to
the storage costs $O(r^2)$ and $O(rn)$, respectively. The computational
complexity is $O(r^2n+r^3)$, mainly due to computing $\widehat{\bW}$ and
the Cholesky decomposition. Note that $r$ could be in the order of $O(p^2)$.
In contrast the KMB scheme described above
only needs to store
$\{\widehat{\bfeta}_t\}_{t=1}^n$ and $\bA$, and draw an $n$-dimensional random vector
$\bg\sim N(\bzero,\bA)$ in each parametric bootstrap sample. This amounts to total
storage cost $O(rn+n^2)$. More significantly, the computational
complexity 
is only $O(n^3)$ which is independent of $r$ and $p$.
\end{remark}

\begin{remark}\label{re:stude}
For the Studentized-type confidence regions
$\mathcal{C}_{\mathcal{S},\alpha,2}$ defined in (\ref{eq:cr2}), we can
choose the diagonal matrix $\widehat{\bD}=\{\diag(\widehat{\bW})\}^{1/2}$
for $\widehat{\bW}$ specified in (\ref{eq:hatW}). Correspondingly, for
$\widehat{\bxi}\sim
N(\bzero,\widehat{\bD}^{-1}\widehat{\bW}\widehat{\bD}^{-1})$, it can be
proved, in the similar manner as that for  Thorem \ref{tm:2}, that
\[
\sup_{x>0}\big|\mathbb{P}\big\{{n}^{1/2}|\widehat{\bD}^{-1} (\widehat{\bOmega}_{\mathcal{S}}-\bOmega_{\mathcal{S}}) |_\infty>x\big\}-\mathbb{P}(|\widehat{\bxi}|_\infty>x|\mathcal{Y}_n)\big|\xrightarrow{p}0~~\textrm{as}~~n\rightarrow\infty.
\]
To approximate the distribution of ${n}^{1/2}|\widehat{\bD}^{-1}
(\widehat{\bOmega}_{\mathcal{S}}-\bOmega_{\mathcal{S}}) |_{\infty}$, we
only need to replace the Step 2 in the KMB procedure by \vspace{-10pt}
\begin{itemize}[leftmargin = 2.35cm, rightmargin=1cm]
\item[{\bf Step 2$^{\prime}$.}] Let $\widehat{\bxi} =
n^{-1/2}\widehat{\bD}^{-1} \widehat{\bH}(\sum_{t=1}^n
g_t\widehat{\bfeta}_t)$ where $\widehat{\bH}$ is defined in
(\ref{eq:hath}).
\end{itemize}
\vspace{-10pt} Based on the i.i.d.\,random vectors
$\widehat{\bxi}_1,\ldots,\widehat{\bxi}_M$ generated by Steps 1 and
$2^{\prime}$, we can estimate $q_{\mathcal{S},\alpha,2}$ via
$\widehat{q}_{\mathcal{S},\alpha,2}$, which is calculated the same as
$\widehat{q}_{\mathcal{S},\alpha,1}$ in (\ref{eq:hatq}). We call the
procedure combining Steps 1 and $2^{\prime}$ as Studentized Kernel based
Multiplier Bootstrap (SKMB).
\end{remark}

\section{Applications}\label{se:app}

\subsection{Testing structures of $\bOmega$}

Many statistical applications require to explore or to detect some
specific structures of the precision matrix
$\bOmega=(\omega_{j_1,j_2})_{p\times p}$. Given an index set
$\mathcal{S}$ of interest and a set of pre-specified constants
$\{c_{j_1,j_2}\}$, we test the hypotheses
\[
H_0:\omega_{j_1,j_2}=c_{j_1,j_2}~~\textrm{for any}~(j_1,j_2)\in\mathcal{S}~~~~~\textrm{vs.}~~~~~H_1:\omega_{j_1,j_2} \neq c_{j_1,j_2}~~\textrm{for some}~(j_1,j_2)\in\mathcal{S}.
\]
Recall that $\bchi(\cdot)=\{\chi_1(\cdot),\chi_2(\cdot)\}$ is a bijective mapping from $\{1,\ldots,r\}$ to $\mathcal{S}$ such that $\bOmega_{\mathcal{S}}=\{\omega_{\bchi(1)},\ldots,\omega_{\bchi(r)}\}^\T$. Let $r=|\mathcal{S}|$ and $\bc=\{c_{\bchi(1)},\ldots,c_{\bchi(r)}\}^\T$.
A usual choice of $\bc$ is the zero vector, corresponding to the test for non-zero structures of $\bOmega$. Given a prescribed level $\alpha\in(0,1)$, define $\Psi_\alpha=\mathbb{I}\{\bc\notin\mathcal{C}_{\mathcal{S}, 1-\alpha, 1}\}$ for $\mathcal{C}_{\mathcal{S}, 1-\alpha, 1}$ specified in (\ref{eq:cr2}). Then, we reject the null hypothesis $H_0$ at level $\alpha$ if $\Psi_\alpha=1$. This procedure is equivalent to the test based on the $L_\infty$-type statistic $n^{1/2}|\widehat{\bOmega}_{\mathcal{S}}-\bc|_\infty$ that rejects $H_0$ if $n^{1/2}|\widehat{\bOmega}_{\mathcal{S}}-\bc|_\infty > \widehat{q}_{\mathcal{S},1-\alpha,1}$. The $L_\infty$-type statistics are widely used in testing high-dimensional means and
covariances. See, for example, \cite{CLX_2013} and \cite{ChangZhouZhou_2014a,ChangZhouZhouWang_2017}. The following corollary gives the empirical size and power of the proposed testing procedure $\Psi_\alpha$.

\begin{cy}\label{cy:1}
Assume conditions of Theorem {\rm\ref{tm:2}} hold. It holds that: {\rm(i)} $\mathbb{P}_{H_0}(\Psi_\alpha=1)\rightarrow\alpha$ as $n\rightarrow\infty$; {\rm(ii)} if $\max_{(j_1,j_2)\in\mathcal{S}}|\omega_{j_1,j_2}-c_{j_1,j_2}|\geq C(n^{-1}\log p)^{1/2}\max_{1\leq j\leq r}w_{j,j}^{1/2}$ where $w_{j,j}$ is the $j$-th component in the diagonal of $\bW$ defined in {\rm(\ref{eq:W})}, and $C$ is a constant larger than $\sqrt{2}$, then $\mathbb{P}_{H_1}(\Psi_\alpha=1)\rightarrow1$ as $n\rightarrow\infty$.
\end{cy}

 Corollary \ref{cy:1} implies that the empirical size of the proposed testing procedure $\Psi_\alpha$ will converge to the nominal level $\alpha$ under $H_0$.
%
The condition $\max_{(j_1,j_2)\in\mathcal{S}}|\omega_{j_1,j_2}-c_{j_1,j_2}|\geq C(n^{-1}\log p)^{1/2}\max_{1\leq j\leq r}w_{j,j}^{1/2}$ specifies the maximal deviation of the precision matrix from the null hypothesis $H_0:\omega_{j_1,j_2}=c_{j_1,j_2}$ for any $(j_1,j_2)\in\mathcal{S}$, which is a commonly used condition for studying the power of the $L_\infty$-type test. See \cite{CLX_2013}  and \cite{ChangZhouZhou_2014a,ChangZhouZhouWang_2017}.
Corollary \ref{cy:1} shows that the power of the proposed test $\Psi_\alpha$ will approach 1 if such condition holds for some constant $C> \sqrt{2}$. A ``Studentized-type" test can be similarly constructed via replacing $n^{1/2}|\widehat{\bOmega}_{\mathcal{S}}-\bc|_\infty$ and $\widehat{q}_{\mathcal{S},1-\alpha,1}$ by $n^{1/2}|\widehat{\bD}^{-1}(\widehat{\bOmega}_{\mathcal{S}}-\bc)|_\infty$ and $\widehat{q}_{\mathcal{S},1-\alpha,2}$ in (\ref{eq:cr2}), respectively.

\subsection{Support recovering of $\bOmega$}

In studying partial correlation networks or GGM, we are interested
in identifying the edges between nodes. This is equivalent to recover the
non-zero components in the associated precision matrix. Let
$\mathcal{M}_0=\{(j_1,j_2):\omega_{j_1,j_2}\neq 0\}$ be the set of
indices with non-zero precision coefficients. Choose
$\mathcal{S}=\{1,\ldots, p\}^2$. Note that
$\mathcal{C}_{\mathcal{S},\alpha,1}$ provides simultaneous confidence
regions for all the entries of $\bOmega$. To recover the set
$\mathcal{M}_0$ consistently, we choose those precision coefficients
whose confidence intervals do not include zero. For any $m$-dimensional
vector $\bu=(u_1,\ldots,u_m)^\T$, let $\supp(\bu)=\{j:u_j\neq 0\}$ be the
support set of $\bu$. Recall
$\bchi(\cdot)=\{\chi_1(\cdot),\chi_2(\cdot)\}$ is a bijective mapping
from $\{1,\ldots,r\}$ to $\mathcal{S}$ such that
$\bOmega_{\mathcal{S}}=\{\omega_{\bchi(1)},\ldots,\omega_{\bchi(r)}\}^\T$.
For any $\alpha\in(0,1)$, let
\[
\widehat{\mathcal{M}}_{n,\alpha}=\bigg\{\bchi^{-1}(l):l\in\bigcap_{\bu\in\mathcal{C}_{\mathcal{S},1-\alpha, 1}}\supp(\bu)\bigg\}
\]
be the estimate of $\mathcal{M}_0$.

In our context, note that the false positive means estimating the zero $\omega_{j_1,j_2}$ as non-zero. Let $\mbox{FP}$ be the number of false positive errors conducted by the estimated signal set $\widehat{\mathcal{M}}_{n,\alpha}$. Let the family wise error rate (FWER) be the probability of conducting any false positive errors, namely, $\mbox{FWER} = \mathbb{P}(\mbox{FP} > 0)$. See \cite{HT_2009} for various types of error rates in multiple testing procedures. Notice that $\mathbb{P}(\mbox{FP} > 0) \leq \mathbb{P}(\bOmega_{\mathcal{S}} \not\in \mathcal{C}_{\mathcal{S},1-\alpha,1}) = \alpha\{1 + o(1)\}$. This shows that the proposed method is able to control family wise error rate at level $\alpha$ for any $\alpha\in(0,1)$. The following corollary further shows the consistency of $\widehat{\mathcal{M}}_{n,\alpha}$.

\begin{cy}\label{cy:2}
Assume conditions of Theorem {\rm\ref{tm:2}} hold, and the signals satisfy $\min_{(j_1,j_2)\in\mathcal{M}_0}|\omega_{j_1,j_2}|\geq C(n^{-1}\log p)^{1/2}\max_{1\leq j\leq r}w_{j,j}^{1/2}$ where $w_{j,j}$ is the $j$-th component in the diagonal of $\bW$ defined in {\rm(\ref{eq:W})}, and $C$ is a constant larger than $\sqrt{2}$. Selecting $\alpha\rightarrow0$ such that $1/\alpha=o(p)$, it holds that $\mathbb{P}(\widehat{\mathcal{M}}_{n,\alpha}=\mathcal{M}_0)\rightarrow1$ as $n\rightarrow\infty$.
\end{cy}

From Corollary \ref{cy:2}, we see that the selected set $\widehat{\mathcal{M}}_{n,\alpha}$ can identify the true set $\mathcal{M}_0$ consistently if the minimum signal strength satisfies $\min_{(j_1,j_2)\in\mathcal{M}_0}|\omega_{j_1,j_2}|\geq C(n^{-1}\log p)^{1/2}\max_{1\leq j\leq r}w_{j,j}^{1/2}$ for some constant $C>\sqrt{2}$.
Notice from Corollary 1 that only the maximum signal is required in the power analysis of the proposed testing procedure.
Compared to signal detection, signal recovery is a more challenging problem. The full support recovery of $\bOmega$ requires all non-zero $|\omega_{l_1,l_2}|$ larger than a specific level.
Similarly, we can also define $\widehat{\mathcal{M}}_{n,\alpha}$ via replacing $\mathcal{C}_{\mathcal{S},1-\alpha,1}$ by its ``Studentized-type'' analogue $\mathcal{C}_{\mathcal{S},1-\alpha,2}$ in (\ref{eq:cr2}).

\section{Numerical study}\label{se:simulation}

In this section, we evaluate the performance of the proposed KMB and SKMB procedures in finite samples. 
Let $\bepsilonb_{1}, \ldots, \bepsilonb_{n}$ be i.i.d.\,$p$-dimensional samples from $N(\bzero, \bSigma)$. The observed data were generated from the model $\by_1=\bepsilonb_1$ and $\by_t=\rho\by_{t-1}+(1-\rho^2)^{1/2}\bepsilonb_t$ for $t\geq2$. The parameter $\rho$ was set to be $0$ and $0.3$, which captures the temporal dependence among observations.
We chose the sample size $n=150$ and $300$, and the dimension $p=100$, $500$ and $1500$ in the simulation. Let $\bSigma=\{\textrm{diag}(\bSigma_*^{-1})\}^{1/2}\bSigma_*\{\textrm{diag}(\bSigma_*^{-1})\}^{1/2}$ based on a  positive definite matrix $\bSigma_*$. The following two settings were considered for $\bSigma_*=(\sigma_{j_1,j_2}^*)_{1\leq j_1,j_2\leq p}$. \vspace{-10pt}
\begin{itemize}[leftmargin = 1.35cm, rightmargin=1cm]
\item[{\bf A}.] Let $\sigma_{j_1,j_2}^* = 0.5^{|j_1-j_2|}$ for any $1\leq j_1,j_2\leq p$. 
\vspace{-5pt}
\item[{\bf B}.] Let $\sigma_{j,j}^*=1$ for any $j=1,\ldots,p$, $\sigma_{j_1,j_2}^*=0.5$ for $5(h-1)+1\leq j_1\neq j_2\leq 5h$, where $h=1,\ldots,p/5$, and $\sigma_{j_1,j_2}^*=0$ otherwise.
\end{itemize}
\vspace{-10pt} Structures A and B lead to, respectively, the banded and block diagonal structures for the precision matrix $\bOmega=\bSigma^{-1}$. Note that, based on such defined covariance $\bSigma$, the diagonal elements of the precision matrix are unit. For each of the precision matrices, we considered two choices for the index set $\mathcal{S}$: (i) all zero components of $\bOmega$, i.e. $\mathcal{S}=\{(j_1,j_2):\omega_{j_1,j_2}=0\}$, and (ii) all the components excluded the ones on the main diagonal, i.e. $\mathcal{S}=\{(j_1,j_2):j_1\neq j_2\}$.
Notice that the sets of all zero components in $\bOmega$ for structures A and B are $\{(j_1,j_2): |j_1-j_2|>1\}$ and $\cap_{h=1}^{p/5}\{(j_1,j_2): 5(h-1)+1\leq j_1, j_2\leq 5h\}^{c}$, respectively.
As we illustrate in the footnote\footnote{It follows from Proposition 1 that $\textrm{Var}\{n^{1/2}(\widehat{\omega}_{j_1,j_2}-\omega_{j_1,j_2})\}=v_{j_1,j_1}^{-2}v_{j_2,j_2}^{-2}\textrm{Var}\{n^{-1/2}\sum_{t=1}^n(\epsilon_{j_1,t}\epsilon_{j_2,t}-v_{j_1,j_2})\}\{1+o(1)\}$, where the term $o(1)$ holds uniformly over $(j_1,j_2)$. Recall $\epsilon_{j,t}=-\balpha_j^\T\by_t$ and $\by_t=(1-\rho^2)^{1/2}\sum_{k=0}^\infty\rho^k\bepsilonb_{t-k}$, if $\omega_{j_1,j_2}=0$ which is equivalent to $v_{j_1,j_2}=0$, then it holds that
$\textrm{Var}(n^{-1/2}\sum_{t=1}^n\epsilon_{j_1,t}\epsilon_{j_2,t})
=n^{-1}(1-\rho^2)^2\sum_{t_1,t_2=1}^n\mathbb{E}\{(\sum_{k=0}^\infty\rho^k\balpha_{j_1}^\T\bepsilonb_{t_1-k})(\sum_{k=0}^\infty\rho^k\balpha_{j_2}^\T\bepsilonb_{t_1-k})(\sum_{k=0}^\infty\rho^k\balpha_{j_1}^\T\bepsilonb_{t_2-k})(\sum_{k=0}^\infty\rho^k\balpha_{j_2}^\T\bepsilonb_{t_2-k})\}.
$
Since $\bepsilonb_t$'s are i.i.d., together with $v_{j_1,j_2}=0$, we have
$
\mathbb{E}\{(\sum_{k=0}^\infty\rho^k\balpha_{j_1}^\T\bepsilonb_{t_1-k})(\sum_{k=0}^\infty\rho^k\balpha_{j_2}^\T\bepsilonb_{t_1-k})(\sum_{k=0}^\infty\rho^k\balpha_{j_1}^\T\bepsilonb_{t_2-k})(\sum_{k=0}^\infty\rho^k\balpha_{j_2}^\T\bepsilonb_{t_2-k})\}=\rho^{2t_2-2t_1}(1-\rho^2)^{-2}\mathbb{E}(\epsilon_{j_1,t}^2\epsilon_{j_2,t}^2)
$ for any $t_2\geq t_1$, which implies $\textrm{Var}(n^{-1/2}\sum_{t=1}^n\epsilon_{j_1,t}\epsilon_{j_2,t})=[1+2(1-\rho^2)^{-2}n^{-1}\{(n-1)\rho^{2n}-(n-2)\rho^{2n+2}-\rho^4\}]\mathbb{E}(\epsilon_{j_1,t}^2\epsilon_{j_2,t}^2)$ for any $(j_1,j_2)$ such that $\omega_{j_1,j_2}=0$. On the other hand, based on the Gaussian assumption, since $v_{j_1,j_2}=\mathbb{E}(\epsilon_{j_1,t}\epsilon_{j_2,t})=0$, we know the two normal distributed random variables $\epsilon_{j_1,t}$ and $\epsilon_{j_2,t}$ are independent, which leads to $\mathbb{E}(\epsilon_{j_1,t}^2\epsilon_{j_2,t}^2)=\mathbb{E}(\epsilon_{j_1,t}^2)\mathbb{E}(\epsilon_{j_2,t}^2)=v_{j_1,j_1}v_{j_2,j_2}$. Therefore, $\textrm{Var}\{n^{1/2}(\widehat{\omega}_{j_1,j_2}-\omega_{j_1,j_2})\}=v_{j_1,j_1}^{-1}v_{j_2,j_2}^{-1}[1+2(1-\rho^2)^{-2}n^{-1}\{(n-1)\rho^{2n}-(n-2)\rho^{2n+2}-\rho^4\}]\{1+o(1)\}$ for any $(j_1,j_2)$ such that $\omega_{j_1,j_2}=0$. Notice that $\omega_{j,j}=1$ in our setting for any $j$, then $v_{j,j}=\omega_{j,j}^{-1}=1$. Hence, the variances of $n^{1/2}(\widehat{\omega}_{j_1,j_2}-\omega_{j_1,j_2})$ for any $(j_1,j_2)$ such that $\omega_{j_1,j_2}=0$ are almost identical.
}, the index sets $\mathcal{S}$ in the setting (i) and (ii) mimic, respectively, the homogeneous and heteroscedastic cases for the variances of $n^{1/2}(\widehat{\omega}_{j_1,j_2}-\omega_{j_1,j_2})$ among $(j_1,j_2)\in\mathcal{S}$.

For each of the cases above, we examined the accuracy of the proposed KMB and SKMB approximations to the distributions of the non-Studentized-type statistic $n^{1/2}|\widehat{\bOmega}_{\mathcal{S}}-\bOmega_{\mathcal{S}}|_\infty$ and the Studentized-type statistic $n^{1/2}|\widehat{\bD}^{-1}(\widehat{\bOmega}_{\mathcal{S}}-\bOmega_{\mathcal{S}})|_\infty$, respectively. Denote by $F_{1n}(\cdot)$ and $F_{2n}(\cdot)$ the distribution functions of $n^{1/2}|\widehat{\bOmega}_{\mathcal{S}}-\bOmega_{\mathcal{S}}|_\infty$ and $n^{1/2}|\widehat{\bD}^{-1}(\widehat{\bOmega}_{\mathcal{S}}-\bOmega_{\mathcal{S}})|_\infty$, respectively. In each of the 1000 independent repetitions, we first draw a sample with size $n$ following the above discussed data generating mechanism, and then computed the associated values of $n^{1/2}|\widehat{\bOmega}_{\mathcal{S}}-\bOmega_{\mathcal{S}}|_\infty$ and $n^{1/2}|\widehat{\bD}^{-1}(\widehat{\bOmega}_{\mathcal{S}}-\bOmega_{\mathcal{S}})|_\infty$ in this sample. Since $F_{1n}(\cdot)$ and $F_{2n}(\cdot)$ are unknown, we used the empirical distributions of $n^{1/2}|\widehat{\bOmega}_{\mathcal{S}}-\bOmega_{\mathcal{S}}|_\infty$ and $n^{1/2}|\widehat{\bD}^{-1}(\widehat{\bOmega}_{\mathcal{S}}-\bOmega_{\mathcal{S}})|_\infty$ over 1000 repetitions, denoted as $F_{1n}^{\ast}(\cdot)$ and $F_{2n}^{\ast}(\cdot)$, to approximate them, respectively.
For each repetition $i$, we applied the KMB and SKMB procedures to estimate the $100(1-\alpha)\%$ quantiles of
$n^{1/2}|\widehat{\bOmega}_{\mathcal{S}}-\bOmega_{\mathcal{S}}|_\infty$ and $n^{1/2}|\widehat{\bD}^{-1}(\widehat{\bOmega}_{\mathcal{S}}-\bOmega_{\mathcal{S}})|_\infty$, denoted as $\widehat{q}_{\mathcal{S}, \alpha, 1}^{(i)}$ and $\widehat{q}_{\mathcal{S}, \alpha, 2}^{(i)}$, respectively, with $M=3000$, and then
computed their associated empirical coverages $ F_{1n}^{\ast}( \widehat{q}_{\mathcal{S}, \alpha, 1}^{(i)} )$ and $ F_{2n}^{\ast}( \widehat{q}_{\mathcal{S}, \alpha, 2}^{(i)} )$. We considered $\alpha=0.075, 0.050$ and $0.025$ in the simulation.
We report the averages and standard deviations of $ \{F_{1n}^{\ast}( \widehat{q}_{\mathcal{S}, \alpha, 1}^{(i)} )\}_{i=1}^{1000}$ and $ \{F_{2n}^{\ast}( \widehat{q}_{\mathcal{S}, \alpha, 2}^{(i)} )\}_{i=1}^{1000}$ in Tables \ref{tb:p100}--\ref{tb:p1500}. Due to the selection of the tuning parameter $\lambda_j$ in (\ref{eq:bestimate}) depends on the standard deviation of the error term $\epsilon_{j,t}$, we adopted the scaled Lasso \citep{SZ_2012} in the simulation which can estimate the regression coefficients and the variance of the error simultaneously. The tuning parameters in scale Lasso were selected according to \cite{Ren_2015}.

It is worth noting that in order to accomplish the  statistical computing for large $p$ under the R environment in high speed,
we programmed  the generation of random numbers and most loops into C functions such that we utilized ``.C()'' routine to call those C functions from R. However, the computation of the two types of statistics involves the fitting of the $p$ node-wise regressions. As a consequence, the simulation for large $p$ still requires a large amount of computation time.
In order to overcome this time-consuming issue,  the computation in this numerical study was undertaken with the assistance of the supercomputer Raijin at the NCI National Facility systems supported by the Australian Government. The supercomputer Raijin comprises 57,864 cores, which helped us parallel process a large number of simulations simultaneously.

From Tables \ref{tb:p100}--\ref{tb:p1500}, we observe that, for both KMB and SKMB procedures, the overall differences between the empirical coverage rates and the corresponding nominal levels are small, which demonstrates that the KMB and SKMB procedures can provide accurate approximations to the distributions of $n^{1/2}|\widehat{\bOmega}_{\mathcal{S}}-\bOmega_{\mathcal{S}}|_\infty$ and $n^{1/2}|\widehat{\bD}^{-1}(\widehat{\bOmega}_{\mathcal{S}}-\bOmega_{\mathcal{S}})|_\infty$, respectively. Also note that the coverage rates improve as $n$ increases. And, our results are robust to the temporal dependence parameter $\rho$, which indicates the proposed procedures are adaptive to time dependent observations.

Comparing the simulation results indicated by KMB and SKMB in the category $\mathcal{S}=\{(j_1,j_2):j_1\neq j_2\}$ of Tables \ref{tb:p100}--\ref{tb:p1500}, when the dimension is less than the sample size ($p=100$, $n=150, 300$), we can see that the SKMB procedure has better accuracy than the KMB procedure if the heteroscedastic issue exists. This finding also exists when the dimension is over the sample size and both of them are large ($n=300$, $p=1500$). For the homogeneous case $\mathcal{S}=\{(j_1,j_2):\omega_{j_1,j_2}=0\}$, the KMB procedure provides better accuracy than the SKMB procedure when sample size is small ($n=150$). However, when the sample size becomes larger $(n=300)$, the accuracy of the SKMB procedure can be significantly improved and it will outperform the KMB procedure. The phenomenon that the SKMB procedure sometimes cannot beat the KMB procedure might be caused by incorporating the estimated standard deviations of $\widehat{\omega}_{j_1,j_2}$'s in the denominator of the Studentized-type statistic, which suffers from high variability when the sample size is small. The simulation results suggest us that: (i) when the dimension is less than the sample size or both the dimension and the sample size are very large, the SKMB procedure should be used to construct the confidence regions of $\bOmega_{\mathcal{S}}$ if the heteroscedastic issue
exists; (ii) if the sample size is small, and we have some previous information that there does not exist heteroscedastic issue, then the KMB procedure should be used to construct the confidence regions of $\bOmega_{\mathcal{S}}$. However, even in the homogeneous case, the SKMB procedure should still be employed when the sample size is large. 


\section{Real data analysis}\label{se:case}

In this section, we follow Example 3 in Section \ref{se:background} to study the partial correlation networks of the Standard and Poors (S\&P) 500 Component Stocks in 2005 (252 trading days, preceding the crisis) and  in 2008 (253 trading days, during the crisis), respectively. The reason to analyze those two periods is to understand the structure and dynamic of financial networks affected by the global financial crisis \citep{Schweitzer_2009}. \cite{Ait-Sahalia_2015} analyzed the data in 2005 and 2008 as well in order to investigate the influence of the financial crisis.

We analyzed the data from {http://quote.yahoo.com/} via the R package \verb"tseries", which contains the daily closing prices of S\&P 500 stocks. The R command \verb"get.hist.quote" can be used to acquire the data. We kept 402 stocks in our analysis whose closing prices were capable of being downloaded by the R command and did not have any missing values during 2005 and 2008.
Let $y_{j,t}$ be the $j$-th stock price at day $t$. We considered the log return of the stocks, which is defined by $\log(y_{j, t}) - \log(y_{j, t-1})$. As kindly pointed out by a referee that the log return data usually exhibit volatility clustering,  we utilized the R package \verb"fGarch" to obtain the conditional standard deviation for the mean centered log return of each stock via fitting a GARCH(1,1) model, and then we standardized the log return by its mean and conditional standard deviation. Ultimately, we had the standardized log returns $\bR_{t} = (R_{1, t}, \ldots, R_{402, t})^\T$ of all the 402 assets at day $t$.

Let $\bOmega=(\omega_{j_1,j_2})_{p\times p}$ be the precision matrix of $\bR_{t}$.
By the relationship between partial correlation and precision matrix, the partial correlation network can be constructed by the non-zero precision coefficients $\omega_{j_1,j_2}$ as demonstrated in Example 3 in Section \ref{se:background}.
To learn the structures of $\bOmega$, we focused on the Global Industry Classification Standard (GICS) sectors and their sub industries of the S\&P 500 companies, and aimed to discover the sub blocks of $\bOmega$ which had nonzero entries. Those blocks could help us build the partial correlation networks of the sectors and  sub industries for the S\&P 500 stocks in 2005 and 2008, respectively.

The advantage of investigating the complex financial network system by partial correlation is to overcome the issue that the marginal correlation between two stocks might be a result of their correlations to other mediating stocks \citep{Kenett_2010}.
For example, if two stocks $R_{j_1,t}$ and $R_{j_2,t}$ are both correlated with some stocks in the set  $\bR_{-(j_1,j_2),t}=\{R_{j,t}: j\neq j_1,j_2 \}$, the partial correlation can suitably remove the linear effect of $\bR_{-(j_1,j_2),t}$ on $R_{j_1,t}$ and $R_{j_2,t}$.
Hence, it measures a ``direct'' relationship between $j_1$ and $j_2$ \citep{DeLaFuente_2004}.
The partial correlation analysis is widely used in the study of financial networks \citep{Shapira_2009,Kenett_2010}, as well as the study of gene networks \citep{DeLaFuente_2004,Reverter_2008,Chen_2009}.

Based on the information on bloomberg and ``List of S\&P 500 companies'' on wikipedia, we identified 10 major sectors with 54 sub industries of the S\&P 500 companies (see Tables \ref{tb:stock1} and \ref{tb:stock2} for detailed categories).
The 10 sectors were Consumer Discretionary, Consumer Staples, Energy, Financials, Health Care, Industrials, Information Technology, Materials, Telecommunication Services and Utilities.
There were one company with the unidentified sector and eight companies with unidentified sub industries due to acquisition or ticket change (represented by ``NA'' in Tables \ref{tb:stock1} and \ref{tb:stock2}).

To explore the partial correlation networks of different sectors and sub industries, we were interested in a set of hypotheses
\be\begin{split}
H_{h_1h_2, 0}:& \  \omega_{j_1,j_2} = 0 \mbox{ \ for any $(j_1, j_2) \in I_{h_1} \times I_{h_2}$ \ vs. \ } \\
H_{h_1h_2, 1}:& \  \omega_{j_1,j_2} \neq 0 \mbox{ \ for some $(j_1, j_2) \in I_{h_1} \times I_{h_2}$}
\end{split}\label{eq:SP500}\ee
for disjoint index sets $\{I_{1}, \ldots, I_{H}\}$, which represented different sub industries.
For each of the hypotheses in (\ref{eq:SP500}), we calculated the Studentized-type statistic ${n}^{1/2}|\widehat{\bD}^{-1}\widehat{\bOmega}_{\mathcal{S}}|_\infty$ in (\ref{eq:cr2}) with $\mathcal{S} = I_{h_1} \times I_{h_2}$ and
apply the SKMB procedure to obtain $M = 10000$ parametric bootstrap samples $\widehat{\bxi}_1,\ldots,\widehat{\bxi}_M$.
The P-value of the hypothesis (\ref{eq:SP500}) was
$$\operatorname{P-value}_{h_1,h_2} = \frac{1}{M} \sum_{m = 1}^{M} \mathbb{I}\{ |\widehat{\bxi}_m|_\infty \geq {n}^{1/2}|\widehat{\bD}^{-1}\widehat{\bOmega}_{\mathcal{S}}|_\infty \} \mbox{ \ for \ } \mathcal{S} = I_{h_1} \times I_{h_2}.$$
To identify the significant blocks, we applied the \citet{Benjamini_1995}'s multiple testing procedure that controls the false discovery rate (FDR) of (\ref{eq:SP500}) at the rate $\alpha=0.1$.
Let $\operatorname{pvalue}_{(1)} \leq \cdots \leq \operatorname{pvalue}_{(K)}$ be the ordered P-values and $H_{(1),0}, \ldots, H_{(K),0}$ be the corresponding null hypotheses, where $K = H(H - 1) / 2$ is the number of hypotheses under our consideration.
Note that 
we had $K = 1431$ for testing sub industry blocks.
We rejected $H_{(1),0}, \ldots, H_{(v),0}$ in (\ref{eq:SP500}) for $v = \max\{1\leq j\leq K : \operatorname{pvalue}_{(j)} \leq \alpha j / K \}$.

\begin{figure}[htp!]
\begin{center}
  \includegraphics[scale=0.24]{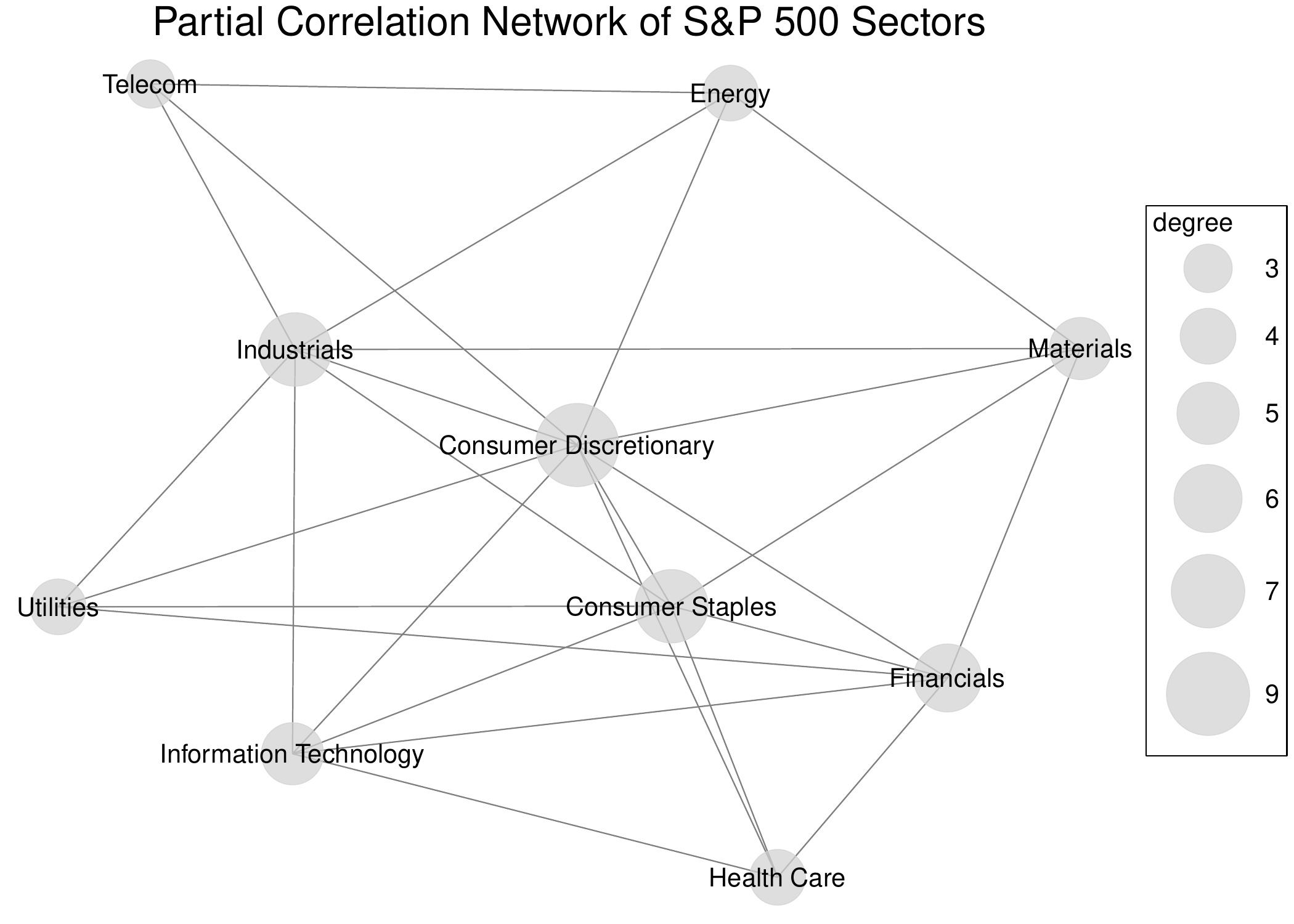}
  \includegraphics[scale=0.24]{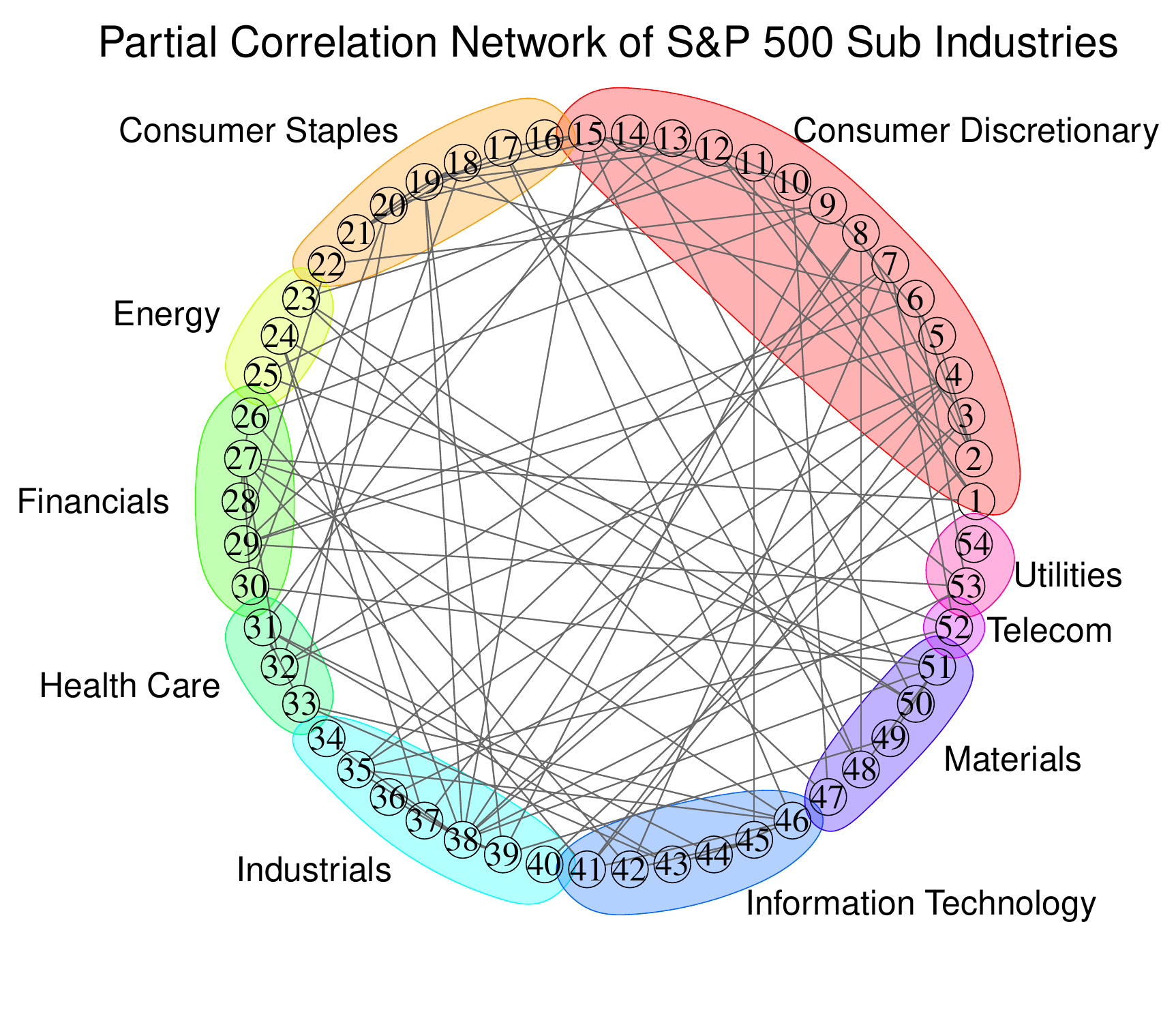}
\end{center}
\caption{Partial correlation networks of S\&P 500 sectors and sub industries in 2005 (preceding the crisis). The detailed information of the sub industries represented by numbers 1-54 in the right panel can be correspondingly found in Tables \ref{tb:stock1} and \ref{tb:stock2}.}
\label{fg:2005}
\end{figure}

\begin{figure}[htp!]
\begin{center}
  \includegraphics[scale=0.24]{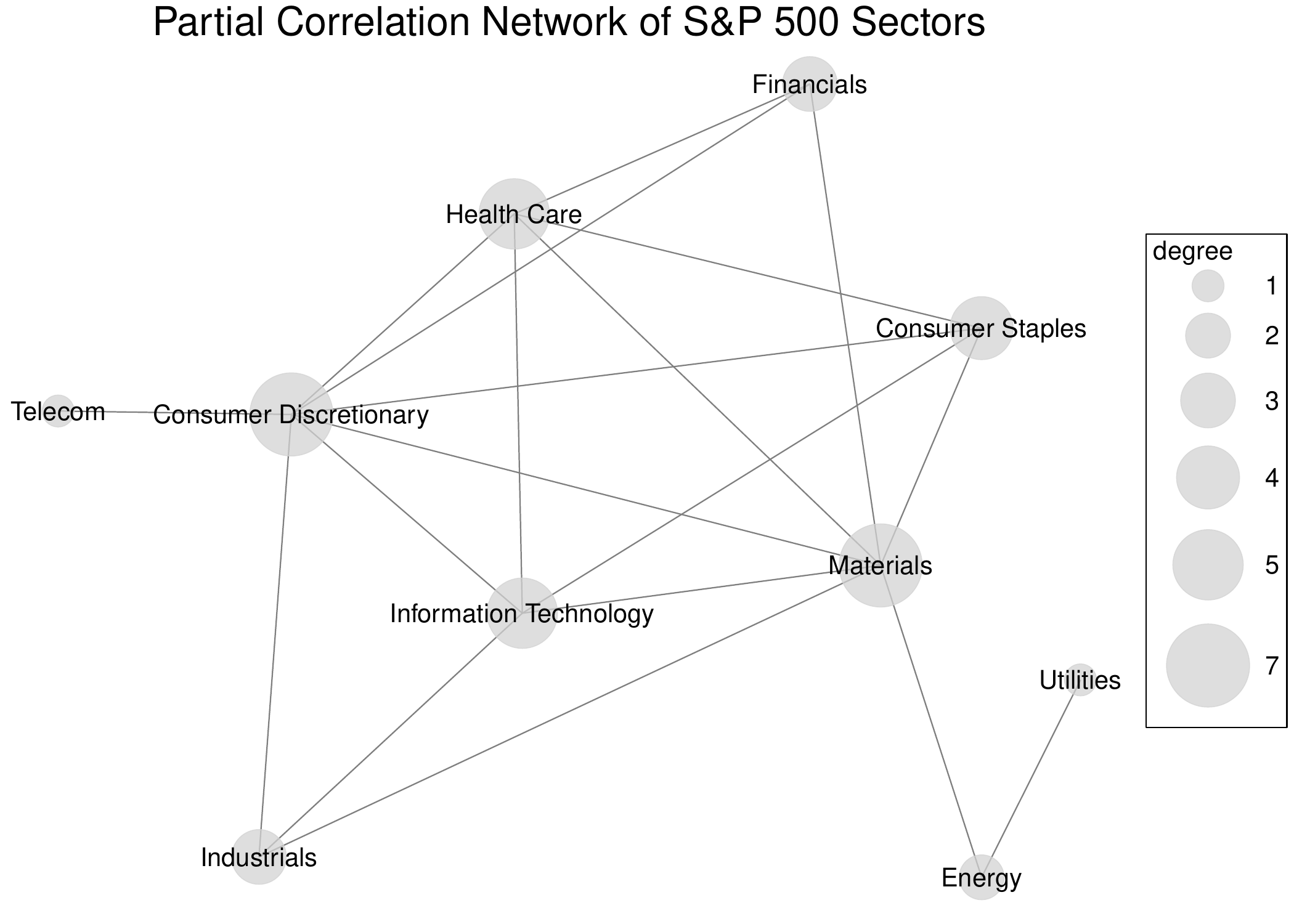}
  \includegraphics[scale=0.24]{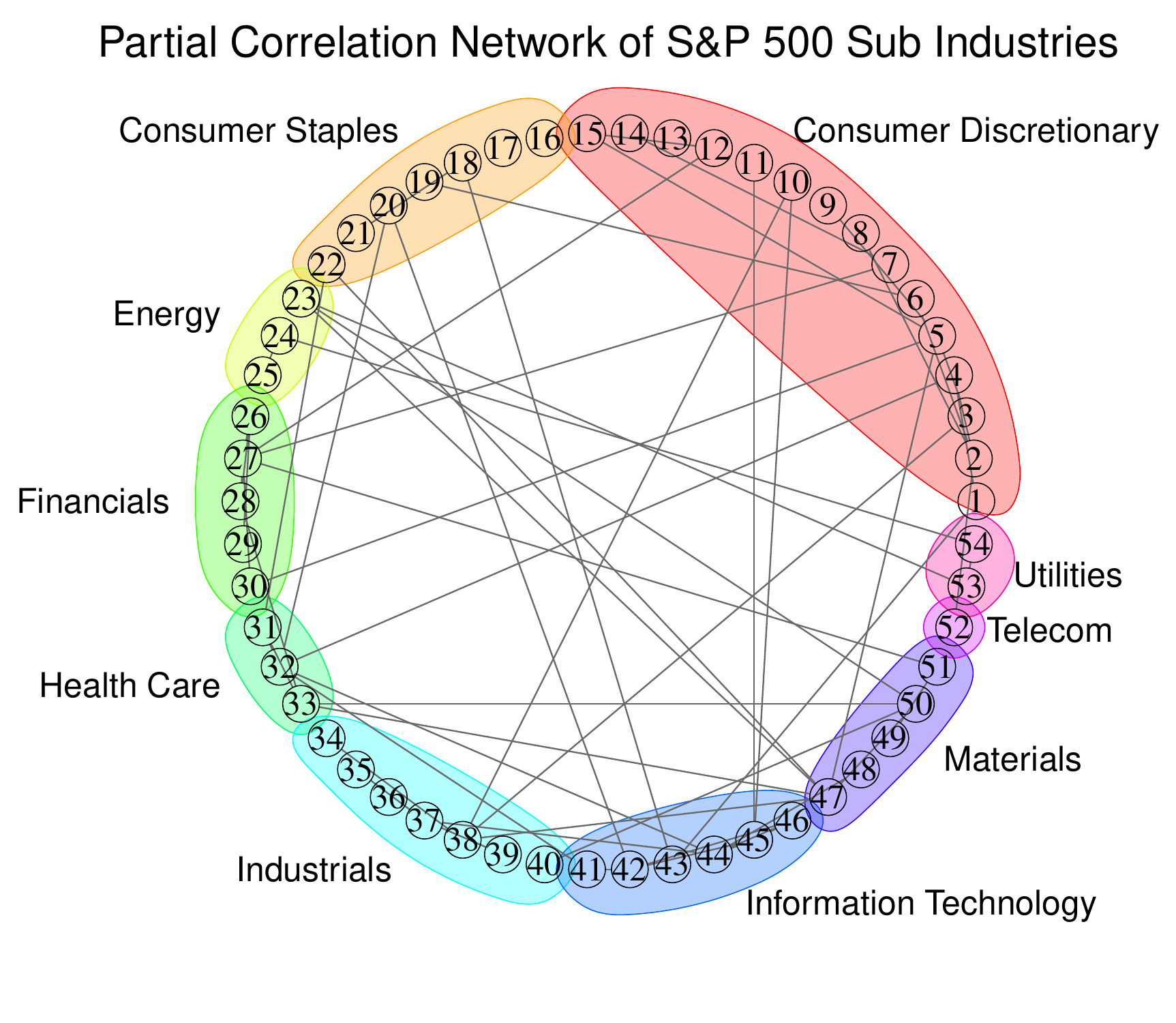}
\end{center}
\caption{Partial correlation networks of S\&P 500 sectors and sub industries in 2008 (during the crisis). The detailed information of the sub industries represented by numbers 1-54 in the right panel can be correspondingly found in Tables \ref{tb:stock1} and \ref{tb:stock2}.}
\label{fg:2008}
\end{figure}

We constructed the partial correlation networks based on the significant blocks from the above multiple testing procedure. The estimated partial correlation networks of the 54 sub industries, labeled by numbers from 1 to 54, are shown in the right panels of Figures \ref{fg:2005} and \ref{fg:2008}, corresponding to 2005 and 2008, respectively. The name of each sub industry and the stocks included can be found in Tables \ref{tb:stock1} and \ref{tb:stock2}. The shaded areas with different colors represent the 10 major sectors, respectively.
The left panels in Figures \ref{fg:2005} and \ref{fg:2008} give the partial correlation networks of the sectors, where the nodes represent the 10 sectors, and two nodes (sectors) $\tilde h_1$ and $\tilde h_2$ are connected if and only if there exists a connection between  one of sub industries belonging to sector $\tilde h_1$ and one of  sub industries belonging to sector $\tilde h_2$ in the right panel.

We observed from the left panel of Figure \ref{fg:2005} that preceding the crisis in 2005, the Consumer Discretionary sector was likely to be a hub connecting to all the other 9 sectors. It was the most influential sector with the largest degree, i.e., the total number of directed links  connecting to the Consumer Discretionary sector  in the network. During the crisis in 2008, the Consumer Discretionary sector was still the most influential sector  as shown by the left panel of Figure \ref{fg:2008}, but it had less connections compared to 2005. The Financials sector was a little bit separated from the other sectors in 2008,  with only half connections in contrast with the network connectivity in 2005. The similar situation also appeared in the partial correlations networks of S\&P 500 sub industries as shown in the right panels of Figures \ref{fg:2005} and \ref{fg:2008}. More specifically, both the numbers of the edges within and between most sectors for the network of S\&P 500 sub industries in 2008 were significantly less than those in 2005 (see Table \ref{tb:degree} for details), which indicated that the market fear in the crisis broke the connections of stock sectors and sub industries. From the perspective of financial network studies, the above analysis confirmed that fear froze the market in the 2008 crisis \citep{Reavis_2012}.

\section*{Acknowledgments}

The authors are grateful to the Co-Editor, an
Associate Editor and two anonymous referees for
constructive comments and suggestions. This research was undertaken with the assistance of resources provided at the NCI National Facility systems at the Australian National University through the ANU Allocation Scheme supported by the Australian Government.
The authors would like to thank Pauline O'Shaughnessy for her great help in getting us started with the supercomputer Raijin.
 Jinyuan Chang was supported in part by the Fundamental Research Funds for the Central Universities
(Grant No. JBK1802069, JBK171121), NSFC (Grant No. 11501462), the Center of Statistical Research
at SWUFE, and the Joint Lab of Data Science and Business Intelligence at SWUFE. Qiwei Yao was supported in part by an EPSRC research grant.

\section*{Appendix}

Throughout the Appendix, let $C$ denote a generic positive constant depending only on the constants specified in Conditions \ref{as:moment}--\ref{as:block}, which may be different in different cases. Let $\rho_1^{-1}=2\gamma_1^{-1}+\gamma_3^{-1}$, $\rho_2^{-1}=2\gamma_2^{-1}+\gamma_3^{-1}$, $\rho_3^{-1}=\gamma_1^{-1}+\gamma_2^{-1}+\gamma_3^{-1}$ and $\rho_4^{-1}=\max\{\rho_2^{-1},\rho_3^{-1}\}+\gamma_3^{-1}$. Define $\zeta=\min\{\rho_1,\rho_2,\rho_3,\rho_4\}$ and $\bDelta=n^{-1}\sum_{t=1}^n\bepsilonbb_t\bepsilonbb_t^\T-\bV=:(\delta_{j_1,j_2})$. 

\begin{lemma}\label{la:1}
Assume Conditions {\rm\ref{as:moment}--\ref{as:betamix}} hold. If $\log p=o\{n^{\zeta/(2-\zeta)}\}$, there exists a uniform constant $A_0>1$ independent of $n$ and $p$ such that
\[
\mathbb{P}\big\{|\widehat{\bSigma}-\bSigma|_\infty>A_1(n^{-1}\log p)^{1/2}\big\}\leq \exp\{-CA_1^{\rho_1}(n\log p)^{\rho_1/2}\}+\exp(-CA_1^2\log p),
\]
\[
\mathbb{P}\big\{|\bDelta|_\infty>A_2(n^{-1}\log p)^{1/2}\big\}\leq \exp\{-CA_2^{\rho_2}(n\log p)^{\rho_2/2}\}+\exp(-CA_2^2\log p),
\]
\[
\sup_{1\leq j\leq p}\mathbb{P}\bigg(\frac{1}{n}\sum_{t=1}^n\epsilon_{j,t}^2>A_3v_{j,j}\bigg)\leq \exp(-CA_3^{\rho_2}n^{\rho_2}),
\]
\[
\sup_{1\leq j\leq p}\mathbb{P}\bigg\{\max_{k\neq j}\bigg|\frac{1}{n}\sum_{t=1}^n\epsilon_{j,t}y_{k,t}\bigg|>A_4(n^{-1}\log p)^{1/2}\bigg\}\leq \exp\{-CA_4^{\rho_3}(n\log p)^{\rho_3/2}\}+\exp(-CA_4^2\log p),
\]
\[
\sup_{1\leq j\leq p}\mathbb{P}\bigg\{\bigg|\frac{1}{n}\sum_{t=1}^n\balpha_{j,-j}^\T\by_{-j,t}\epsilon_{j,t}\bigg|>A_5(n^{-1}\log p)^{1/2}\bigg\}\leq \exp\{-CA_5^{\rho_4}(n\log p)^{\rho_4/2}\}+\exp(-CA_5^2\log p)
\]
for any $A_1,A_2,A_3, A_4, A_5>A_0$.
\end{lemma}

\noindent {\bf Proof:}  For any given $j_1$ and $j_2$, based on the first part of Condition \ref{as:moment}, Lemma 2 of \cite{ChangTangWu_2013} leads to
\[
\sup_{1\leq t\leq n}\mathbb{P}\big(|y_{j_1,t}y_{j_2,t}-\sigma_{j_1,j_2}|>x\big)\leq C\exp(-Cx^{\gamma_1/2})~~\textrm{for any}~x>0.
\]
Hence, for any $x>0$ such that $nx\rightarrow\infty$, Theorem 1 of \cite{Merlevedeetaj_2011} leads to
\[
\mathbb{P}\bigg(\bigg|\frac{1}{n}\sum_{t=1}^ny_{j_1,t}y_{j_2,t}-\sigma_{j_1,j_2}\bigg|>x\bigg)\leq n\exp(-Cn^{\rho_1}x^{\rho_1})+\exp(-Cnx^2).
\]
By Bonferroni inequality, we have
\[
\mathbb{P}\big(|\widehat{\bSigma}-\bSigma|_\infty>x\big)\leq np^2\exp(-Cn^{\rho_1}x^{\rho_1})+p^2\exp(-Cnx^2).
\]
Let $x=A_1(n^{-1}\log p)^{1/2}$, we obtain the first conclusion. Following the same arguments, we can establish the other inequalities. $\hfill\Box$

\begin{lemma}\label{la:lasso}
Assume Conditions {\rm\ref{as:moment}--\ref{as:betamix}} hold. Let $s=\max_{1\leq j\leq p}|\balpha_j|_0$. For some suitable $\lambda_j\asymp (n^{-1}\log p)^{1/2}$ for each $j=1,\ldots,p$, we have
\[
\max_{1\leq j\leq p}|\widehat{\balpha}_j-\balpha_j|_1=o_p\{(\log p)^{-1}\}~~~\textrm{and}~~~\max_{1\leq j\leq p}|\widehat{\balpha}_j-\balpha_j|_2=o_p\{(n\log p)^{-1/4}\}
\]
provided that $\log p=o\{n^{\zeta/(2-\zeta)}\}$ and $s^2(\log p)^3n^{-1}=o(1)$.
\end{lemma}

\noindent {\bf Proof:} Define
\[
\mathscr{T}=\bigg\{\max_{1\leq j\leq p}\max_{k\neq j}\bigg|\frac{1}{n}\sum_{t=1}^n\epsilon_{j,t}y_{k,t}\bigg|\leq A_4(n^{-1}\log p)^{1/2}\bigg\}
\]
for some $A_4>A_0$, where $A_0$ is given in Lemma \ref{la:1}. Selecting $\lambda_j\geq 4A_4(n^{-1}\log p)^{1/2}$ for any $j$, Theorem 6.1 and Corollary 6.8 of \cite{BuhlmannvandeGeer_2011} imply that, restricted on $\mathscr{T}$, we have
\begin{equation}\label{eq:l1}
\max_{1\leq j\leq p}|\widehat{\balpha}_j-\balpha_j|_1 \leq Cs(n^{-1}\log p)^{1/2}
\end{equation}
and
\begin{equation}\label{eq:l2}
(\widehat{\balpha}_j-\balpha_j)^\T\widehat{\bSigma}_{-j,-j}(\widehat{\balpha}_j-\balpha_j)\leq Csn^{-1}\log p
\end{equation}
with probability approaching 1. By Bonferroni inequality and Lemma \ref{la:1},
\[
\begin{split}
\mathbb{P}(\mathscr{T}^c)\leq&~\sum_{j=1}^p\mathbb{P}\bigg\{\sum_{k\neq j}\bigg|\frac{1}{n}\sum_{t=1}^n\epsilon_{j,t}y_{k,t}\bigg|>A_4(n^{-1}\log p)^{1/2}\bigg\}\\
\leq&~p\exp\{-CA_4^{\rho_3}(n\log p)^{\rho_3/2}\}+p\exp(-CA_4^2\log p).
\end{split}
\]
For suitable selection of $A_4$, we have $\mathbb{P}(\mathscr{T}^c)\rightarrow0$ as $n\rightarrow\infty$. Thus, from (\ref{eq:l1}), it holds that
\begin{equation}\label{eq:m1}
\begin{split}
\max_{1\leq j\leq p}|\widehat{\balpha}_j-\balpha_j|_1=&~O_p\{s(n^{-1}\log p)^{1/2}\}\\
=&~o_p\{(\log p)^{-1}\}.
\end{split}
\end{equation}
On the other hand, notice that
\[
\begin{split}
(\widehat{\balpha}_j-\balpha_j)^\T\widehat{\bSigma}_{-j,-j}(\widehat{\balpha}_j-\balpha_j)\geq&~\lambda_{\min}(\bSigma_{-j,-j})|\widehat{\balpha}_j-\balpha_j|_2^2\\
&-|\widehat{\bSigma}_{-j,-j}-\bSigma_{-j,-j}|_\infty|\widehat{\balpha}_j-\balpha_j|_1^2,
\end{split}
\]
by Condition \ref{as:cov}, Lemma \ref{la:1}, (\ref{eq:l2}) and (\ref{eq:m1}), we have
\[
\begin{split}
\max_{1\leq j\leq p}|\widehat{\balpha}_j-\balpha_j|_2=&~O_p\{(sn^{-1}\log p)^{1/2}\}\\
=&~o_p\{(n\log p)^{-1/4}\}.
\end{split}
\]
Hence, we complete the proof. $\hfill\Box$

\begin{lemma}\label{la:bias}
Assume the conditions for Lemmas {\rm\ref{la:1}} and {\rm\ref{la:lasso}} hold, then
\[
\begin{split}
&\frac{1}{n}\sum_{t=1}^n\widehat{\epsilon}_{j_1,t}\widehat{\epsilon}_{j_2,t}-\frac{1}{n}\sum_{t=1}^n\epsilon_{j_1,t}\epsilon_{j_2,t}\\
=&-(\widehat{\alpha}_{j_1,j_2}-\alpha_{j_1,j_2})\bigg(\frac{1}{n}\sum_{t=1}^n\epsilon_{j_2,t}^2\bigg)\mathbb{I}(j_1\neq j_2)\\
&-(\widehat{\alpha}_{j_2,j_1}-\alpha_{j_2,j_1})\bigg(\frac{1}{n}\sum_{t=1}^n\epsilon_{j_1,t}^2\bigg)\mathbb{I}(j_1\neq j_2)\\
&+o_p\{(n\log p)^{-1/2}\}.
\end{split}
\]
Here the remainder term $o_p\{(n\log p)^{-1/2}\}$ is uniform over all $j_1$ and $j_2$.
\end{lemma}

\noindent {\bf Proof:} Notice that $\epsilon_{j,t}=-\balpha_{j}^\T\by_t$ and $\widehat{\epsilon}_{j,t}=-\widehat{\balpha}_j^\T\by_t$ for any $t$, then
\[
\begin{split}
\frac{1}{n}\sum_{t=1}^n\widehat{\epsilon}_{j_1,t}\widehat{\epsilon}_{j_2,t}-\frac{1}{n}\sum_{t=1}^n\epsilon_{j_1,t}\epsilon_{j_2,t}=&-\frac{1}{n}\sum_{t=1}^n(\widehat{\balpha}_{j_1}-\balpha_{j_1})^\T\by_t\epsilon_{j_2,t}\\
&-\frac{1}{n}\sum_{t=1}^n(\widehat{\balpha}_{j_2}-\balpha_{j_2})^\T\by_t\epsilon_{j_1,t}\\
&+\frac{1}{n}\sum_{t=1}^n(\widehat{\balpha}_{j_1}-\balpha_{j_1})^\T\by_t\by_t^\T(\widehat{\balpha}_{j_2}-\balpha_{j_2}).
\end{split}
\]
Condition \ref{as:cov}, Lemmas \ref{la:1} and \ref{la:lasso} imply that
\[
\begin{split}
&\max_{1\leq j_1,j_2\leq p}\bigg|\frac{1}{n}\sum_{t=1}^n(\widehat{\balpha}_{j_1}-\balpha_{j_1})^\T\by_t\by_t^\T(\widehat{\balpha}_{j_2}-\balpha_{j_2})\bigg|\\
\leq&\max_{1\leq j_1,j_2\leq p}|(\widehat{\balpha}_{j_1}-\balpha_{j_1})^\T\bSigma (\widehat{\balpha}_{j_2}-\balpha_{j_2})|\\
&+\max_{1\leq j_1,j_2\leq p}|(\widehat{\balpha}_{j_1}-\balpha_{j_1})^\T(\widehat{\bSigma}-\bSigma) (\widehat{\balpha}_{j_2}-\balpha_{j_2})|\\
\leq&~C\max_{1\leq j\leq p}|\widehat{\balpha}_j-\balpha_j|_2^2+|\widehat{\bSigma}-\bSigma|_\infty\max_{1\leq j\leq p}|\widehat{\balpha}_j-\balpha_j|_1^2\\
=&~o_p\{(n\log p)^{-1/2}\}.
\end{split}
\]
Meanwhile, by Lemma \ref{la:1}, we have
$
\max_{1\leq j\leq p}\max_{k\neq j}|n^{-1}\sum_{t=1}^n\epsilon_{j,t}y_{k,t}|=O_p\{(n^{-1}\log p)^{1/2}\},
$
which implies that
\[
\begin{split}
&\max_{1\leq j_1,j_2\leq p}\bigg|\sum_{k\neq j_1,j_2}(\widehat{\alpha}_{j_1,k}-\alpha_{j_1,k})\bigg(\frac{1}{n}\sum_{t=1}^ny_{k,t}\epsilon_{j_2,t}\bigg)\bigg|\\
\leq& \max_{1\leq j\leq p}|\widehat{\balpha}_j-\balpha_j|_1\cdot\max_{1\leq j\leq p}\max_{k\neq j}\bigg|\frac{1}{n}\sum_{t=1}^ny_{k,t}\epsilon_{j,t}\bigg|\\
=&~o_p\{(n\log p)^{-1/2}\}.
\end{split}
\]
Therefore, we have
\begin{equation}\label{eq:eq1}
\begin{split}
&~\frac{1}{n}\sum_{t=1}^n(\widehat{\balpha}_{j_1}-\balpha_{j_1})^\T\by_t\epsilon_{j_2,t}\\
=&~(\widehat{\alpha}_{j_1,j_2}-\alpha_{j_1,j_2})\bigg(\frac{1}{n}\sum_{t=1}^ny_{j_2,t}\epsilon_{j_2,t}\bigg)\mathbb{I}(j_1\neq j_2)\\
&+\sum_{k\neq j_1,j_2}(\widehat{\alpha}_{j_1,k}-\alpha_{j_1,k})\bigg(\frac{1}{n}\sum_{t=1}^ny_{k,t}\epsilon_{j_2,t}\bigg)\\
=&~(\widehat{\alpha}_{j_1,j_2}-\alpha_{j_1,j_2})\bigg(\frac{1}{n}\sum_{t=1}^ny_{j_2,t}\epsilon_{j_2,t}\bigg)\mathbb{I}(j_1\neq j_2)\\
&+o_p\{(n\log p)^{-1/2}\}.
\end{split}
\end{equation}
Here the remainder term is uniform over any $j_1$ and $j_2$. On the other hand,
$
{n}^{-1}\sum_{t=1}^ny_{j,t}\epsilon_{j,t}={n}^{-1}\sum_{t=1}^n\epsilon_{j,t}^2+{n}^{-1}\sum_{t=1}^n\balpha_{j,-j}^\T\by_{-j,t}\epsilon_{j,t}.
$
By the fourth result of Lemma \ref{la:1}, it yields that
$
{n}^{-1}\sum_{t=1}^ny_{j,t}\epsilon_{j,t}={n}^{-1}\sum_{t=1}^n\epsilon_{j,t}^2+O_p\{(n^{-1}\log p)^{1/2}\}.
$
Here the remainder term is uniform over all $j$. Together with (\ref{eq:eq1}), we have
\[
\begin{split}
&~\frac{1}{n}\sum_{t=1}^n(\widehat{\balpha}_{j_1}-\balpha_{j_1})^\T\by_t\epsilon_{j_2,t}\\
=&~(\widehat{\alpha}_{j_1,j_2}-\alpha_{j_1,j_2})\bigg(\frac{1}{n}\sum_{t=1}^n\epsilon_{j_2,t}^2\bigg)\mathbb{I}(j_1\neq j_2)\\
&+o_p\{(n\log p)^{-1/2}\}.
\end{split}
\]
Here the remainder term is also uniform over all $j_1$ and $j_2$. Hence,
\[
\begin{split}
&\frac{1}{n}\sum_{t=1}^n\widehat{\epsilon}_{j_1,t}\widehat{\epsilon}_{j_2,t}-\frac{1}{n}\sum_{t=1}^n\epsilon_{j_1,t}\epsilon_{j_2,t}\\
=&-(\widehat{\alpha}_{j_1,j_2}-\alpha_{j_1,j_2})\bigg(\frac{1}{n}\sum_{t=1}^n\epsilon_{j_2,t}^2\bigg)\mathbb{I}(j_1\neq j_2)\\
&-(\widehat{\alpha}_{j_2,j_1}-\alpha_{j_2,j_1})\bigg(\frac{1}{n}\sum_{t=1}^n\epsilon_{j_1,t}^2\bigg)\mathbb{I}(j_1\neq j_2)\\
&+o_p\{(n\log p)^{-1/2}\}.
\end{split}
\]
We complete the proof. $\hfill\Box$

\bigskip

\noindent {\bf Proof of Proposition \ref{pro:1}:} Notice that $v_{j_1,j_2} = \frac{\omega_{j_1,j_2}}{\omega_{j_1,j_1}\omega_{j_2,j_2}}$,  $\alpha_{j_1,j_2} = -\frac{\omega_{j_1,j_2}}{\omega_{j_1,j_1}}$ and $\widetilde{v}_{j_1,j_2}=n^{-1}\sum_{t=1}^n\widehat{\epsilon}_{j_1,t}\widehat{\epsilon}_{j_2,t}$ for any $j_1$ and $j_2$, Lemma \ref{la:bias} implies that
\[
\begin{split}
-\widehat{v}_{j_1,j_2}+v_{j_1,j_2}=&~\widetilde{v}_{j_1,j_2}+\frac{\widehat{\alpha}_{j_1,j_2}}{n}\sum_{t=1}^n\widehat{\epsilon}_{j_2,t}^2+\frac{\widehat{\alpha}_{j_2,j_1}}{n}\sum_{t=1}^n\widehat{\epsilon}_{j_1,t}^2+v_{j_1,j_2}\\
=&~\frac{1}{n}\sum_{t=1}^n(\epsilon_{j_1,t}\epsilon_{j_2,t}-v_{j_1,j_2})+\frac{\alpha_{j_1,j_2}}{n}\sum_{t=1}^n(\epsilon_{j_2,t}^2-v_{j_2,j_2})\\
&+\frac{\alpha_{j_2,j_1}}{n}\sum_{t=1}^n(\epsilon_{j_1,t}^2-v_{j_1,j_1})+o_p\{(n\log p)^{-1/2}\}\\
\end{split}
\]
for any $j_1\neq j_2$. Recall $\bDelta=n^{-1}\sum_{t=1}^n\bepsilonbb_t\bepsilonbb_t^\T-\bV=:(\delta_{j_1,j_2})$. It follows from Lemma \ref{la:1} that
$
\max_{1\leq j_1,j_2\leq p}|\delta_{j_1,j_2}|=O_p\{(n^{-1}\log p)^{1/2}\}.
$
Recall $\widehat{\omega}_{j_1,j_2}=\frac{\widehat{v}_{j_1,j_2}}{\widehat{v}_{j_1,j_1}\widehat{v}_{j_2,j_2}}$, if $\log p=o\{n^{\zeta/(2-\zeta)}\}$ for $\zeta$ specified in Lemma \ref{la:1} and $s^2(\log p)^3n^{-1}=o(1)$, it holds that
\[
\begin{split}
\widehat{\omega}_{j_1,j_2}-\omega_{j_1,j_2}=&~\frac{v_{j_1,j_2}-\delta_{j_1,j_2}-\alpha_{j_1,j_2}\delta_{j_2,j_2}-\alpha_{j_2,j_1}\delta_{j_1,j_1}+o_p\{(n\log p)^{-1/2}\}}{[v_{j_1,j_1}+\delta_{j_1,j_1}+o_p\{(n\log p)^{-1/2}\}][v_{j_2,j_2}+\delta_{j_2,j_2}+o_p\{(n\log p)^{-1/2}\}]}-\frac{v_{j_1,j_2}}{v_{j_1,j_1}v_{j_2,j_2}}\\
=&~-\frac{\delta_{j_1,j_2}}{v_{j_1,j_1}v_{j_2,j_2}}+o_p\{(n\log p)^{-1/2}\}
\end{split}
\]
for any $j_1\neq j_2$. Meanwhile, by the same arguments, for each $j=1,\ldots,p$, it holds that
$
\widehat{\omega}_{j,j}-\omega_{j,j}=-\frac{\delta_{j,j}}{v_{j,j}^2}+o_p\{(n\log p)^{-1/2}\}.
$
This proves Proposition \ref{pro:1}. $\hfill\Box$

\bigskip

\noindent {\bf Proof of Theorem \ref{tm:1}:} Define
$
d_1=\sup_{x>0}|\mathbb{P}({n}^{1/2}|\bPi_{\mathcal{S}}|_{\infty}>x)-\mathbb{P}(|\bxi|_\infty>x)|.
$
For any $x>0$ and $\varepsilon_1>0$, it yields that
\begin{equation*}
\begin{split}
&~\mathbb{P}\big({n}^{1/2}|\widehat{\bOmega}_{\mathcal{S}}-\bOmega_{\mathcal{S}}|_\infty>x\big)\\
\leq&~ \mathbb{P}({n}^{1/2}|\bPi_{\mathcal{S}}|_{\infty}>x-\varepsilon_1)+\mathbb{P}({n}^{1/2}|\bUpsilon_{\mathcal{S}}|_\infty>\varepsilon_1)\\
\leq&~\mathbb{P}(|\bxi|_\infty>x-\varepsilon_1)+d_1+\mathbb{P}({n}^{1/2}|\bUpsilon_{\mathcal{S}}|_\infty>\varepsilon_1)\\
=&~\mathbb{P}(|\bxi|_\infty>x)+\mathbb{P}(x-\varepsilon_1<|\bxi|_\infty\leq x)+d_1\\
&+\mathbb{P}({n}^{1/2}|\bUpsilon_{\mathcal{S}}|_\infty>\varepsilon_1).
\end{split}
\end{equation*}
On the other hand, notice that $\mathbb{P}\big({n}^{1/2}|\widehat{\bOmega}_{\mathcal{S}}-\bOmega_{\mathcal{S}}|_\infty>x\big)\geq \mathbb{P}({n}^{1/2}|\bPi_{\mathcal{S}}|_{\infty}>x+\varepsilon_1)-\mathbb{P}({n}^{1/2}|\bUpsilon_{\mathcal{S}}|_\infty>\varepsilon_1)$, following the same arguments, we have
\begin{equation}\label{eq:bound1}
\begin{split}
&~\sup_{x>0}\big|\mathbb{P}\big({n}^{1/2}|\widehat{\bOmega}_{\mathcal{S}}-\bOmega_{\mathcal{S}}|_\infty>x\big)-\mathbb{P}(|\bxi|_\infty>x)\big|\\
\leq&~ d_1+\sup_{x>0}\mathbb{P}(x-\varepsilon_1<|\bxi|_\infty\leq x+\varepsilon_1)+\mathbb{P}({n}^{1/2}|\bUpsilon_{\mathcal{S}}|_\infty>\varepsilon_1).
\end{split}
\end{equation}
By the Anti-concentration inequality for Gaussian random vector [Corollary 1 of \cite{CCK_2015}], it holds that
\begin{equation}\label{eq:anti}
\sup_{x>0}\mathbb{P}(x-\varepsilon_1<|\bxi|_\infty\leq x+\varepsilon_1)\leq C\varepsilon_1(\log p)^{1/2}
\end{equation}
for any $\varepsilon_1\rightarrow0$. From the proofs of Lemmas \ref{la:lasso} and \ref{la:bias}, we know ${n}^{1/2}|\bUpsilon_{\mathcal{S}}|_\infty=O_p(sn^{-1/2}\log p)$. Thus, if $s^2(\log p)^3n^{-1}=o(1)$, we can select a suitable $\varepsilon_1$ to guarantee $\varepsilon_1(\log p)^{1/2}\rightarrow0$ and ${n}^{1/2}|\bUpsilon_{\mathcal{S}}|_\infty=o_p(\varepsilon_1)$. Therefore, for such selected $\varepsilon_1$, (\ref{eq:bound1}) leads to
\begin{equation}
\sup_{x>0}\big|\mathbb{P}\big({n}^{1/2}|\widehat{\bOmega}_{\mathcal{S}}-\bOmega_{\mathcal{S}}|_\infty>x\big)-\mathbb{P}(|\bxi|_\infty>x)\big|\leq d_1+o(1).
\end{equation}
To prove Theorem \ref{tm:1}, it suffices to show $d_1\rightarrow0$ as $n\rightarrow\infty$. We will show this below.

Write $\bPi_{\mathcal{S}}=-(\bar{\varsigma}_1,\ldots,\bar{\varsigma}_r)^\T$ where $\bar{\varsigma}_j=n^{-1}\sum_{t=1}^n\varsigma_{j,t}$ and $\bxi=(\xi_1,\ldots,\xi_r)^\T$. 
Given a $D_n\rightarrow\infty$, define
$
\varsigma_{j,t}^{+}=\varsigma_{j,t}\mathbb{I}\{|\varsigma_{j,t}|\leq D_n\}-\mathbb{E}[\varsigma_{j,t}\mathbb{I}\{|\varsigma_{j,t}|\leq D_n\}]$ and $
\varsigma_{j,t}^{-}=\varsigma_{j,t}\mathbb{I}\{|\varsigma_{j,t}|> D_n\}-\mathbb{E}[\varsigma_{j,t}\mathbb{I}\{|\varsigma_{j,t}|> D_n\}].
$
Write $\bvarsigma_t^+=(\varsigma_{1,t}^+,\ldots,\varsigma_{r,t}^+)^\T$ and $\bvarsigma_t^-=(\varsigma_{1,t}^-,\ldots,\varsigma_{r,t}^-)^\T$ for each $t$. The diverging rate of $D_n$ will be specified later. Let $L$ be a positive integer satisfying $L\leq n/2$, $L\rightarrow\infty$ and $L=o(n)$. We decompose the sequence $\{1,\ldots,n\}$ to the following $m+1$ blocks where $m=\lfloor n/L\rfloor$ and $\lfloor\cdot\rfloor$ is the integer truncation operator: $\mathcal{G}_{\ell}=\{(\ell-1)L+1,\ldots,\ell L\}$ $(\ell=1,\ldots,m)$ and $\mathcal{G}_{m+1}=\{mL+1,\ldots,n\}$. Additionally, let $b>h$ be two positive integers such that $L=b+h$, $h\rightarrow\infty$ and $h=o(b)$. We decompose each $\mathcal{G}_{\ell}$ $(\ell=1,\ldots,m)$ to a ``large" block with length $b$ and a ``small" block with length $h$. Specifically, $\mathcal{I}_{\ell}=\{(\ell-1)L+1,\ldots,(\ell-1)L+b\}$ and $\mathcal{J}_{\ell}=\{(\ell-1)L+b+1,\ldots,\ell L\}$ for any $\ell=1,\ldots,m$, and
$\mathcal{J}_{m+1}=\mathcal{G}_{m+1}$. Assume $\bu$ is a centered normal random vector such that
\[
\bu=({u}_1,\ldots,{u}_{r})^\T\sim N\bigg[\bzero,\frac{1}{mb}\sum_{\ell=1}^m\mathbb{E}\bigg\{\bigg(\sum_{t\in\mathcal{I}_\ell}\bvarsigma_{t}^+\bigg)\bigg(\sum_{t\in\mathcal{I}_\ell}\bvarsigma_t^+\bigg)^\T\bigg\}\bigg].
\]
Our following proof includes two steps. The first step is to show
\begin{equation}\label{eq:step1}
d_2:=\sup_{x>0}\big|\mathbb{P}\big({n}^{1/2}|\bPi_{\mathcal{S}}|_\infty>x\big)-\mathbb{P}(|\bu|_\infty>x)\big|=o(1). \end{equation} And the second step is to show
\begin{equation}\label{eq:step2}
\sup_{x>0}\big|\mathbb{P}(|\bu|_\infty>x)-\mathbb{P}(|\bxi|_\infty>x)\big|=o(1).
\end{equation}
From (\ref{eq:step1}) and (\ref{eq:step2}), we have $d_1=o(1)$.

We first show (\ref{eq:step1}). Define
$
d_3=\sup_{x>0}|\mathbb{P}(|n^{-1/2}\sum_{t=1}^n\bvarsigma_t^+|_\infty>x)-\mathbb{P}(|\bu|_\infty>x)|.
$
Notice that ${n}^{1/2}\bPi_{\mathcal{S}}=n^{-1/2}\sum_{t=1}^n\bvarsigma_t^++n^{-1/2}\sum_{t=1}^n\bvarsigma_t^-$, by the triangle inequality, it holds that
$
|{n}^{1/2}|\bPi_{\mathcal{S}}|_\infty-|{n}^{-1/2}\sum_{t=1}^n\bvarsigma_t^+|_\infty|\leq |{n}^{-1/2}\sum_{t=1}^n\bvarsigma_t^-|_\infty.
$
Similar to (\ref{eq:bound1}), we have
\begin{equation}\label{eq:d2}
d_2\leq d_3+\sup_{x>0}\mathbb{P}(x-\varepsilon_2<|\bu|_\infty\leq x+\varepsilon_2)+\mathbb{P}\bigg(\bigg|\frac{1}{{n}^{1/2}}\sum_{t=1}^n\bvarsigma_t^-\bigg|_\infty>\varepsilon_2\bigg)
\end{equation}
for any $\varepsilon_2>0$. For each $j$, it follows from Davydov inequality \citep{Davydov_1968} that
\[
\begin{split}
\mathbb{E}\bigg(\bigg|\frac{1}{\sqrt{n}}\sum_{t=1}^n\varsigma_{j,t}^-\bigg|^2\bigg)=&~\frac{1}{n}\sum_{t=1}^n\mathbb{E}\{(\varsigma_{j,t}^-)^2\}+\frac{1}{n}\sum_{t_1\neq t_2}\mathbb{E}(\varsigma_{j,t_1}^-\varsigma_{j,t_2}^-)\\
\leq&~\frac{1}{n}\sum_{t=1}^n\mathbb{E}\{(\varsigma_{j,t}^-)^2\}+\frac{C}{n}\sum_{t_1\neq t_2}[\mathbb{E}\{(\varsigma_{j,t_1}^-)^4\}]^{1/4}[\mathbb{E}\{(\varsigma_{j,t_2}^-)^4\}]^{1/4}\exp(-C|t_1-t_2|^{\gamma_3})\\
\end{split}
\]
Applying Lemma 2 of \cite{ChangTangWu_2013}, Conditions \ref{as:moment} and \ref{as:block} imply that $\sup_{j,t}\mathbb{P}(|\varsigma_{j,t}|>x)\leq C\exp(-Cx^{\gamma_2/2})$ for any $x>0$. Then
\begin{equation}\label{eq:tailbound}
\begin{split}
\mathbb{E}\{\varsigma_{j,t}^4\mathbb{I}(|\varsigma_{j,t}|>D_n)\}=&~4\int_0^{D_n}x^3\mathbb{P}(|\varsigma_{j,t}|>D_n)~dx+4\int_{D_n}^\infty x^3\mathbb{P}(|\varsigma_{j,t}|>x)~dx\\
\leq&~CD_n^4\exp(-CD_n^{\gamma_2/2}).
\end{split}
\end{equation}
By the triangle inequality and Jensen's inequality,
\begin{equation}\label{eq:1}
\begin{split}
\mathbb{E}\{(\varsigma_{j,t}^-)^4\}\leq&~ C\mathbb{E}\{\varsigma_{j,t}^4\mathbb{I}(|\varsigma_{j,t}|>D_n)\}+C[\mathbb{E}\{\varsigma_{j,t}\mathbb{I}(|\varsigma_{j,t}|>D_n)\}]^4\\
\leq&~C\mathbb{E}\{\varsigma_{j,t}^4\mathbb{I}(|\varsigma_{j,t}|>D_n)\}\\
\leq&~CD_n^4\exp(-CD_n^{\gamma_2/2}),
\end{split}
\end{equation}
which implies that
\[
\begin{split}
\sup_{1\leq j\leq r}\mathbb{E}\bigg(\bigg|\frac{1}{{n}^{1/2}}\sum_{t=1}^n\varsigma_{j,t}^-\bigg|^2\bigg)\leq&~ CD_n^2\exp(-CD_n^{\gamma_2/2})+CD_n^2\exp(-CD_n^{\gamma_2/2})\sum_{k=1}^{n-1}\exp(-Ck^{\gamma_3})\\
\leq&~CD_n^2\exp(-CD_n^{\gamma_2/2}).
\end{split}
\]
Thus, it follows from Markov inequality that
\[
\begin{split}
\mathbb{P}\bigg(\bigg|\frac{1}{{n}^{1/2}}\sum_{t=1}^n\bvarsigma_t^-\bigg|_\infty>\varepsilon_2\bigg)\leq&~ \frac{r}{\varepsilon_2^2}\sup_{1\leq j\leq r}\mathbb{E}\bigg(\bigg|\frac{1}{{n}^{1/2}}\sum_{t=1}^n\varsigma_{j,t}^-\bigg|^2\bigg)\\
\leq&~Cr\varepsilon_2^{-2}D_n^2\exp(-CD_n^{\gamma_2/2}).
\end{split}
\]
Similar to (\ref{eq:anti}), it holds that
$
\sup_{x>0}\mathbb{P}(x-\varepsilon_2<|\bu|_\infty\leq x+\varepsilon_2)\leq C\varepsilon_2(\log p)^{1/2}
$ for any $\varepsilon_2\rightarrow0$.
If we choose $\varepsilon_2=(\log p)^{-1}$ and $D_n=C(\log p)^{2/\gamma_2}$ for some sufficiently large $C$, then
$
\sup_{x>0}\mathbb{P}(x-\varepsilon_2<|\bu|_\infty\leq x+\varepsilon_2)+\mathbb{P}(|{n}^{-1/2}\sum_{t=1}^n\bvarsigma_t^-|_\infty>\varepsilon_2)=o(1).
$
Therefore, (\ref{eq:d2}) implies $d_2\leq d_3+o(1)$. To show (\ref{eq:step1}) that $d_2=o(1)$, it suffices to prove $d_3=o(1)$.

Let $\bvarsigma_{t}^{+,\textrm{ext}}=(\bvarsigma_t^{+,\T},-\bvarsigma_t^{+,\T})^\T=(\varsigma_{1,t}^{+,\textrm{ext}},\ldots,\varsigma_{2r,t}^{+,\textrm{ext}})^\T$ and $\bu^{\textrm{ext}}=(\bu^{\T},-\bu^{\T})^\T=(u_1^{\textrm{ext}},\ldots,u_{2r}^{\textrm{ext}})^\T$. To prove $d_3=\sup_{x>0}|\mathbb{P}(|n^{-1/2}\sum_{t=1}^n\bvarsigma_t^+|_\infty>x)-\mathbb{P}(|\bu|_\infty>x)|\rightarrow0$, it is equivalent to show $\sup_{x>0}|\mathbb{P}(\max_{1\leq j\leq 2r}n^{-1/2}\sum_{t=1}^n\varsigma_{j,t}^{+,\textrm{ext}}>x)-\mathbb{P}(\max_{1\leq j\leq 2r}u_{j}^{\textrm{ext}}>x)|\rightarrow0$. From Theorem B.1 of Chernozhukov, Chetverikov and Kato (2014),
$\sup_{z\in\mathbb{R}}|\mathbb{P}(\max_{1\leq j\leq 2r}n^{-1/2}\sum_{t=1}^n\varsigma_{j,t}^{+,\textrm{ext}}>z)-\mathbb{P}(\max_{1\leq j\leq 2r}u_{j}^{\textrm{ext}}>z)|\rightarrow0$ if $|\textrm{Var}(n^{-1/2}\sum_{t=1}^n\bvarsigma_{t}^{+,\textrm{ext}})-\textrm{Var}(\bu^{\textrm{ext}})|_\infty\rightarrow0$. Notice that $|\textrm{Var}(n^{-1/2}\sum_{t=1}^n\bvarsigma_{t}^{+,\textrm{ext}})-\textrm{Var}(\bu^{\textrm{ext}})|_\infty=|\textrm{Var}(n^{-1/2}\sum_{t=1}^n\bvarsigma_{t}^+)-\textrm{Var}(\bu)|_\infty$, thus to show $d_3=o(1)$, it suffices to show
\[
d_4:=\sup_{z\in\mathbb{R}}\bigg|\mathbb{P}\bigg(\max_{1\leq j\leq r}n^{-1/2}\sum_{t=1}^n\varsigma_{j,t}^{+}>z\bigg)-\mathbb{P}\bigg(\max_{1\leq j\leq r}u_{j}>z\bigg)\bigg|\rightarrow0.
\]
By Theorem B.1 of \cite{CCK_2014}, it holds that
$
d_4\leq Cn^{-C}+Cm\exp(-Ch^{\gamma_3})
$
provided that \begin{equation} \label{eq:rest} hb^{-1}(\log p)^2\leq Cn^{-\varpi}~~\textrm{and}~~b^2D_n^2\log p+bD_n^2(\log p)^7\leq Cn^{1-2\varpi}
\end{equation}
for some $\varpi\in(0,1/4)$. As we mentioned above, $D_n\asymp (\log p)^{2/\gamma_2}$. To make $p$ diverge as fast as possible, we can take $h\asymp (\log n)^{\vartheta}$ for some $\vartheta>0$. Then (\ref{eq:rest}) becomes
\begin{equation*}
\left\{ \begin{aligned}
         C(\log n)^{\vartheta}n^{\varpi}(\log p)^2 &\leq b; \\
         C(\log n)^{2\vartheta}(\log p)^{4/\gamma_2+5}&\leq n^{1-4\varpi};\\
         C(\log n)^{\vartheta}(\log p)^{4/\gamma_2+9}&\leq n^{1-3\varpi}.
                          \end{aligned} \right.
\end{equation*}
Therefore, $ \log p=o(n^\varphi)~~\textrm{where}~~\varphi=\min\big\{\tfrac{(1-4\varpi)\gamma_2}{4+5\gamma_2},\tfrac{(1-3\varpi)\gamma_2}{4+9\gamma_2}\big\}. $ Notice that $\varphi$ takes the supremum when $\varpi=0$. Hence, if $\log p=o\{n^{\gamma_2/(4+9\gamma_2)}\}$, it holds that $d_4\rightarrow0$. Then we construct the result (\ref{eq:step1}).

Analogously, to show (\ref{eq:step2}), it suffices to show $\sup_{z\in\mathbb{R}}|\mathbb{P}(\max_{1\leq j\leq r}u_j> z)-\mathbb{P}(\max_{1\leq j\leq r}\xi_j> z)|\rightarrow0$. Write $\widetilde{\bW}$ as the covariance of $\bu$. Recall $\bW$ denotes the covariance of $\bxi$. Lemma 3.1 of \cite{CCK_2013} leads to
\begin{equation}\label{eq:2}
\begin{split}
&~\sup_{z\in\mathbb{R}}\bigg|\mathbb{P}\bigg(\max_{1\leq j\leq r}u_j> z\bigg)-\mathbb{P}\bigg(\max_{1\leq j\leq r}\xi_j> z\bigg)\bigg|\\
\leq&~ C|\widetilde{\bW}-\bW|_\infty^{1/3}\{1\vee \log(r/|\widetilde{\bW}-\bW|_\infty)\}^{2/3}.
\end{split}
\end{equation}
We will specify the convergence rate of $|\widetilde{\bW}-\bW|_\infty$ below. Notice that, for any $1\leq j_1,j_2\leq r$, we have
\[
\begin{split}
&\frac{1}{mb}\sum_{\ell=1}^m\mathbb{E}\bigg\{\bigg(\sum_{t\in\mathcal{I}_\ell}\varsigma_{j_1,t}^+\bigg)\bigg(\sum_{t\in\mathcal{I}_\ell}\varsigma_{j_2,t}^+\bigg)\bigg\}\\
&~~~~~~~~~~~~~~-\frac{1}{mb}\sum_{\ell=1}^m\mathbb{E}\bigg\{\bigg(\sum_{t\in\mathcal{I}_\ell}\varsigma_{j_1,t}\bigg)\bigg(\sum_{t\in\mathcal{I}_\ell}\varsigma_{j_2,t}\bigg)\bigg\}\\
=&-\frac{1}{mb}\sum_{\ell=1}^m\mathbb{E}\bigg\{\bigg(\sum_{t\in\mathcal{I}_\ell}\varsigma_{j_1,t}^-\bigg)\bigg(\sum_{t\in\mathcal{I}_\ell}\varsigma_{j_2,t}^-\bigg)\bigg\}\\
&-\frac{1}{mb}\sum_{\ell=1}^m\mathbb{E}\bigg\{\bigg(\sum_{t\in\mathcal{I}_\ell}\varsigma_{j_1,t}^+\bigg)\bigg(\sum_{t\in\mathcal{I}_\ell}\varsigma_{j_2,t}^-\bigg)\bigg\}\\
&-\frac{1}{mb}\sum_{\ell=1}^m\mathbb{E}\bigg\{\bigg(\sum_{t\in\mathcal{I}_\ell}\varsigma_{j_1,t}^-\bigg)\bigg(\sum_{t\in\mathcal{I}_\ell}\varsigma_{j_2,t}^+\bigg)\bigg\}.
\end{split}
\]
The triangle inequality yields
\[
\begin{split}
&~\bigg|\frac{1}{mb}\sum_{\ell=1}^m\mathbb{E}\bigg\{\bigg(\sum_{t\in\mathcal{I}_\ell}\varsigma_{j_1,t}^+\bigg)\bigg(\sum_{t\in\mathcal{I}_\ell}\varsigma_{j_2,t}^+\bigg)\bigg\}\\
&~~~~~~~~~~~~~~~~~~~-\frac{1}{mb}\sum_{\ell=1}^m\mathbb{E}\bigg\{\bigg(\sum_{t\in\mathcal{I}_\ell}\varsigma_{j_1,t}\bigg)\bigg(\sum_{t\in\mathcal{I}_\ell}\varsigma_{j_2,t}\bigg)\bigg\}\bigg|\\
\leq&~\frac{1}{mb}\sum_{\ell=1}^m\bigg|\mathbb{E}\bigg\{\bigg(\sum_{t\in\mathcal{I}_\ell}\varsigma_{j_1,t}^-\bigg)\bigg(\sum_{t\in\mathcal{I}_\ell}\varsigma_{j_2,t}^-\bigg)\bigg\}\bigg|\\
&+\frac{1}{mb}\sum_{\ell=1}^m\bigg|\mathbb{E}\bigg\{\bigg(\sum_{t\in\mathcal{I}_\ell}\varsigma_{j_1,t}^+\bigg)\bigg(\sum_{t\in\mathcal{I}_\ell}\varsigma_{j_2,t}^-\bigg)\bigg\}\bigg|\\
&~+\frac{1}{mb}\sum_{\ell=1}^m\bigg|\mathbb{E}\bigg\{\bigg(\sum_{t\in\mathcal{I}_\ell}\varsigma_{j_1,t}^-\bigg)\bigg(\sum_{t\in\mathcal{I}_\ell}\varsigma_{j_2,t}^+\bigg)\bigg\}\bigg|.
\end{split}
\]
For each $\ell=1,\ldots,m$, the following identities hold:
\[
\begin{split}
\mathbb{E}\bigg\{\bigg(\sum_{t\in\mathcal{I}_\ell}\varsigma_{j_1,t}^-\bigg)\bigg(\sum_{t\in\mathcal{I}_\ell}\varsigma_{j_2,t}^-\bigg)\bigg\}=&~\sum_{t\in\mathcal{I}_\ell}\mathbb{E}(\varsigma_{j_1,t}^{-}\varsigma_{j_2,t}^{-})+\sum_{t_1\neq t_2}\mathbb{E}(\varsigma_{j_1,t_1}^{-}\varsigma_{j_2,t_2}^{-}), \\
\mathbb{E}\bigg\{\bigg(\sum_{t\in\mathcal{I}_\ell}\varsigma_{j_1,t}^+\bigg)\bigg(\sum_{t\in\mathcal{I}_\ell}\varsigma_{j_2,t}^-\bigg)\bigg\}=&~\sum_{t\in\mathcal{I}_\ell}\mathbb{E}(\varsigma_{j_1,t}^{+}\varsigma_{j_2,t}^{-})+\sum_{t_1\neq t_2}\mathbb{E}(\varsigma_{j_1,t_1}^{+}\varsigma_{j_2,t_2}^{-}),\\
\mathbb{E}\bigg\{\bigg(\sum_{t\in\mathcal{I}_\ell}\varsigma_{j_1,t}^-\bigg)\bigg(\sum_{t\in\mathcal{I}_\ell}\varsigma_{j_2,t}^+\bigg)\bigg\}=&~\sum_{t\in\mathcal{I}_\ell}\mathbb{E}(\varsigma_{j_1,t}^{-}\varsigma_{j_2,t}^{+})+\sum_{t_1\neq t_2}\mathbb{E}(\varsigma_{j_1,t_1}^{-}\varsigma_{j_2,t_2}^{+}).
\end{split}
\]
Together with the triangle inequality, Davydov inequality and Cauchy-Schwarz inequality, we have
\[
\begin{split}
\bigg|\mathbb{E}\bigg\{\bigg(\sum_{t\in\mathcal{I}_\ell}\varsigma_{j_1,t}^-\bigg)\bigg(\sum_{t\in\mathcal{I}_\ell}\varsigma_{j_2,t}^-\bigg)\bigg\}\bigg|\leq&~ Cb\sup_{j,t}[\mathbb{E}\{(\varsigma_{j,t}^{-})^4\}]^{1/2},\\
\bigg|\mathbb{E}\bigg\{\bigg(\sum_{t\in\mathcal{I}_\ell}\varsigma_{j_1,t}^+\bigg)\bigg(\sum_{t\in\mathcal{I}_\ell}\varsigma_{j_2,t}^-\bigg)\bigg\}\bigg|\leq&~ Cb\sup_{j,t}[\mathbb{E}\{(\varsigma_{j,t}^+)^4\}]^{1/4}\sup_{j,t}[\mathbb{E}\{(\varsigma_{j,t}^-)^4\}]^{1/4},\\
\bigg|\mathbb{E}\bigg\{\bigg(\sum_{t\in\mathcal{I}_\ell}\varsigma_{j_1,t}^-\bigg)\bigg(\sum_{t\in\mathcal{I}_\ell}\varsigma_{j_2,t}^+\bigg)\bigg\}\bigg|\leq&~ Cb\sup_{j,t}[\mathbb{E}\{(\varsigma_{j,t}^+)^4\}]^{1/4}\sup_{j,t}[\mathbb{E}\{(\varsigma_{j,t}^-)^4\}]^{1/4}.\\
\end{split}
\]
From (\ref{eq:1}), it holds that
\[
\begin{split}
&~\sup_{1\leq j_1,j_2\leq r}\bigg|\frac{1}{mb}\sum_{\ell=1}^m\mathbb{E}\bigg\{\bigg(\sum_{t\in\mathcal{I}_\ell}\varsigma_{j_1,t}^+\bigg)\bigg(\sum_{t\in\mathcal{I}_\ell}\varsigma_{j_2,t}^+\bigg)\bigg\}\\
&~~~~~~~~~~~~~~~~~-\frac{1}{mb}\sum_{\ell=1}^m\mathbb{E}\bigg\{\bigg(\sum_{t\in\mathcal{I}_\ell}\varsigma_{j_1,t}\bigg)\bigg(\sum_{t\in\mathcal{I}_\ell}\varsigma_{j_2,t}\bigg)\bigg\}\bigg|\leq CD_n\exp(-CD_n^{\gamma_2/2}).
\end{split}
\]
By the proof of Lemma 2 in \cite{ChangChenChen_2015}, we can prove that
\begin{equation}\label{eq:toprove}
\begin{split}
&~\sup_{1\leq j_1,j_2\leq r}\bigg|\frac{1}{mb}\sum_{\ell=1}^m\mathbb{E}\bigg\{\bigg(\sum_{t\in\mathcal{I}_\ell}\varsigma_{j_1,t}\bigg)\bigg(\sum_{t\in\mathcal{I}_\ell}\varsigma_{j_2,t}\bigg)\bigg\}\\
&~~~~~~~~~~~~~~~~~-\frac{1}{n}\mathbb{E}\bigg\{\bigg(\sum_{t=1}^n\varsigma_{j_1,t}\bigg)\bigg(\sum_{t=1}^n\varsigma_{j_2,t}\bigg)\bigg\}\bigg|\leq Ch^{1/2}b^{-1/2}+Cbn^{-1}.
\end{split}
\end{equation}
Specifically, notice that
\begin{equation}\label{eq:p1}
\begin{split}
&~\mathbb{E}\bigg\{\bigg(\sum_{t=1}^n\varsigma_{j_1,t}\bigg)\bigg(\sum_{t=1}^n\varsigma_{j_2,t}\bigg)\bigg\}\\
=&~\sum_{\ell=1}^m\mathbb{E}\bigg\{\bigg(\sum_{t\in\mathcal{I}_\ell}\varsigma_{j_1,t}\bigg)\bigg(\sum_{t\in\mathcal{I}_\ell}\varsigma_{j_2,t}\bigg)\bigg\}+\sum_{\ell_1\neq \ell_2}\mathbb{E}\bigg\{\bigg(\sum_{t\in\mathcal{I}_{\ell_1}}\varsigma_{j_1,t}\bigg)\bigg(\sum_{t\in\mathcal{I}_{\ell_2}}\varsigma_{j_2,t}\bigg)\bigg\}\\
&+\sum_{\ell=1}^{m+1}\mathbb{E}\bigg\{\bigg(\sum_{t\in\mathcal{I}_\ell}\varsigma_{j_1,t}\bigg)\bigg(\sum_{t\in\mathcal{J}_\ell}\varsigma_{j_2,t}\bigg)\bigg\}+\sum_{\ell_1\neq \ell_2}\mathbb{E}\bigg\{\bigg(\sum_{t\in\mathcal{I}_{\ell_1}}\varsigma_{j_1,t}\bigg)\bigg(\sum_{t\in\mathcal{J}_{\ell_2}}\varsigma_{j_2,t}\bigg)\bigg\}\\
&+\sum_{\ell=1}^{m+1}\mathbb{E}\bigg\{\bigg(\sum_{t\in\mathcal{J}_\ell}\varsigma_{j_1,t}\bigg)\bigg(\sum_{t\in\mathcal{I}_\ell}\varsigma_{j_2,t}\bigg)\bigg\}+\sum_{\ell_1\neq \ell_2}\mathbb{E}\bigg\{\bigg(\sum_{t\in\mathcal{J}_{\ell_1}}\varsigma_{j_1,t}\bigg)\bigg(\sum_{t\in\mathcal{I}_{\ell_2}}\varsigma_{j_2,t}\bigg)\bigg\}\\
&+\sum_{\ell=1}^{m+1}\mathbb{E}\bigg\{\bigg(\sum_{t\in\mathcal{J}_\ell}\varsigma_{j_1,t}\bigg)\bigg(\sum_{t\in\mathcal{J}_\ell}\varsigma_{j_2,t}\bigg)\bigg\}+\sum_{\ell_1\neq \ell_2}\mathbb{E}\bigg\{\bigg(\sum_{t\in\mathcal{J}_{\ell_1}}\varsigma_{j_1,t}\bigg)\bigg(\sum_{t\in\mathcal{J}_{\ell_2}}\varsigma_{j_2,t}\bigg)\bigg\}\\
\end{split}
\end{equation}
where we set $\mathcal{I}_{m+1}=\emptyset$. By Cauchy-Schwarz inequality and Davydov inequality, we have
\[
\begin{split}
&~\bigg|\frac{1}{mb}\sum_{\ell=1}^m\mathbb{E}\bigg\{\bigg(\sum_{t\in\mathcal{I}_\ell}\varsigma_{j_1,t}\bigg)\bigg(\sum_{t\in\mathcal{I}_\ell}\varsigma_{j_2,t}\bigg)\bigg\}\\
&~~~~~~~~~~~~~~~~~~~~~~-\frac{1}{n}\sum_{\ell=1}^m\mathbb{E}\bigg\{\bigg(\sum_{t\in\mathcal{I}_\ell}\varsigma_{j_1,t}\bigg)\bigg(\sum_{t\in\mathcal{I}_\ell}\varsigma_{j_2,t}\bigg)\bigg\}\bigg|\\
=&~\frac{n-mb}{nm}\sum_{\ell=1}^m\bigg|\mathbb{E}\bigg\{\bigg(\frac{1}{\sqrt{b}}\sum_{t\in\mathcal{I}_\ell}\varsigma_{j_1,t}\bigg)\bigg(\frac{1}{\sqrt{b}}\sum_{t\in\mathcal{I}_\ell}\varsigma_{j_2,t}\bigg)\bigg\}\bigg|\\
\leq&~\frac{mh+b}{nm}\times Cm\leq Chb^{-1}+Cbn^{-1},
\end{split}
\]
\[
\begin{split}
&~\bigg|\frac{1}{n}\sum_{\ell_1\neq \ell_2}\mathbb{E}\bigg\{\bigg(\sum_{t\in\mathcal{I}_{\ell_1}}\varsigma_{j_1,t}\bigg)\bigg(\sum_{t\in\mathcal{I}_{\ell_2}}\varsigma_{j_2,t}\bigg)\bigg\}\bigg|\\
\leq&~\frac{b}{n}\sum_{\ell_1\neq\ell_2}\bigg|\mathbb{E}\bigg\{\bigg(\frac{1}{\sqrt{b}}\sum_{t\in\mathcal{I}_{\ell_1}}\varsigma_{j_1,t}\bigg)\bigg(\frac{1}{\sqrt{b}}\sum_{t\in\mathcal{I}_{\ell_2}}\varsigma_{j_2,t}\bigg)\bigg\}\bigg|\\
\leq&~Cbn^{-1}\sum_{\ell_1\neq\ell_2}\exp\{-C|(\ell_1-\ell_2)b|^{\gamma_3}\}\leq Cbn^{-1}.
\end{split}
\]
Similarly, we can bound the other terms in (\ref{eq:p1}). Therefore, we have (\ref{eq:toprove}) holds which implies that $ |\widetilde{\bW}-\bW|_\infty\leq Ch^{1/2}b^{-1/2}+Cbn^{-1}+CD_n\exp(-CD_n^{\gamma_2/2})$. For $b, h$ and $D_n$ specified above, (\ref{eq:2}) implies $\sup_{z\in\mathbb{R}}|\mathbb{P}(\max_{1\leq j\leq r}u_j> z)-\mathbb{P}(\max_{1\leq j\leq r}\xi_j> z)|\rightarrow0$. Then we construct the result (\ref{eq:step2}). Hence, we complete the proof of Theorem \ref{tm:1}. $\hfill\Box$

\begin{lemma}\label{la:4}
Assume Conditions {\rm\ref{as:moment}} and {\rm\ref{as:betamix}} hold, the kernel function $\mathcal{K}(\cdot)$ satisfies $|\mathcal{K}(x)|\asymp |x|^{-\tau}$ as $x\rightarrow\infty$ for some $\tau>1$, and the bandwidth $S_n\asymp n^{\rho}$ for some $0<\rho<\min\{\tfrac{\tau-1}{3\tau},\tfrac{\gamma_3}{2\gamma_3+1}\}$. Let $\kappa=\max\big\{\tfrac{1}{2\gamma_3+1},\tfrac{\rho \tau-\rho+2}{\tau+1+\gamma_3},\tfrac{\rho\tau+1}{\tau}\big\}$, and $\alpha_0$ be the maximizer for the function $f(\alpha)=\min\{1-\alpha-2\rho,2(\alpha-\rho)\tau-2\}$ over $\kappa<\alpha<1-2\rho$. Then
\[
\bigg|\sum_{k=0}^{n-1}\mathcal{K}\bigg(\frac{k}{S_n}\bigg)\bigg[\frac{1}{n}\sum_{t=k+1}^n\{\bfeta_t\bfeta_{t-k}^\T-\mathbb{E}(\bfeta_t\bfeta_{t-k}^\T)\}\bigg]\bigg|_\infty=O_p\big(\{\log (pn)\}^{4/\gamma_2}n^{-f(\alpha_0)/2}\big)
\]
provided that $\log p \leq Cn^{C\delta}$ where $\delta=\min[\tfrac{\gamma_2}{\gamma_2+8}(2\alpha_0\gamma_3+\alpha-1), \tfrac{\gamma_2}{8}\{(\alpha_0-\rho)\tau+\alpha_0+\alpha_0\gamma_3+\rho-2\}].$
\end{lemma}

\noindent {\bf Proof:} We first construct an upper bound for
$
\sup_{1\leq j_1,j_2\leq r}\mathbb{P}\{|\sum_{k=0}^{n-1}\mathcal{K}(k/S_n)[n^{-1}\sum_{t=k+1}^n\{\eta_{j_1,t}\eta_{j_2,t-k}-\mathbb{E}(\eta_{j_1,t}\eta_{j_2,t-k})\}]|>x\}
$.
 For any $j_1$ and $j_2$, it holds that
\begin{equation}\label{eq:ss1}
\begin{split}
&~\mathbb{P}\bigg\{\bigg|\sum_{k=0}^{n-1}\mathcal{K}\bigg(\frac{k}{S_n}\bigg)\bigg[\frac{1}{n}\sum_{t=k+1}^n\{\eta_{j_1,t}\eta_{j_2,t-k}-\mathbb{E}(\eta_{j_1,t}\eta_{j_2,t-k})\}\bigg]\bigg|>x\bigg\}\\
\leq&~\mathbb{P}\bigg\{\sum_{k=0}^{\lfloor Cn^{\alpha}\rfloor}\bigg|\mathcal{K}\bigg(\frac{k}{S_n}\bigg)\bigg|\bigg|\frac{1}{n}\sum_{t=1}^{n-k}\psi_{t,k}\bigg|>\frac{x}{2}\bigg\}\\
&+\mathbb{P}\bigg\{\sum_{k=\lfloor Cn^{\alpha}\rfloor +1}^{n-1}\bigg|\mathcal{K}\bigg(\frac{k}{S_n}\bigg)\bigg|\bigg|\frac{1}{n}\sum_{t=1}^{n-k}\psi_{t,k}\bigg|>\frac{x}{2}\bigg\}
\end{split}
\end{equation}
for any $\alpha\in(0,1)$, where $\psi_{t,k}=\eta_{j_1,t+k}\eta_{j_2,t}-\mathbb{E}(\eta_{j_1,t+k}\eta_{j_2,t})$. Following Lemma 2 of \cite{ChangTangWu_2013}, it holds that
\begin{equation}\label{eq:t1}
\sup_{0\leq k\leq n-1}\sup_{1\leq t\leq n-k}\mathbb{P}\left(|\psi_{t,k}|>x\right)\leq C\exp(-Cx^{\gamma_2/4})
\end{equation}
for any $x>0$. Notice that $S_n\asymp n^\rho$, we have $\max_{\lfloor Cn^{\alpha}\rfloor+1\leq k\leq n-1}|\mathcal{K}(k/S_n)|\leq Cn^{-(\alpha-\rho)\tau}$ if $\alpha>\rho$. Then, (\ref{eq:t1}) leads to
\begin{equation}\label{eq:term1}
\begin{split}
&~\mathbb{P}\bigg\{\sum_{k=\lfloor Cn^{\alpha}\rfloor +1}^{n-1}\bigg|\mathcal{K}\bigg(\frac{k}{S_n}\bigg)\bigg|\bigg|\frac{1}{n}\sum_{t=1}^{n-k}\psi_{t,k}\bigg|>\frac{x}{2}\bigg\}\\
\leq&~\sum_{k=\lfloor Cn^\alpha\rfloor +1}^{n-1}\mathbb{P}\bigg\{\bigg|\frac{1}{n}\sum_{t=1}^{n-k}\psi_{t,k}\bigg|>Cxn^{(\alpha-\rho)\tau-1}\bigg\}\\
\leq&~\sum_{k=\lfloor Cn^\alpha\rfloor +1}^{n-1}\sum_{t=1}^{n-k}\mathbb{P}\big\{|\psi_{t,k}|>Cxn^{(\alpha-\rho)\tau-1}\big\}\\
\leq&~Cn^2\exp[-C\{xn^{(\alpha-\rho)\tau-1}\}^{\gamma_2/4}].
\end{split}
\end{equation}
We will specify the upper bound for $\mathbb{P}\{\sum_{k=0}^{\lfloor Cn^{\alpha}\rfloor}|\mathcal{K}(k/S_n)||{n}^{-1}\sum_{t=1}^{n-k}\psi_{t,k}|>{x}/{2}\}$ below. Similar to (\ref{eq:t1}), we have that
\begin{equation}\label{eq:s1}
\sup_{1\leq j_1,j_2\leq r}\sup_{0\leq k\leq n-1}\sup_{1\leq t\leq n-k}\mathbb{P}(|\eta_{j_1,t+k}\eta_{j_2,t}|>x)\leq C\exp(-Cx^{\gamma_2/4})
\end{equation}
for any $x>0$. Denote by $\mathcal{T}$ the event $\{\sup_{0\leq k\leq n-1}\sup_{1\leq t\leq n-k}|\eta_{j_1,t+k}\eta_{j_2,t}|>M\}$. For each $k=0,\ldots,\lfloor Cn^\alpha\rfloor$, let $\psi_{t,k}^+=\eta_{j_1,t+k}\eta_{j_2,t}\mathbb{I}\{|\eta_{j_1,t+k}\eta_{j_2,t}|\leq M\}-\mathbb{E}[\eta_{j_1,t+k}\eta_{j_2,t}\mathbb{I}\{|\eta_{j_1,t+k}\eta_{j_2,t}|\leq M\}]$ for $t=1,\ldots,n-k$. Write $D=\sum_{k=0}^{\lfloor Cn^{\alpha}\rfloor}|\mathcal{K}(k/S_n)|$, then
\begin{equation}\label{eq:term2}
\begin{split}
&~\mathbb{P}\bigg\{\sum_{k=0}^{\lfloor Cn^{\alpha}\rfloor}\bigg|\mathcal{K}\bigg(\frac{k}{S_n}\bigg)\bigg|\bigg|\frac{1}{n}\sum_{t=1}^{n-k}\psi_{t,k}\bigg|>\frac{x}{2}\bigg\}\\
\leq&~\sum_{k=0}^{\lfloor Cn^{\alpha} \rfloor}\mathbb{P}\bigg(\bigg|\frac{1}{n}\sum_{t=1}^{n-k}\psi_{t,k}\bigg|>\frac{x}{2D},~\mathcal{T}^c\bigg) +\mathbb{P}(\mathcal{T})\\
\leq&~\sum_{k=0}^{\lfloor Cn^{\alpha} \rfloor}\mathbb{P}\bigg(\bigg|\frac{1}{n}\sum_{t=1}^{n-k}\psi_{t,k}^+\bigg|>\frac{x}{4D}\bigg) +\mathbb{P}(\mathcal{T})\\
&+\sum_{k=0}^{\lfloor Cn^{\alpha} \rfloor}\mathbb{P}\bigg(\frac{1}{n}\sum_{t=1}^{n-k}\mathbb{E}[|\eta_{j_1,t+k}\eta_{j_2,t}|\mathbb{I}\{|\eta_{j_1,t+k}\eta_{j_2,t}|> M\}]>\frac{x}{4D}\bigg).
\end{split}
\end{equation}
From (\ref{eq:s1}), we have $\mathbb{P}(\mathcal{T})\leq Cn^2\exp(-CM^{\gamma_2/4})$. Similar to (\ref{eq:tailbound}), we have
\[
\sup_{1\leq j_1,j_2\leq r}\sup_{0\leq k\leq n-1}\sup_{1\leq t\leq n-k}\mathbb{E}[|\eta_{j_1,t+k}\eta_{j_2,t}|\mathbb{I}\{|\eta_{j_1,t+k}\eta_{j_2,t}|> M\}]\leq CM\exp(-CM^{\gamma_2/4}).
\]
If $DMx^{-1}\exp(-CM^{\gamma_2/4})\rightarrow0$, then (\ref{eq:term2}) yields that
\begin{equation}\label{eq:term3}
\begin{split}
&~\mathbb{P}\bigg\{\sum_{k=0}^{\lfloor Cn^{\alpha}\rfloor}\bigg|\mathcal{K}\bigg(\frac{k}{S_n}\bigg)\bigg|\bigg|\frac{1}{n}\sum_{t=1}^{n-k}\psi_{t,k}\bigg|>\frac{x}{2}\bigg\}\\
\leq&~\sum_{k=0}^{\lfloor Cn^{\alpha} \rfloor}\mathbb{P}\bigg(\bigg|\frac{1}{n}\sum_{t=1}^{n-k}\psi_{t,k}^+\bigg|>\frac{x}{4D}\bigg) +Cn^{2}\exp(-CM^{\gamma_2/4}).
\end{split}
\end{equation}
For each $k=0,\ldots,\lfloor Cn^{\alpha} \rfloor$, we first consider $\mathbb{P}\{n^{-1}\sum_{t=1}^{n-k}\psi_{t,k}^+>x/(4D)\}$. By Markov inequality, it holds that
\begin{equation}\label{eq:u2}
\mathbb{P}\bigg(\frac{1}{n}\sum_{t=1}^{n-k}\psi_{t,k}^+>\frac{x}{4D}\bigg)\leq \exp\bigg(-\frac{unx}{4D}\bigg)\mathbb{E}\bigg\{\exp\bigg(\sum_{t=1}^{n-k}u\psi_{t,k}^+\bigg)\bigg\}
\end{equation} for any $u>0$. Let $L$ be a positive integer such that $L\asymp n^{\alpha}$ and $L\geq 3\lfloor Cn^{\alpha}\rfloor$ for $C$ specified in (\ref{eq:ss1}). We decompose the sequence $\{1,\ldots, n-k\}$ to the following $m+1$ blocks where $m = \lfloor (n-k)/L\rfloor$: $\mathcal{G}_\ell= \{(\ell-1)L+1,\ldots, \ell L\}$ $(\ell= 1, \ldots ,m)$ and $\mathcal{G}_{m+1} = \{mL+1,\ldots, n-k\}$. Additionally, let $b=\lfloor L/2\rfloor$ and $h=L-b$. We then decompose each $\mathcal{G}_\ell$ $(\ell=1,\ldots,m)$ to a block with length $b$ and a block with length $h$. Specifically, $\mathcal{I}_\ell=\{(\ell-1)L+1,\ldots,(\ell-1)L+b\}$ and $\mathcal{J}_\ell=\{(\ell-1)L+b+1,\ldots,\ell L\}$ for any $\ell=1,\ldots,m$, and
$\mathcal{I}_{m+1}=\mathcal{G}_{m+1}$. Based on these notations and Cauchy-Schwarz inequality, it holds that
\[
\begin{split}
\mathbb{E}\bigg\{\exp\bigg(\sum_{t=1}^{n-k}u\psi_{t,k}^+\bigg)\bigg\}\leq&~ \bigg[\mathbb{E}\bigg\{\exp\bigg(\sum_{\ell=1}^{m+1}\sum_{t\in\mathcal{I}_\ell}2u\psi_{t,k}^+\bigg)\bigg\}\bigg]^{1/2}\\
&~~\times\bigg[\mathbb{E}\bigg\{\exp\bigg(\sum_{\ell=1}^{m}\sum_{t\in\mathcal{J}_\ell}2u\psi_{t,k}^+\bigg)\bigg\}\bigg]^{1/2}.
\end{split}
\]
By Lemma 2 of \cite{Merlevedeetaj_2011}, noticing that $b(m+1)\leq 2n$, we have
\begin{equation}\label{eq:u1}
\begin{split}
\mathbb{E}\bigg\{\exp\bigg(\sum_{\ell=1}^{m+1}\sum_{t\in\mathcal{I}_\ell}2u\psi_{t,k}^+\bigg)\bigg\}\leq&~ \prod_{\ell=1}^{m+1}\mathbb{E}\bigg\{\exp\bigg(\sum_{t\in\mathcal{I}_\ell}2u\psi_{t,k}^+\bigg)\bigg\}\\&+CuMn\exp(8uMn-C|b-k|_+^{\gamma_3}).
\end{split}
\end{equation}
Following the inequality $e^x\leq 1+x+{x^2}e^{x\vee0}/2$ for any $x\in\mathbf{R}$, we have that
\[
\begin{split}
\mathbb{E}\bigg\{\exp\bigg(\sum_{t\in\mathcal{I}_\ell}2u\psi_{t,k}^+\bigg)\bigg\}\leq&~ 1+2u^2\mathbb{E}\bigg\{\bigg(\sum_{t\in\mathcal{I}_\ell}\psi^+_{t,k}\bigg)^2\bigg\}\exp(4ubM)\\
\leq&~ 1+Cu^2b^2\exp(4ubM).
\end{split}
\]
Together with (\ref{eq:u1}), following the inequality $(1+x)^{m+1}\leq e^{(m+1)x}$ for any $x>0$, and $bm\leq n/2$, it holds that
\[
\begin{split}
\mathbb{E}\bigg\{\exp\bigg(\sum_{\ell=1}^{m+1}\sum_{t\in\mathcal{I}_\ell}2u\psi_{t,k}^+\bigg)\bigg\}
\leq&~\exp\{Cu^2nb\exp(4ubM)\}\\
&+CuMn\exp(8uMn-C|b-k|_+^{\gamma_3}).
\end{split}
\]
Similarly, we can obtain the same upper bound for $\mathbb{E}\{\exp(\sum_{\ell=1}^{m}\sum_{t\in\mathcal{J}_\ell}2u\psi_{t,k}^+)\}$. Hence,
\[
\begin{split}
\mathbb{E}\bigg\{\exp\bigg(\sum_{t=1}^{n-k}u\psi_{t,k}^+\bigg)\bigg\}\leq&~\exp\{Cu^2nb\exp(4ubM)\}\\
&+CuMn\exp\{8uMn-C|b-k|_+^{\gamma_3}).
\end{split}
\]
We restrict $|ubM|\leq C$. Notice that $b-k\geq \lfloor Cn^{\alpha}\rfloor/2-1$, then
\[
\mathbb{E}\bigg\{\exp\bigg(\sum_{t=1}^{n-k}u\psi_{t,k}^+\bigg)\bigg\}\leq C\exp(Cu^2nb)+CuMn\exp(8uMn-Cn^{\alpha \gamma_3}).
\]
Together with (\ref{eq:u2}), notice that $D\asymp S_n\asymp n^{\rho}$ and $b\asymp n^{\alpha}$, it holds that
\begin{equation}\label{eq:upper1}
\begin{split}
\mathbb{P}\bigg(\frac{1}{n}\sum_{t=1}^{n-k}\psi_{t,k}^+>\frac{x}{4D}\bigg)\leq&~ C\exp(-Cun^{1-\rho}x+Cu^2n^{1+\alpha})\\
&+CuMn\exp(-Cun^{1-\rho}x+8uMn-Cn^{\alpha\gamma_3}).
\end{split}
\end{equation}
To make the upper bound in above inequality decay to zero for some $x\rightarrow0^+$ and $M\rightarrow\infty$, we need to require $ uMn^{1-\alpha\gamma_3}\leq C. $ For the first term on the right-hand side of above inequality, the optimal selection of $u$ is $u\asymp xn^{-\alpha-\rho}$. Therefore, (\ref{eq:upper1}) can be simplified to
\[
\begin{split}
\mathbb{P}\bigg(\frac{1}{n}\sum_{t=1}^{n-k}\psi_{t,k}^+>\frac{x}{4D}\bigg)\leq&~C\exp(-Cn^{1-\alpha-2\rho}x^2)+C\exp(-Cn^{\alpha\gamma_3})
\end{split}
\]
if $xMn^{1-\alpha-\alpha\gamma_3-\rho}\leq C$. The same inequality also hold for $\mathbb{P}\{{n}^{-1}\sum_{t=1}^{n-k}\psi_{t,k}^+<-{x}/(4D)\}$. Combining with (\ref{eq:ss1}), (\ref{eq:term1}) and (\ref{eq:term3}),
\[
\begin{split}
&~\mathbb{P}\bigg\{\bigg|\sum_{k=0}^{n-1}\mathcal{K}\bigg(\frac{k}{S_n}\bigg)\bigg[\frac{1}{n}\sum_{t=k+1}^n\{\eta_{j_1,t}\eta_{j_2,t-k}-\mathbb{E}(\eta_{j_1,t}\eta_{j_2,t-k})\}\bigg]\bigg|>x\bigg\}\\
\leq&~Cn^\alpha\exp(-Cn^{1-\alpha-2\rho}x^2)+Cn^{\alpha}\exp(-Cn^{\alpha\gamma_3})\\
&+Cn^2\exp[-C\{xn^{(\alpha-\rho)\tau-1}\}^{\gamma_2/4}]+Cn^{2}\exp(-CM^{\gamma_2/4})
\end{split}
\]
for any $x>0$ such that $xMn^{1-\alpha-\alpha\gamma_3-\rho}\leq C$. Notice that above inequality is uniformly for any $j_1$ and $j_2$, thus
\[
\begin{split}
&~\mathbb{P}\bigg\{\bigg|\sum_{k=0}^{n-1}\mathcal{K}\bigg(\frac{k}{S_n}\bigg)\bigg[\frac{1}{n}\sum_{t=k+1}^n\{\bfeta_t\bfeta_{t-k}^\T-\mathbb{E}(\bfeta_t\bfeta_{t-k}^\T)\}\bigg]\bigg|_\infty>x\bigg\}\\
 \leq&~Cp^2n^\alpha\exp(-Cn^{1-\alpha-2\rho}x^2)+Cp^2n^{\alpha}\exp(-Cn^{\alpha\gamma_3})\\
 &+Cp^2n^2\exp[-C\{xn^{(\alpha-\rho)\tau-1}\}^{\gamma_2/4}]+Cp^2n^{2}\exp(-CM^{\gamma_2/4}).
\end{split}
\]
To make the upper bound of above inequality converge to zero, $x$ and $M$ should satisfy the following restrictions:
\begin{equation} \label{eq:restr2}
\left\{ \begin{aligned}
         x\geq &~C\bigg[\sqrt{\frac{\log (pn)}{n^{1-\alpha-2\rho}}} \vee \frac{\{\log(pn)\}^{4/\gamma_2}}{n^{(\alpha-\rho)\tau-1}}\bigg], \\
                  M\geq&~C\{\log(pn)\}^{4/\gamma_2}.
                          \end{aligned} \right.
                          \end{equation}
Notice that $xMn^{1-\alpha-\alpha\gamma_3-\rho}\leq C$, (\ref{eq:restr2}) implies that $\log p\leq Cn^{C\delta} $ where $\delta=\min\{\tfrac{\gamma_2}{\gamma_2+8}(2\alpha\gamma_3+\alpha-1),\tfrac{\gamma_2}{8}\{(\alpha-\rho)\tau+\alpha+\alpha\gamma_3+\rho-2\}\}. $ To make $x$ can decay to zero and $p$ can diverge at exponential rate of $n$, we need to assume $0<\rho<\min\{\frac{\tau-1}{3\tau},\frac{\gamma_3}{2\gamma_3+1}\}$ and $ \kappa<\alpha<1-2\rho. $ Let $f(\alpha)=\min\{1-\alpha-2\rho,2(\alpha-\rho)\tau-2\}$ and $\alpha_0=\arg\max_{\kappa<\alpha<1-2\rho}f(\alpha)$. We select $\alpha=\alpha_0$ and $x=C\{\log(pn)\}^{4/\gamma_2}n^{-f(\alpha_0)/2}$, then
\[
\mathbb{P}\bigg\{\bigg|\sum_{k=0}^{n-1}\mathcal{K}\bigg(\frac{k}{S_n}\bigg)\bigg[\frac{1}{n}\sum_{t=k+1}^n\{\bfeta_{t}\bfeta_{t-k}^\T-\mathbb{E}(\bfeta_{t}\bfeta_{t-k}^\T)\}\bigg]\bigg|_\infty>x\bigg\}\rightarrow0.
\]
Hence, we complete the proof of Lemma \ref{la:4}. $\hfill\Box$

\bigskip

\noindent{\bf Proof of Theorem \ref{tm:2}:} Similar to the proof of (\ref{eq:step2}), it suffices to prove $|\widehat{\bW}-\bW|_\infty=o_p(1)$. By Lemmas \ref{la:1} and \ref{la:bias}, we have $\max_{1\leq j\leq p}|\widehat{v}_{j,j}-v_{j,j}|=O_p\{(n^{-1}\log p)^{1/2}\}$. Notice that $v_{j,j}$'s are uniformly bounded away from zero, then $\widehat{v}_{j,j}^{-1}$'s are uniformly bounded away from infinity with probability approaching one. Thus,
\begin{equation}
\begin{split}
|\widehat{\bW}-\bW|_\infty\leq&~C|\widehat{\bXi}-\bXi|_\infty+C|\widehat{\bH}-\bH|_\infty\\
=&~C|\widehat{\bXi}-\bXi|_\infty+O_p\{(n^{-1}\log p)^{1/2}\}.
\end{split}
\end{equation}
We will show $|\widehat{\bXi}-\bXi|_\infty=o_p(1)$ below.

Define
\[
\widetilde{\bXi}=\sum_{k=-n+1}^{n-1}\mathcal{K}\bigg(\frac{k}{S_n}\bigg)\bGamma_k
\]
where
\[
{\bGamma}_k= \left\{ \begin{aligned}
         \frac{1}{n}\sum_{t=k+1}^n\mathbb{E}({\bfeta}_t{\bfeta}_{t-k}^\T),~~~ &k\geq0; \\
                  \frac{1}{n}\sum_{t=-k+1}^n\mathbb{E}({\bfeta}_{t+k}{\bfeta}_t^\T),~~&k<0.
                          \end{aligned} \right.
\]
We will specify the convergence rates of $|\widehat{\bXi}-\widetilde{\bXi}|_\infty$ and $|\widetilde{\bXi}-\bXi|_\infty$, respectively. Notice that
\[
\begin{split}
\widehat{\bXi}-\widetilde{\bXi}=&~\sum_{k=0}^{n-1}\mathcal{K}\bigg(\frac{k}{S_n}\bigg)\big(\widehat{\bGamma}_k-\bGamma_k\big)\\
&~+\sum_{k=-n+1}^{-1}\mathcal{K}\bigg(\frac{k}{S_n}\bigg)\big(\widehat{\bGamma}_k-\bGamma_k\big).
\end{split}
\]
For any $k\geq0$, it holds that
\[
\begin{split}
\widehat{\bGamma}_k=&~\frac{1}{n}\sum_{t=k+1}^n\bfeta_t\bfeta_{t-k}^\T+\frac{1}{n}\sum_{t=k+1}^n\big(\widehat{\bfeta}_t-\bfeta_t\big)\bfeta_{t-k}^\T\\
&+\frac{1}{n}\sum_{t=k+1}^n\bfeta_t\big(\widehat{\bfeta}_{t-k}-\bfeta_{t-k}\big)^\T\\
&+\frac{1}{n}\sum_{t=k+1}^n\big(\widehat{\bfeta}_t-\bfeta_t\big)\big(\widehat{\bfeta}_{t-k}-\bfeta_{t-k}\big)^\T,
\end{split}
\]
which implies
\begin{equation}\label{eq:bound}
\begin{split}
\sum_{k=0}^{n-1}\mathcal{K}\bigg(\frac{k}{S_n}\bigg)\big(\widehat{\bGamma}_k-\bGamma_k\big)=&~\sum_{k=0}^{n-1}\mathcal{K}\bigg(\frac{k}{S_n}\bigg)\bigg[\frac{1}{n}\sum_{t=k+1}^n\{\bfeta_t\bfeta_{t-k}^\T-\mathbb{E}(\bfeta_t\bfeta_{t-k}^\T)\}\bigg]\\
&+\sum_{k=0}^{n-1}\mathcal{K}\bigg(\frac{k}{S_n}\bigg)\bigg\{\frac{1}{n}\sum_{t=k+1}^n\big(\widehat{\bfeta}_t-\bfeta_t\big)\bfeta_{t-k}^\T\bigg\}\\
&+\sum_{k=0}^{n-1}\mathcal{K}\bigg(\frac{k}{S_n}\bigg)\bigg\{\frac{1}{n}\sum_{t=k+1}^n\bfeta_t\big(\widehat{\bfeta}_{t-k}-\bfeta_{t-k}\big)^\T\bigg\}\\
&+\sum_{k=0}^{n-1}\mathcal{K}\bigg(\frac{k}{S_n}\bigg)\bigg\{\frac{1}{n}\sum_{t=k+1}^n\big(\widehat{\bfeta}_t-\bfeta_t\big)\big(\widehat{\bfeta}_{t-k}-\bfeta_{t-k}\big)^\T\bigg\}.
\end{split}
\end{equation}
We will prove the $|\cdot|_\infty$-norm of the last three terms on the right-hand side of above identity are $O_p\{sS_n(n^{-1}\log p)^{1/2}\}$. We only need to show this rate for one of them and the proofs for the other two are similar. For any $j$ and $t$,
\[
\begin{split}
\widehat{\eta}_{j,t}-\widehat{\eta}_{j,t}=&~\big\{\widehat{\epsilon}_{\chi_1(j),t}\widehat{\epsilon}_{\chi_2(j),t}-\epsilon_{\chi_1(j),t}\epsilon_{\chi_2(j),t}\big\}-\big\{\widehat{v}_{\bchi(j)}-v_{\bchi(j)}\big\}\\
=&~\widehat{\epsilon}_{\chi_1(j),t}\widehat{\epsilon}_{\chi_2(j),t}-\epsilon_{\chi_1(j),t}\epsilon_{\chi_2(j),t}+O_p\{(n^{-1}\log p)^{1/2}\}\\
=&~\big\{\widehat{\balpha}_{\chi_1(j)}-\balpha_{\chi_1(j)}\big\}^\T\by_t\by_t^\T\big\{\widehat{\balpha}_{\chi_2(j)}-\balpha_{\chi_2(j)}\big\}\\
&-\epsilon_{\chi_2(j),t}\big\{\widehat{\balpha}_{\chi_1(j)}-\balpha_{\chi_1(j)}\big\}^\T\by_t\\
&-\epsilon_{\chi_1(j),t}\big\{\widehat{\balpha}_{\chi_2(j)}-\balpha_{\chi_2(j)}\big\}^\T\by_t\\
&+O_p\{(n^{-1}\log p)^{1/2}\}.
\end{split}
\]
Here the term $O_p\{(n^{-1}\log p)^{1/2}\}$ is uniform for any $j$ and $t$. Then the $(j_1,j_2)$-th component of $\sum_{k=0}^{n-1}\mathcal{K}(k/S_n)\{n^{-1}\sum_{t=k+1}^n(\widehat{\bfeta}_t-\bfeta_t)\bfeta_{t-k}^\T\}$ is
\begin{equation}\label{eq:l1l2}
\begin{split}
&\big\{\widehat{\balpha}_{\chi_1(j_1)}-\balpha_{\chi_1(j_1)}\big\}^\T\bigg\{\sum_{k=0}^{n-1}\mathcal{K}\bigg(\frac{k}{S_n}\bigg)\bigg(\frac{1}{n}\sum_{t=k+1}^n\eta_{j_2,t-k}\by_t\by_t^\T\bigg)\bigg\}\big\{\widehat{\balpha}_{\chi_2(j_2)}-\balpha_{\chi_2(j_2)}\big\}\\
-&\big\{\widehat{\balpha}_{\chi_1(j_1)}-\balpha_{\chi_1(j_1)}\big\}^\T\bigg\{\sum_{k=0}^{n-1}\mathcal{K}\bigg(\frac{k}{S_n}\bigg)\bigg(\frac{1}{n}\sum_{t=k+1}^n\by_t\eta_{j_2,t-k}\epsilon_{\chi_2(j_1),t}\bigg)\bigg\}\\
-&\big\{\widehat{\balpha}_{\chi_2(j_1)}-\balpha_{\chi_2(j_1)}\big\}^\T\bigg\{\sum_{k=0}^{n-1}\mathcal{K}\bigg(\frac{k}{S_n}\bigg)\bigg(\frac{1}{n}\sum_{t=k+1}^n\by_t\eta_{j_2,t-k}\epsilon_{\chi_1(j_1),t}\bigg)\bigg\}\\
+&\widetilde{R}_{j_1,j_2},
\end{split}
\end{equation}
where
\[
\begin{split}
|\widetilde{R}_{j_1,j_2}|\leq&~ \bigg\{\sum_{k=0}^{n-1}\bigg|\mathcal{K}\bigg(\frac{k}{S_n}\bigg)\bigg|\bigg(\frac{1}{n}\sum_{t=k+1}^n|\eta_{j_2,t-k}|\bigg)\bigg\}\cdot O_p\{(n^{-1}\log p)^{1/2}\}\\
\leq&~\bigg\{\sum_{k=0}^n\bigg|\mathcal{K}\bigg(\frac{k}{S_n}\bigg)\bigg|\bigg\}\bigg(\frac{1}{n}\sum_{t=1}^n|\eta_{j_2,t}|\bigg)\cdot O_p\{(n^{-1}\log p)^{1/2}\}\\
=&~O_p\{S_n(n^{-1}\log p)^{1/2}\}.
\end{split}
\]
Here the term $O_p\{S_n(n^{-1}\log p)^{1/2}\}$ is uniform for any $j_1$ and $j_2$. Following the same arguments, we have
\[
\begin{split}
\sup_{1\leq j_1,j_2\leq p}\bigg|\sum_{k=0}^{n-1}\mathcal{K}\bigg(\frac{k}{S_n}\bigg)\bigg(\frac{1}{n}\sum_{t=k+1}^n\eta_{j_2,t-k}\by_t\by_t^\T\bigg)\bigg|_\infty\leq&~CS_n,\\
\sup_{1\leq j_1,j_2\leq p}\bigg|\sum_{k=0}^{n-1}\mathcal{K}\bigg(\frac{k}{S_n}\bigg)\bigg(\frac{1}{n}\sum_{t=k+1}^n\by_t\eta_{j_2,t-k}\epsilon_{\chi_2(j_1),t}\bigg)\bigg|_\infty\leq&~CS_n,\\
\sup_{1\leq j_1,j_2\leq p}\bigg|\sum_{k=0}^{n-1}\mathcal{K}\bigg(\frac{k}{S_n}\bigg)\bigg(\frac{1}{n}\sum_{t=k+1}^n\by_t\eta_{j_2,t-k}\epsilon_{\chi_1(j_1),t}\bigg)\bigg|_\infty\leq&~CS_n.
\end{split}
\]
Therefore, the $(j_1,j_2)$-th component of $\sum_{k=0}^{n-1}\mathcal{K}(k/S_n)\{n^{-1}\sum_{t=k+1}^n(\widehat{\bfeta}_t-\bfeta_t)\bfeta_{t-k}^\T\}$ can be bounded by
$
CS_n\sup_{1\leq j\leq p}|\widehat{\balpha}_j-\balpha_j|_1+O_p\{S_n(n^{-1}\log p)^{1/2}\}=O_p\{sS_n(n^{-1}\log p)^{1/2}\},
$
where the last identity in above equation is based on (\ref{eq:m1}). Therefore, from (\ref{eq:bound}), by Lemma \ref{la:4}, we have
\[
\begin{split}
&~\bigg|\sum_{k=0}^{n-1}\mathcal{K}\bigg(\frac{k}{S_n}\bigg)\big(\widehat{\bGamma}_k-\bGamma_k\big)\bigg|_\infty\\
\leq&~\bigg|\sum_{k=0}^{n-1}\mathcal{K}\bigg(\frac{k}{S_n}\bigg)\bigg[\frac{1}{n}\sum_{t=k+1}^n\{\bfeta_t\bfeta_{t-k}^\T-\mathbb{E}(\bfeta_t\bfeta_{t-k}^\T)\}\bigg]\bigg|_\infty\\
&+O_p\{sS_n(n^{-1}\log p)^{1/2}\}\\
=&~O_p[\{\log (pn)\}^{4/\gamma_2}n^{-f(\alpha_0)/2}]+O_p\{sS_n(n^{-1}\log p)^{1/2}\}.
\end{split}
\]
Analogously, we can prove the same result for $|\sum_{k=-n+1}^{-1}\mathcal{K}({k}/{S_n})(\widehat{\bGamma}_k-\bGamma_k)|_\infty$. Therefore,
$
|\widehat{\bXi}-\widetilde{\bXi}|_\infty=O_p[\{\log (pn)\}^{4/\gamma_2}n^{-f(\alpha_0)/2}]+O_p\{sS_n(n^{-1}\log p)^{1/2}\}.
$
Repeating the the proof of Proposition 1(b) in \cite{Andrews_1991}, we know the convergence in Proposition 1(b) is uniformly for each component of $\widetilde{\bXi}-\bXi$. Thus, $|\widetilde{\bXi}-\bXi|_\infty=o(1)$. Then
$
|\widehat{\bXi}-{\bXi}|_\infty=o_p(1).
$
Similar to (\ref{eq:2}), we complete the proof. $\hfill\Box$

\bigskip

\noindent {\bf Proof of Corollary \ref{cy:1}:} From Theorem \ref{tm:2}, it holds that $\mathbb{P}_{H_0}(\bc\in\mathcal{C}_{\mathcal{S},1-\alpha,1})\rightarrow1-\alpha$. Therefore, $\mathbb{P}_{H_0}(\Psi_{\alpha}=1)=\mathbb{P}_{H_0}(\bc\notin\mathcal{C}_{\mathcal{S},1-\alpha,1})\rightarrow\alpha$ which establishes part (i). For part (ii), the following standard results on Gaussian maximum hold:
\[
\mathbb{E}\big(|\widehat{\bxi}|_\infty | \mathcal{Y}_n\big)\leq\{1+(2\log p)^{-1}\}(2\log p)^{1/2}\max_{1\leq j\leq r}\widehat{w}_{j,j}^{1/2}
\]
and
\[
\mathbb{P}\big\{|\widehat{\bxi}|_\infty\geq \mathbb{E}\big(|\widehat{\bxi}|_\infty|\mathcal{Y}_n\big)+u | \mathcal{Y}_n\big\}\leq \exp\bigg(-\frac{u^2}{2\max_{1\leq j\leq p}\widehat{w}_{j,j}}\bigg)
\]
for any $u>0$. Then,
$
\widehat{q}_{\mathcal{S},1-\alpha,1}\leq[\{1+(2\log p)^{-1}\}(2\log p)^{1/2}+\{2\log(1/\alpha)\}^{1/2}]\max_{1\leq j\leq r}\widehat{w}_{j,j}^{1/2}.
$
 Let $\mathscr{T}_\varepsilon=\{\max_{1\leq j\leq r}|\widehat{w}_{j,j}^{1/2}-w_{j,j}^{1/2}|/w_{j,j}^{1/2}\leq \varepsilon\}$ for some $\varepsilon>0$. Restricted on $\mathscr{T}_\varepsilon$,
$
\widehat{q}_{\mathcal{S},1-\alpha,1}\leq(1+\varepsilon)[\{1+(2\log p)^{-1}\}(2\log p)^{1/2}+\{2\log(1/\alpha)\}^{1/2}]\max_{1\leq j\leq r}{w}_{j,j}^{1/2}.
$
Let $(\tilde{j}_{1},\tilde{j}_{2})=\arg\max_{(j_1,j_2)\in\mathcal{S}}|\omega_{j_1,j_2}-c_{j_1,j_2}|$. Without lose of generality, we assume $\omega_{\tilde{j}_{1},\tilde{j}_{2}}-c_{\tilde{j}_{1},\tilde{j}_{2}}>0$. Therefore,
\[
\begin{split}
\mathbb{P}_{H_1}(\Psi_\alpha=1)=&~\mathbb{P}_{H_1}\bigg\{\max_{(j_1,j_2)\in\mathcal{S}}{n}^{1/2}|\widehat{\omega}_{j_1,j_2}-c_{j_1,j_2}|>\widehat{q}_{\mathcal{S},1-\alpha,1}\bigg\}\\
\geq&~\mathbb{P}_{H_1}\Big\{{n}^{1/2}(\widehat{\omega}_{\tilde{j}_1,\tilde{j}_2}-c_{\tilde{j}_1,\tilde{j}_2})>\widehat{q}_{\mathcal{S},1-\alpha,1}\Big\}\\
=&~1-\mathbb{P}_{H_1}\Big\{{n}^{1/2}(\widehat{\omega}_{\tilde{j}_1,\tilde{j}_2}-c_{\tilde{j}_1,\tilde{j}_2})\leq\widehat{q}_{\mathcal{S},1-\alpha,1},~\mathscr{T}_\varepsilon\Big\}\\
&~-\mathbb{P}(\mathscr{T}_\varepsilon^c).
\end{split}
\]
Restricted on $\mathscr{T}_\varepsilon$, if $\varepsilon\rightarrow0$, it holds that
$
\widehat{q}_{\mathcal{S},1-\alpha,1}-(\omega_{\tilde{j}_1,\tilde{j}_2}-c_{\tilde{j}_1,\tilde{j}_2})\leq-C(\log p)^{1/2}\max_{1\leq j\leq r}w_{j,j}^{1/2}
$
for some $C>0$, which implies
\[
\begin{split}
&~\mathbb{P}_{H_1}\Big\{{n}^{1/2}(\widehat{\omega}_{\tilde{j}_1,\tilde{j}_2}-c_{\tilde{j}_1,\tilde{j}_2})\leq\widehat{q}_{\mathcal{S},1-\alpha,1},~\mathscr{T}_\varepsilon\Big\}\\
\leq&~ \mathbb{P}_{H_1}\Big\{{n}^{1/2}(\widehat{\omega}_{\tilde{j}_1,\tilde{j}_2}-\omega_{\tilde{j}_1,\tilde{j}_2})\leq -C(\log p)^{1/2}\max_{1\leq j\leq r}w_{j,j}^{1/2}\Big\}\\
\rightarrow&~0.
\end{split}
\]
From Lemma \ref{la:4}, we know that $\max_{1\leq j\leq r}|\widehat{w}_{j,j}-w_{j,j}|=o_p(1)$ which also implies that $\max_{1\leq j\leq r}|\widehat{w}_{j,j}^{1/2}-w_{j,j}^{1/2}|/w_{j,j}^{1/2}=o_p(1)$. Then we can choose suitable $\varepsilon\rightarrow0$ such that $\mathbb{P}(\mathscr{T}_\varepsilon^c)\rightarrow0$. Hence, we complete part (ii). $\hfill\Box$

\noindent {\bf Proof of Corollary \ref{cy:2}:} Our proof includes two steps: (i) to show $\mathbb{P}(\widehat{\mathcal{M}}_{n,\alpha}\subset\mathcal{M}_0)\rightarrow1$, and (ii) to show $\mathbb{P}(\mathcal{M}_0\subset\widehat{\mathcal{M}}_{n,\alpha})\rightarrow1$. Result (i) is equivalent to $\mathbb{P}(\mathcal{M}_0^c\subset\widehat{\mathcal{M}}_{n,\alpha}^c)\rightarrow1$. The latter one is equivalent to $\mathbb{P}\{\max_{(j_1,j_2)\in\mathcal{M}_0^c}{n}^{1/2}|\widehat{\omega}_{j_1,j_2}|\geq \widehat{q}_{\mathcal{S},1-\alpha,1}\}\rightarrow0$. Notice that $\mathcal{S}=\{1,\ldots,p\}^2$, it holds that
\[
\begin{split}
&~\mathbb{P}\bigg\{\max_{(j_1,j_2)\in\mathcal{M}_0^c}{n}^{1/2}|\widehat{\omega}_{j_1,j_2}|\geq\widehat{q}_{\mathcal{S},1-\alpha,1}\bigg\}\\
\leq&~\mathbb{P}\bigg\{\max_{(j_1,j_2)\in\mathcal{S}}{n}^{1/2}|\widehat{\omega}_{j_1,j_2}-\omega_{j_1,j_2}|\geq\widehat{q}_{\mathcal{S},1-\alpha,1}\bigg\}\\
\leq&~\alpha+o(1),
\end{split}
\]
which implies $\mathbb{P}\{\max_{(j_1,j_2)\in\mathcal{M}_0^c}{n}^{1/2}|\widehat{\omega}_{j_1,j_2}|\geq\widehat{q}_{\mathcal{S},1-\alpha,1}\}\rightarrow0$. Then we construct result (i). Result (ii) is equivalent to $\mathbb{P}\{\min_{(j_1,j_2)\in\mathcal{M}_0}{n}^{1/2}|\widehat{\omega}_{j_1,j_2}|\leq \widehat{q}_{\mathcal{S},1-\alpha,1}\}\rightarrow0$. Let $(\tilde{j}_1,\tilde{j}_2)=\arg\min_{(j_1,j_2)\in\mathcal{M}_0}|\omega_{j_1,j_2}|$. Without lose of generality, we assume $\omega_{\tilde{j}_1,\tilde{j}_2}>0$. Notice that
\[
\begin{split}
&~\mathbb{P}\bigg\{\min_{(j_1,j_2)\in\mathcal{M}_0}{n}^{1/2}|\widehat{\omega}_{j_1,j_2}|\leq \widehat{q}_{\mathcal{S},1-\alpha,1}\bigg\}\\
\leq&~\mathbb{P}\big\{{n}^{1/2}(\widehat{\omega}_{\tilde{j}_1,\tilde{j}_2}-\omega_{\tilde{j}_1,\tilde{j}_2})\leq \widehat{q}_{\mathcal{S},1-\alpha,1}-{n}^{1/2}\omega_{\tilde{j}_1,\tilde{j}_2}\big\},
\end{split}
\]
we can construct result (ii) following the arguments for the proof of Corollary \ref{cy:1}. $\hfill\Box$

\newpage

\begin{sidewaystable}\footnotesize
\begin{center}
\caption{Averages of empirical coverages and their standard deviations (in parentheses) for $p = 100$.}
\label{tb:p100}
\begin{tabular}{c|c|c|c|c|c|c|c|c|c|c}    \hline\hline
\multirow{3}{*} {\parbox{1.6cm}{Covariance Structure}} & &  & \multicolumn{4}{c|}{$n=150$}   & \multicolumn{4}{c}{$n=300$} \\
\cline{4-11}
 & $\rho$& $1-\alpha$  & \multicolumn{2}{c|}{$\mathcal{S} = \{(j_1,j_2): \omega_{j_1,j_2} = 0\}$} & \multicolumn{2}{c|}{$\mathcal{S} = \{(j_1,j_2): j_1 \neq j_2\}$}  & \multicolumn{2}{c|}{$\mathcal{S} = \{(j_1,j_2): \omega_{j_1,j_2} = 0\}$} & \multicolumn{2}{c}{$\mathcal{S} = \{(j_1,j_2): j_1 \neq j_2\}$}\\ \cline{4-10}
 \cline{5-11}
 &  &   & KMB  & SKMB  & KMB & SKMB & KMB & SKMB & KMB & SKMB\\
  \cline{1-11}
\multirow{6}{*}{A}&  & 0.925 & 0.963(0.013) & 0.919(0.005) & 0.885(0.022) & 0.906(0.007) & 0.954(0.011) & 0.939(0.004) & 0.915(0.014) & 0.937(0.004) \\
& 0 & 0.950 & 0.978(0.008) & 0.949(0.007) & 0.913(0.016) & 0.941(0.007) & 0.972(0.007) & 0.956(0.002) & 0.941(0.011) & 0.954(0.002) \\
&   & 0.975 & 0.991(0.004) & 0.978(0.003) & 0.950(0.014) & 0.976(0.003) & 0.985(0.003) & 0.982(0.002) & 0.963(0.006) & 0.981(0.002) \\
 \cline{2-11}
&   & 0.925 & 0.950(0.014) & 0.888(0.014) & 0.835(0.029) & 0.875(0.014) & 0.955(0.012) & 0.920(0.009) & 0.890(0.019) & 0.916(0.010) \\
&0.3& 0.950 & 0.967(0.010) & 0.930(0.008) & 0.876(0.025) & 0.920(0.009) & 0.973(0.007) & 0.956(0.005) & 0.924(0.013) & 0.952(0.006) \\
&   & 0.975 & 0.987(0.006) & 0.966(0.004) & 0.923(0.017) & 0.958(0.005) & 0.987(0.004) & 0.979(0.003) & 0.956(0.010) & 0.978(0.003) \\
  \cline{1-11}
\multirow{6}{*}{B}&  & 0.925 & 0.953(0.016) & 0.927(0.005) & 0.812(0.036) & 0.874(0.008) & 0.950(0.009) & 0.931(0.003) & 0.894(0.014) & 0.917(0.004) \\
& 0 & 0.950 & 0.973(0.010) & 0.957(0.005) & 0.863(0.028) & 0.918(0.007) & 0.969(0.008) & 0.956(0.006) & 0.925(0.013) & 0.947(0.007) \\
&   & 0.975 & 0.986(0.004) & 0.979(0.002) & 0.918(0.020) & 0.965(0.004) & 0.989(0.004) & 0.981(0.004) & 0.961(0.008) & 0.978(0.004) \\
 \cline{2-11}
&   & 0.925 & 0.950(0.019) & 0.898(0.011) & 0.772(0.039) & 0.815(0.018) & 0.950(0.011) & 0.933(0.005) & 0.880(0.017) & 0.915(0.007) \\
&0.3& 0.950 & 0.971(0.011) & 0.930(0.007) & 0.826(0.031) & 0.873(0.012) & 0.968(0.007) & 0.956(0.003) & 0.913(0.012) & 0.943(0.004) \\
&   & 0.975 & 0.987(0.004) & 0.970(0.005) & 0.885(0.021) & 0.938(0.008) & 0.985(0.004) & 0.972(0.003) & 0.943(0.010) & 0.964(0.004) \\
 \cline{1-11}
\hline
\hline
\end{tabular}

\end{center}
\end{sidewaystable}

\begin{sidewaystable}\footnotesize
\begin{center}
\label{tb:p500}
\caption{Averages of empirical coverages and their standard deviations (in parentheses) for $p = 500$.}
\begin{tabular}{c|c|c|c|c|c|c|c|c|c|c}    \hline\hline
\multirow{3}{*} {\parbox{1.6cm}{Covariance Structure}}&  & & \multicolumn{4}{c|}{$n=150$} & \multicolumn{4}{c}{$n=300$}\\ \cline{4-11}
 & $\rho$ & $1-\alpha$  & \multicolumn{2}{c|}{$\mathcal{S} = \{(j_1,j_2): \omega_{j_1,j_2} = 0\}$} & \multicolumn{2}{c|}{$\mathcal{S} = \{(j_1,j_2): j_1 \neq j_2\}$} & \multicolumn{2}{c|}{$\mathcal{S} = \{(j_1,j_2): \omega_{j_1,j_2} = 0\}$} & \multicolumn{2}{c}{$\mathcal{S} = \{(j_1,j_2): j_1 \neq j_2\}$}\\
 \cline{4-11}
 \cline{4-11}
&  & & KMB  & SKMB  & KMB  & SKMB  & KMB  & SKMB  & KMB  & SKMB \\
  \cline{1-11}
\multirow{6}{*}{A}&  & 0.925 & 0.967(0.006) & 0.891(0.010) & 0.872(0.017) & 0.873(0.009) & 0.971(0.003) & 0.935(0.003) & 0.924(0.008) & 0.799(0.003) \\
& 0 & 0.950 & 0.978(0.004) & 0.934(0.007) & 0.903(0.011) & 0.923(0.009) & 0.977(0.002) & 0.954(0.002) & 0.939(0.006) & 0.822(0.004) \\
&   & 0.975 & 0.987(0.003) & 0.975(0.003) & 0.933(0.012) & 0.968(0.004) & 0.983(0.002) & 0.977(0.002) & 0.956(0.004) & 0.856(0.004) \\
 \cline{2-11}
&   & 0.925 & 0.961(0.010) & 0.871(0.010) & 0.786(0.027) & 0.833(0.011) & 0.973(0.004) & 0.937(0.005) & 0.867(0.011) & 0.905(0.007) \\
&0.3& 0.950 & 0.979(0.006) & 0.918(0.010) & 0.842(0.021) & 0.890(0.011) & 0.982(0.004) & 0.959(0.003) & 0.899(0.011) & 0.934(0.004) \\
&   & 0.975 & 0.991(0.004) & 0.966(0.005) & 0.890(0.014) & 0.949(0.006) & 0.991(0.001) & 0.973(0.003) & 0.936(0.007) & 0.950(0.003) \\
 \cline{1-11}
\multirow{6}{*}{B}&  & 0.925 & 0.961(0.007) & 0.884(0.009) & 0.713(0.027) & 0.746(0.015) & 0.966(0.006) & 0.921(0.005) & 0.884(0.011) & 0.814(0.007) \\
& 0 & 0.950 & 0.974(0.004) & 0.934(0.008) & 0.780(0.030) & 0.831(0.015) & 0.980(0.003) & 0.938(0.004) & 0.915(0.009) & 0.840(0.006) \\
&   & 0.975 & 0.985(0.003) & 0.974(0.004) & 0.869(0.019) & 0.912(0.010) & 0.988(0.002) & 0.970(0.003) & 0.952(0.006) & 0.887(0.007) \\
 \cline{2-11}
&   & 0.925 & 0.954(0.007) & 0.856(0.014) & 0.641(0.034) & 0.673(0.019) & 0.964(0.005) & 0.928(0.004) & 0.853(0.016) & 0.850(0.007) \\
&0.3& 0.950 & 0.968(0.006) & 0.908(0.008) & 0.716(0.036) & 0.767(0.018) & 0.979(0.004) & 0.950(0.003) & 0.900(0.012) & 0.889(0.006) \\
&   & 0.975 & 0.983(0.004) & 0.954(0.005) & 0.821(0.028) & 0.878(0.014) & 0.988(0.002) & 0.971(0.002) & 0.940(0.010) & 0.925(0.005) \\
\hline \hline
\end{tabular}
\end{center}
\end{sidewaystable}

\begin{sidewaystable}\footnotesize
\begin{center}
\caption{Averages of empirical coverages and their standard deviations (in parentheses) for $p = 1500$.}
\label{tb:p1500}
\begin{tabular}{c|c|c|c|c|c|c|c|c|c|c}    \hline\hline
\multirow{3}{*} {\parbox{1.6cm}{Covariance Structure}}&  & & \multicolumn{4}{c|}{$n=150$} & \multicolumn{4}{c}{$n=300$}\\ \cline{4-11}
 & $\rho$ & $1-\alpha$  & \multicolumn{2}{c|}{$\mathcal{S} = \{(j_1,j_2): \omega_{j_1,j_2} = 0\}$} & \multicolumn{2}{c|}{$\mathcal{S} = \{(j_1,j_2): j_1 \neq j_2\}$} & \multicolumn{2}{c|}{$\mathcal{S} = \{(j_1,j_2): \omega_{j_1,j_2} = 0\}$} & \multicolumn{2}{c}{$\mathcal{S} = \{(j_1,j_2): j_1 \neq j_2\}$}\\
 \cline{4-11}
 \cline{4-11}
&  & & KMB  & SKMB  & KMB  & SKMB  & KMB  & SKMB  & KMB  & SKMB \\
  \cline{1-11}
\multirow{6}{*}{A}&  & 0.925 & 0.976(0.005) & 0.854(0.013) & 0.826(0.017) & 0.834(0.013) & 0.979(0.002) & 0.959(0.003) & 0.913(0.009) & 0.948(0.004) \\
& 0 & 0.950 & 0.987(0.003) & 0.908(0.010) & 0.866(0.011) & 0.892(0.010) & 0.991(0.002) & 0.974(0.001) & 0.945(0.007) & 0.963(0.001) \\
&   & 0.975 & 0.991(0.002) & 0.954(0.005) & 0.903(0.009) & 0.944(0.006) & 0.997(0.001) & 0.987(0.003) & 0.967(0.003) & 0.979(0.003) \\
 \cline{2-11}
&   & 0.925 & 0.967(0.010) & 0.823(0.013) & 0.674(0.031) & 0.758(0.016) & 0.981(0.002) & 0.951(0.004) & 0.822(0.011) & 0.933(0.004) \\
&0.3& 0.950 & 0.983(0.004) & 0.887(0.011) & 0.754(0.030) & 0.840(0.012) & 0.987(0.002) & 0.972(0.004) & 0.861(0.012) & 0.958(0.005) \\
&   & 0.975 & 0.994(0.002) & 0.952(0.010) & 0.841(0.019) & 0.922(0.011) & 0.996(0.001) & 0.988(0.002) & 0.926(0.010) & 0.978(0.003) \\
 \cline{1-11}
\multirow{6}{*}{B}&  & 0.925 & 0.964(0.008) & 0.852(0.013) & 0.638(0.031) & 0.631(0.019) & 0.973(0.004) & 0.944(0.005) & 0.882(0.010) & 0.912(0.006) \\
& 0 & 0.950 & 0.981(0.004) & 0.915(0.008) & 0.729(0.031) & 0.738(0.021) & 0.987(0.003) & 0.967(0.003) & 0.915(0.009) & 0.946(0.004) \\
&   & 0.975 & 0.991(0.002) & 0.961(0.007) & 0.831(0.017) & 0.860(0.015) & 0.995(0.001) & 0.984(0.001) & 0.952(0.006) & 0.968(0.003) \\
 \cline{2-11}
&   & 0.925 & 0.958(0.008) & 0.781(0.025) & 0.528(0.047) & 0.417(0.031) & 0.978(0.003) & 0.930(0.006) & 0.813(0.015) & 0.867(0.010) \\
&0.3& 0.950 & 0.977(0.006) & 0.870(0.009) & 0.643(0.040) & 0.564(0.023) & 0.985(0.002) & 0.956(0.005) & 0.866(0.013) & 0.912(0.009) \\
&   & 0.975 & 0.989(0.002) & 0.939(0.007) & 0.787(0.031) & 0.737(0.023) & 0.997(0.001) & 0.980(0.002) & 0.932(0.011) & 0.954(0.005) \\
\hline \hline
\end{tabular}
\end{center}
\end{sidewaystable}

\begin{sidewaystable}[ht]\scriptsize
\caption{Sectors and sub industries of the 402 S\&P 500 stocks. ``NA'' represents that the sector or sub industry of the corresponding stock cannot be identified due to acquisition or ticket change.}
\label{tb:stock1}
\centering
\begin{tabular}{llclc}
  \hline
  \hline
 Stock Symbols & Sectors & Sector No. & Sub Industries & Industry No. \\
  \hline
 IPG & Consumer Discretionary &   1 & Advertising &   1 \\
 \hline
ANF, COH, NKE, TIF, VFC & Consumer Discretionary &   1 & Apparel, Accessories \& Luxury Goods &   2 \\
   \hline
  F, HOG, JCI & Consumer Discretionary &   1 & Auto Parts \& Equipment &   3 \\
   \hline
   CBS, CMCSA, DIS, DTV, TWC, TWX & Consumer Discretionary &   1 & Broadcasting \& Cable TV &   4 \\
   \hline
   IGT, WYNN & Consumer Discretionary &   1 & Casinos \& Gaming &   5 \\
   \hline
 JCP, JWN, KSS, M & Consumer Discretionary &   1 & Department Stores &   6 \\
   \hline
   APOL, DV & Consumer Discretionary &   1 & Educational Services &   7 \\
   \hline
   DHI, KBH, LEN, LOW, PHM & Consumer Discretionary &   1 & Homebuilding &   8 \\
   \hline
   EXPE, HOT, MAR, WYN & Consumer Discretionary &   1 & Hotels, Resorts \& Cruise Lines &   9 \\
   \hline
   BDK, NWL, SNA, SWK, WHR & Consumer Discretionary &   1 & Household Appliances &  10 \\
   \hline
   AMZN & Consumer Discretionary &   1 & Internet Retail &  11 \\
   \hline
   HAS, MAT, ODP, RRD & Consumer Discretionary &   1 & Printing Services &  12 \\
   \hline
   GCI, MDP, NYT & Consumer Discretionary &   1 & Publishing &  13 \\
   \hline
   DRI, SBUX, YUM & Consumer Discretionary &   1 & Restaurants &  14 \\
   \hline
   AN, AZO, BBBY, GPC, GPS, HAR, LTD, SPLS & Consumer Discretionary &   1 & Specialty Stores &  15 \\
   \hline
   FPL, WPO & Consumer Discretionary &   1 & NA & NA \\
   \hline
   ADM & Consumer Staples &   2 & Agricultural Products &  16 \\
   \hline
   CVS, SVU, SWY, WAG & Consumer Staples &   2 & Food \& Drug Stores &  17 \\
   \hline
   AVP, CL, KMB & Consumer Staples &   2 & Household Products &  18 \\
   \hline
  TGT, FDO, WMT & Consumer Staples &   2 & Hypermarkets \& Super Centers &  19 \\
   \hline
   CAG, CCE, CPB, DF, GIS, HNZ, HRL, HSY, K, KFT,& \multirow{2}{*}{Consumer Staples} &   \multirow{2}{*}{2} & \multirow{2}{*}{Packaged Food} &  \multirow{2}{*}{20} \\
       KO, MKC, PBG, PEP, SJM, SLE, STZ, TAP, TSN  &  &   &  &   \\
     \hline
   EL, PG & Consumer Staples &   2 & Personal Products &  21 \\
   \hline
   MO, RAI & Consumer Staples &   2 & Tobacco &  22 \\
   \hline
   BTU, CNX, MEE & Energy &   3 & Coal Operations &  23 \\
   \hline
   APA, CHK, COG, COP, CTX, CVX, DNR, DO, DVN, & \multirow{3}{*}{Energy} &   \multirow{3}{*}{3} & \multirow{3}{*}{Oil \& Gas Exploration \& Production} &  \multirow{3}{*}{24} \\
   EOG, EP, EQT, ESV, FO, HES, MRO, MUR, NBL, &  &    &   &   \\
    OXY, PXD, RRC, SE, SWN, TSO, VLO, WMB, XTO &  &    &    &   \\
     \hline
  BHI, BJS, CAM, FTI, NBR, NOV, RDC, SII, SLB & Energy &   3 & Oil \& Gas Equipment \& Services &  25 \\
   \hline

 BAC, BBT, BK, C, CIT, CMA, COF, FHN, FITB, & \multirow{3}{*}{Financials} &   \multirow{3}{*}{4} & \multirow{3}{*}{Banks} &  \multirow{3}{*}{26} \\

 HCBK, HRB, IVZ, KEY, LM, MI, MTB, NTRS, PNC,&  &    & &   \\

SLM, STI, USB, WFC &  &    &  &   \\
  \hline
  CME, EFX, ICE, NYX, PFG, PRU, RF, STT, TROW, & \multirow{2}{*}{Financials} &   \multirow{2}{*}{4} & \multirow{2}{*}{Diversified Financial Services} &  \multirow{2}{*}{27} \\

  UNM, VTR &  &    &  &   \\
    \hline

   ETFC, FII, JNS, LUK, MS, SCHW & Financials &   4 & Investment Banking \& Brokerage &  28 \\
    \hline
  AFL, AIG, AIZ, CB, CINF, GNW, HIG, L, LNC, & \multirow{2}{*}{Financials} &   \multirow{2}{*}{4} & \multirow{2}{*}{Property \& Casualty Insurance} &  \multirow{2}{*}{29} \\

   MBI, MET, MMC, PGR, TMK, TRV, XL &  &    &  &   \\

 \hline

  AMT, AVB, BXP, CBG, HCN, HCP, HST, IRM,& \multirow{2}{*}{Financials} &   \multirow{2}{*}{4} & \multirow{2}{*}{REITs} &  \multirow{2}{*}{30} \\

    KIM, PBCT, PCL, PSA, SPG, VNO, WY &  &    &  &   \\

 \hline
   \hline
\end{tabular}
\end{sidewaystable}

\begin{sidewaystable}[ht]\scriptsize
\caption{Sectors and sub industries of the 402 S\&P 500 stocks (continued). ``NA'' represents that the sector or sub industry of the corresponding stock cannot be identified due to acquisition or ticket change.}
\label{tb:stock2}
\centering
\begin{tabular}{llclc}
  \hline
  \hline
 Stock Symbols & Sectors & Sector No. & Sub Industries & Industry No. \\

      \hline
 AOC & Financials &   4 & NA & NA \\
      \hline
  AMGN, BIIB, CELG, FRX, GENZ, GILD, HSP,& \multirow{2}{*}{Health Care} &   \multirow{2}{*}{5} & \multirow{2}{*}{Pharmaceuticals} &  \multirow{2}{*}{31} \\

 KG, LIFE, LLY, MRK, MYL, WPI &  &    &  &   \\

      \hline

 ABC, AET, BMY, CAH, CI, DGX, DVA, ESRX, HUM,& \multirow{2}{*}{Health Care} &   \multirow{2}{*}{5} & \multirow{2}{*}{Health Care Supplies} &  \multirow{2}{*}{32} \\

 MCK, MHS, PDCO, THC, UNH, WAT, WLP, XRAY &  &    &  &   \\

      \hline

 ABT, BAX, BCR, BDX, ISRG, JNJ, MDT, MIL, & \multirow{2}{*}{Health Care} &   \multirow{2}{*}{5} & \multirow{2}{*}{Health Care Equipment \& Services} &  \multirow{2}{*}{33} \\

  PKI, PLL, STJ, SYK, TMO, VAR & &  &  &   \\

      \hline

 BA, RTN & Industrials &   6 & Aerospace \& Defense &  34 \\
      \hline
 CHRW, EXPD, FDX, UPS & Industrials &   6 & Air Freight \& Logistics &  35 \\
      \hline
 LUV & Industrials &   6 & Airlines &  36 \\
      \hline
  DE, FAST, GLW, MAS, MTW, PCAR & Industrials &   6 & Construction \& Farm Machinery \& Heavy Trucks &  37 \\

      \hline
 COL, EMR, ETN, GE, HON, IR, JEC, LEG,& \multirow{2}{*}{Industrials} &   \multirow{2}{*}{6} & \multirow{2}{*}{Industrial Conglomerates} &  \multirow{2}{*}{38} \\

LLL, MMM, PH, ROK, RSG, TXT, TYC &  &   &  &   \\
      \hline
 CMI, DHR, DOV, FLS, GWW, ITT, ITW & Industrials &   6 & Industrial Machinery &  39 \\
      \hline
 CSX, NSC, UNP & Industrials &   6 & Railroads &  40 \\
      \hline
 ACS, CTAS, FLR, RHI & Industrials &   6 & NA  & NA  \\
      \hline
 CBE, MOLX, JBL, LXK & Information Technology &   7 & Office Electronics &  41 \\
      \hline

  ADBE, ADSK, BMC, CA, ERTS, MFE, MSFT, NOVL,& \multirow{2}{*}{Information Technology} &   \multirow{2}{*}{7} & \multirow{2}{*}{Application Software} &  \multirow{2}{*}{42} \\

 ORCL, TDC &  &    &  &   \\

      \hline
 CIEN, HRS, JDSU, JNPR, MOT & Information Technology &   7 & Communications Equipment &  43 \\
      \hline
  AAPL, AMD, HPQ, JAVA, QLGC, SNDK & Information Technology &   7 & Computer Storage \& Peripherals &  44 \\
      \hline

 ADP, AKAM, CRM, CSC, CTSH, CTXS, CVG, EBAY,& \multirow{3}{*}{Information Technology} &   \multirow{3}{*}{7} & \multirow{3}{*}{Information Services} &  \multirow{3}{*}{45} \\

FIS, GOOG, IBM, INTU, MA, MWW, PAYX, TSS,&  &    &  &   \\

  XRX, YHOO, DNB &  &    &  &   \\
      \hline

 ALTR, AMAT, BRCM, INTC, KLAC, LLTC, LSI, MCHP,& \multirow{2}{*}{Information Technology} &   \multirow{2}{*}{7} & \multirow{2}{*}{Semiconductors} &  \multirow{2}{*}{46} \\

 MU, NSM, NVDA, NVLS, QCOM, TXN, XLNX &  &    &  &   \\

      \hline

 ATI, BLL, FCX, NEM, OI & Materials &   8 & Metal \& Glass Containers &  47 \\
      \hline
 DD, DOW, ECL, EMN, IFF, MON, PPG, PX, SHW, SIAL & Materials &   8 & Specialty Chemicals &  48 \\
      \hline
 BMS, MWV, PTV & Materials &   8 & Containers \& Packaging &  49 \\
      \hline
 AKS, TIE, X & Materials &   8 & Iron \& Steel &  50 \\
      \hline
 AVY, IP, SEE & Materials &   8 & Paper Packaging &  51 \\
      \hline
  VMC & Materials &   8 &  NA& NA \\
      \hline
 CTL, EQ, FTR, Q, S, T, VZ, WIN & Telecommunications Services &   9 & Telecom Carriers &  52 \\
      \hline
 AEE, AEP, AES, AYE, CMS, CNP, D, DYN, ETR,& \multirow{2}{*}{Utilities} &  \multirow{2}{*}{10} & \multirow{2}{*}{MultiUtilities} &  \multirow{2}{*}{53} \\

    FE, PEG, POM, PPL, SCG, SO, SRE, TE, WEC, XEL & &   &  &   \\

       \hline
   STR, TEG & Utilities &  10 & Utility Networks &  54 \\
      \hline
 RX &  NA& NA & NA & NA \\
   \hline
   \hline
\end{tabular}
\end{sidewaystable}

\begin{table}[ht]
\caption{The numbers of edges within and between sectors for the partial correlation networks of the S\&P 500 sub industries in Figures \ref{fg:2005} and \ref{fg:2008}.}
\label{tb:degree}
\centering
\begin{tabular}{l|cc|cc}
  \hline
  \hline
\multirow{2}{*}{Sectors}  & \multicolumn{2}{c|}{2005} &\multicolumn{2}{c}{2008} \\
\cline{2-3}
\cline{4-5}

 & Within & Between & Within & Between \\
  \hline
Consumer Discretionary & 13 & 37 & 9 & 12 \\
Consumer Staples & 4 & 16 & 1 & 6 \\
Energy & 0 & 8 & 1 & 4 \\
  Financials & 3 & 14 & 5 & 5 \\
Health Care & 2 & 10 & 2 & 8 \\
  Industrials & 5 & 19 & 3 & 5 \\
  Information Technology & 5 & 13 & 6 & 9 \\
  Materials & 2 & 12 & 2 & 10 \\
  Telecommunication Services & 0 & 3 & 0 & 1 \\
  Utilities & 0 & 4 & 1 & 2 \\
   \hline
   \hline
\end{tabular}
\end{table}

\end{document}